%% file: main-arxiv.tex
\documentclass[acmtog,screen=true]{acmartArxiv}

\usepackage{booktabs} %
\usepackage{arydshln} %
\usepackage{mdframed}
\usepackage{tikz}
\usepackage{pgfplots}
\usepgfplotslibrary{fillbetween}
\usepackage{subfigure}
\usepackage[most]{tcolorbox}
\usepackage{wrapfig}
\usepackage{lipsum}
\usepackage{overpic}
\usepackage{multirow}
\usepackage{tikz}
\usepackage{listings}
\usepackage[normalem]{ulem}
\usetikzlibrary{decorations.pathreplacing}
\usepackage{todonotes}
\usepackage[ruled]{algorithm2e} %

\SetAlFnt{\small}
\SetAlCapFnt{\small}
\SetAlCapNameFnt{\small}
\SetAlCapHSkip{0pt}

\usepackage{xr}
\newcommand{\epsi}{\varepsilon}

\author{Bastien Doignies}
\affiliation{%
  \institution{Univ. Lyon}
  \country{France}
 }
\author{Nicolas Bonneel}
\orcid{0000-0001-5243-4810}
\affiliation{%
  \institution{CNRS, Univ. Lyon}
  \country{France}}
\author{David Coeurjolly}
\orcid{0000-0003-3164-8697}
\affiliation{%
  \institution{CNRS, Univ. Lyon}
  \country{France}}
\author{Julie Digne}
\orcid{0000-0003-0905-0840}
\affiliation{%
  \institution{CNRS, Univ. Lyon}
  \country{France}}
\author{Loïs Paulin}
\orcid{0000-0001-9170-8407}
\affiliation{%
  \institution{Univ. Lyon}
  \country{France}}
\author{Jean-Claude Iehl}
\orcid{0000-0001-6877-2398}
\affiliation{%
  \institution{Univ. Lyon}
  \country{France}}
\author{Victor Ostromoukhov}
\affiliation{%
  \institution{Univ. Lyon}
  \country{France}}

\input{macros}

\definecolor{HighlightColor}{rgb}{0.462745098, 0.725490196, 0.000000000}
\definecolor{HighlightColor2}{rgb}{0.000000000, 0.584313725, 0.466666667}
\definecolor{HighlightColor3}{rgb}{0.447058824, 0.168627451, 0.580392157}
\definecolor{HighlightColor4}{rgb}{0.043137255, 0.298039216, 0.788235294}
\definecolor{HighlightColor5}{rgb}{0.360784314, 0.360784314, 0.360784314}
\definecolor{IntelColor}{rgb}{0.000000000, 0.525490196, 0.815686275}
\definecolor{AMDColor}{rgb}{0.505882353, 0.505882353, 0.505882353}
\definecolor{AppleColor}{rgb}{0.752941176, 0.752941176, 0.752941176}
\definecolor{GoogleColor}{rgb}{0.894117647, 0.28627451, 0.270588235}
\definecolor{AmazonColor}{rgb}{1.000000000, 0.662745098, 0.000000000}
\definecolor{DarkGray}{rgb}{0.360784314, 0.360784314, 0.360784314}
\definecolor{MediumGray}{rgb}{0.505882353, 0.505882353, 0.505882353}
\definecolor{LightGray}{rgb}{0.752941176, 0.752941176, 0.752941176}
\definecolor{VeryLightGray}{rgb}{0.905882353, 0.905882353, 0.905882353}

\begin{document}

\title{Example-Based Sampling with Diffusion Models}

\begin{abstract}

Much effort has been put into developing samplers with specific properties, such as producing 
blue noise, low-discrepancy, lattice or Poisson disk samples. These samplers 
can be slow if they rely on optimization processes, may rely on a wide range 
of numerical methods, are not always differentiable.
 The success of recent diffusion models for image generation suggests that these models could be appropriate 
for learning how to generate point sets from examples. However, their convolutional nature makes these 
methods impractical for dealing with scattered data such as point sets.
We propose a generic way to produce 2-d point sets imitating existing samplers from observed point sets using a diffusion model. 
We address the problem of convolutional layers by leveraging neighborhood information from 
an optimal transport matching to a uniform grid, that 
allows us to benefit from fast convolutions on grids, and to support the example-based 
learning of non-uniform sampling patterns.
We demonstrate how the differentiability of our approach can be used to optimize 
point sets to enforce properties. 

\end{abstract}

\ccsdesc[300]{Computing methodologies~Rendering}
\ccsdesc[300]{Mathematics of computing~Quadrature}
\ccsdesc[300]{Computing methodologies~Neural networks}

\keywords{Path tracing, quasi-Monte Carlo integration, low discrepancy sequences, generator matrices, integer linear programming.}

\maketitle

\section{Introduction}

A wide range of samplers have been designed in the past, for
quasi-Monte Carlo integration, rendering, image stippling,
positionning objects or generally, to uniformly or non-uniformly cover
some space. The generated samples can have various properties, such as
being low discrepancy or stratified, having a blue noise spectrum,
producing low integration error, with high packing density, satisfying
a Poisson disk criterion, or high inter-point distances \cite{Pharr2016Physically,singh2019analysis}. Generating these
samples can come at significant cost, especially when points are
obtained from complex optimization schemes~\cite{Fattal:2011:Bluenoise,de2012blue,paulin2020sliced,GBN22,oztireli2012analysis,roveri2017general}. In addition, satisfying multiple properties at the same time is difficult, and is the focus of entire methods -- e.g., generating low discrepancy sequences with blue noise properties. Differentiability can also be desirable in contexts involving further optimizations, but may be problematic for specific samplers, for instance when considered in a differential renderer~\cite{Mitsuba3}. The large set of available samplers makes sample generation little generic, with methods involving smooth non-convex optimization, integer linear programming, number theory, bruteforce approaches with clever data structures, etc.

Recently, diffusion models have become extremely popular in the context of image generation~\cite{Sohl15,Ho20,Rombach22}. By learning how to denoise an image that initially only contains random values, these models have been able to produce impressive results, i.e., to learn the very fine structure of the manifold of realistic images. It hence seems judicious to take advantage of these models to learn the very fine structure of sample points produced by existing samplers. However, these models heavily rely on convolutions, which makes it impractical to efficiently handle point sets.

In this paper, we propose to learn the distribution of 2-d samples produced by a wide range of samplers using a diffusion model. When point sets are not stratified, we resort to an optimal transport matching to a uniform grid that mostly preserves neighborhood information so as to benefit from efficient convolutional layers. We demonstrate that a single architecture is able to learn sample points produced by different methods, and even allows to reproduce non-uniform point sets. The differentiability of our network allows us to add properties to a given samplers, e.g., allowing to add low discrepancy properties to a given optimal transport-based sampler. 
While our network is currently limited to generating 2-d samples, it produces samples beyond the range of samples count it has been trained on. We provide trained networks alongside the paper and believe this exciting step will open the door to further conditioning.
Code is provided in supplementary material.

\section{Related works}

Existing samplers have a wide range of properties. We enumerate importants classes of samplers below.

\paragraph{Blue Noise.} Blue noise samples have a characteristic
``ring-like'' Fourier power spectrum, with low frequencies converging to zero. 
They are interesting for Monte Carlo integration
purposes~\cite{Subr2013FASSSABVI,Pilleboue:2015:Variance}, digital halftoning~\cite{Ulichney:1987:Digital} or
stippling~\cite{Deussen00} and well describe arrangements of natural phenomenas
that have been optimized through evolution such as the retinal
distribution of cones~\cite{YellottJr19821205}. They are often costly obtained through
optimization, for instance using kernel
approaches~\cite{Fattal:2011:Bluenoise,GBN22}, pair-correlation function~\cite{oztireli2012analysis}
or optimal transport~\cite{de2012blue,Qin:2017:Wasserstein,paulin2020sliced},
though fast approximations exist~\cite{nader2018instant}. Tile-based
approaches pre-compute tiles for fast synthesis, but are
memory
demanding~\cite{Ostromoukhov:2004:FHI,Kopf2006RWTRTBN,wachtel2014fast}.

\paragraph{Poisson Disk.} Poisson disk samples have the property that no point fall within a distance smaller than a threshold from another point~\cite{yuksel2015sample,gamito2009accurate,wei2008parallel,bridson2007fast,dunbar2006spatial}. Their spectra resemble those of blue noise distributions, except that they do not decrease towards zero as the frequency decreases \cite{Pilleboue:2015:Variance}. They naturally occur in other natural process such as the placement of trees in a forest. In low dimensions, they are relatively fast to compute.

\paragraph{Low Discrepancy Sequences.} Discrepancy is a uniformity
measure directly related to Monte Carlo integration error. Low
discrepancy sequences (LDS) thus have several advantages. First they
are sequences, so that samples can be progressively added. Second,
they are low discrepancy, hence guaranteeing good numerical
integration error \cite{Niederreiter1992,Lemieux2009}. Samplers
achieving low discrepancy usually rely on arithmetic and number theory
constructions leading to extremely fast
generators (\emph{e.g.} in base 2, the $i$-th sample using
\cite{Sobol67} is given by a matrix/vector multiplication in $GF(2)$ on the
bitwise representation of $i$).
Alternatively, lattices produce low discrepancy sequences. A rank-1
lattice repeatedly translates an initial point by a given amount in a
given direction in a toric domain \cite{keller2004stratification}. Rank-n lattices similarly use multiple independent vectors. Good lattices can be similarly hard to optimize for~\cite{latnetbuilder}.

\paragraph{Designing Complex Point Processes.} Aside global point set
  properties such as blue-noise, Poisson disk or low discrepancy, the
  problem of designing a point process matching some exemplars or
  satisfying additional constraints has been addressed in several
  ways. One can design sampler mixing global properties such as low
  discrepancy and blue-noise
  \cite{Ahmed:2016:Lowdiscrepancy,DyadicOptimized,perrier2018sequences},
  we can use a profile based approach to generate LDS samplers 
  with adjustable or with scriptable properties (\emph{e.g.} blue-noise properties,
  stratification on some projections\ldots)
  \cite{matbuilder,latnetbuilder}. Mixing point process properties can
  also be achieved by interpolating their high order statistics such
  as their pair-correlation functions
  \cite{oztireli2012analysis}. Focusing on spectral properties,
  \cite{leimkuhler2019deep} have proposed a neural network approach to
  target specific profiles defined as combinations of radial power spectra.

\paragraph{Point sets through deep learning.}
Perhaps the closest to our work is that of~\cite{leimkuhler2019deep}. They learn arbitrary dimensional point sets by matching power spectra. There is a number of important differences with respect to our work. First, they require a power spectrum as input while we require examples from a given sampler. This allows us to capture all characteristics of samplers and not just spectra. Second, our network is able to produce point sets of significantly different sizes without re-training. Third, we propose a way to benefit from efficient convolutions on grids. While this restricts us to low-dimensional settings (we demonstrate our approach in two dimensions), this allows us to use thousands of convolution layers at different scales and to benefit from recent advances in diffusion models.  These differences allow us to finely capture the structure of point sets (see Sec.~\ref{sec:prop}).

In the context of Monte Carlo integration, deep learning has been used to learn a control variate~\cite{muller2020neural}, though this does not directly address the location of point samples. Deep learning has also been used for importance sampling~\cite{muller2019neural}. %

\paragraph{Probabilistic Denoising Diffusion} Our method is based on Probabilistic Denoising Diffusion, a concept introduced by \cite{Sohl15} in the context of unsupervised learning.
The core idea of Denoising diffusion is to gradually remove any structure in the image by progressively adding noise and to train a neural network to invert the degradation process. This allows to capture the data distribution and sample from it.
This idea has been extensively used for image synthesis~\cite{Ho20} with impressive results, either by working directly in pixel space or in the latent space~\cite{Rombach22}.
In this paper, we propose to exploit the capacity of these networks to learn structure from a set of examples to learn point distributions.

\section{Denoising Diffusion model}

\subsection{Architecture}
The denoising process involves a sequence of denoising operations which operate at given timesteps.
Each denoising is achieved by a forward pass in a single denoising network $\varepsilon_\theta$, which takes as input both the noisy image $\tilde{x}_t$ and the embedded timestep $t$.

Our network architecture is very similar to the one of \cite{Ho20}. It
corresponds to a U-Net~\cite{ronneberger2015u}, where each level is composed of two convolutional residual blocks (ResNet) and the feature maps are downsampled by a factor $2$ between each level.
While the original architecture only included attention blocks between
the two convolutional blocks of the $16 \times 16$ level, we add attention to
all levels, which we found to work better in practice. Unless specified otherwise, we used 1000 diffusion time steps.
The overall architecture design is detailed  in supplementary material.

The network learns a time-dependent noise model $\varepsilon_\theta(\tilde{x}_t, t)$ given a noise $\varepsilon_t$ added to the input data, $\tilde{x}_t = x_t + \varepsilon_t$ at each time step $t$. In our setting, $x_0$ is the offset between strata centers and the input point set as obtained in Sec~\ref{sec:convolution}. The network thus predicts noise, that can then be progressively removed from a white noise point set to denoise it according to the learned data distribution.

\subsection{Convolutions on grids}
\label{sec:convolution}
\begin{figure}[!tbh]
  \centering
   \includegraphics[width=5cm]{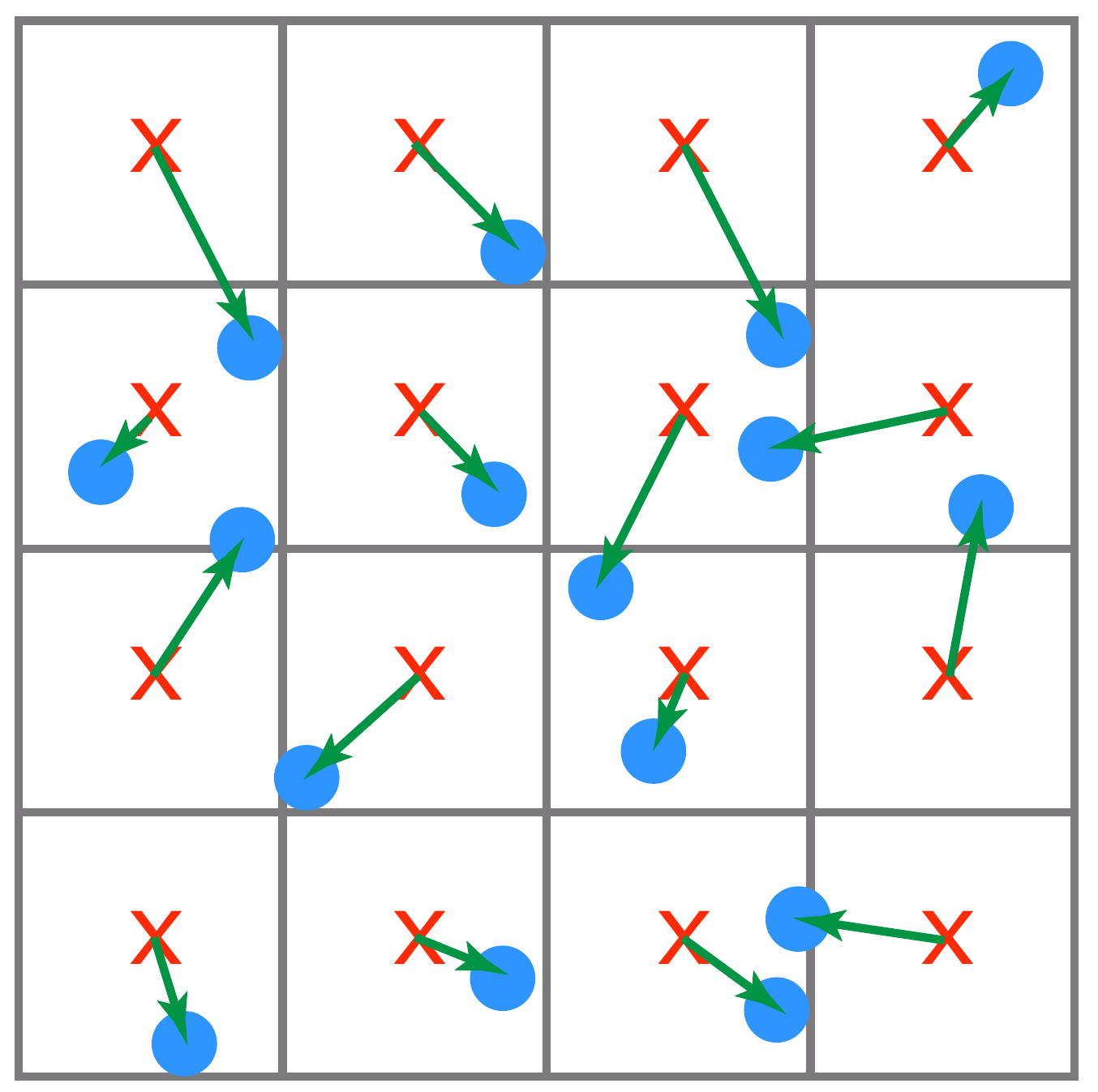}
   \caption{When input point sets are not stratified, we compute a linear assignment problem between strata centers (red) and sample points (blue) using optimal transport. Each stratum stores its assigned point offset (green arrows). The grid thus serves as an approximate nearest neighbor acceleration data structure and benefits from efficient convolutions.}
   \label{fig:OT}
\end{figure}

While computing the required convolutions used in the diffusion model
is possible on unstructured point sets~\cite{groh2019flex,simonovsky2017dynamic,hua2018pointwise}, this comes at a
prohibitive cost in our context, due to the large number of convolutions involved. Fortunately, our point sets are not arbitrary but
may uniformly cover the unit square. In certain cases, they can be
stratified, i.e., each stratum of size $\frac{1}{\sqrt{n}} \times
\frac{1}{\sqrt{n}}$ contains a single sample. This is notably the case for the large class
of $(0,m,s)$-nets samplers~\cite{Niederreiter1992}. In that case, we use a pixel grid of $\sqrt{n} \times \sqrt{n}$ pixels, and store in each pixel the 2-d offset between the stratum center and its corresponding sample location. When this is not the case, we compute a linear assignment using optimal transport between the strata centers and the set of samples (Fig.~\ref{fig:OT})~\cite{bonneel2011displacement}, and similarly store in each pixel the 2-d offset between the stratum center and its corresponding sample location. Doing so allows to work on 2-d grids and benefit from optimized convolutions. In our settings, the grid acts as an approximate nearest neighbor acceleration data structure, such that, when a convolution is performed, neighboring samples approximately correspond to neighboring pixels, and are thus appropriately weighted. We evaluate this property with non-uniform sampling in Sec.~\ref{sec:nonuniform}. This remapping further allows to remain invariant under re-ordering of samples.

\subsection{Training}

The benefit of a convolutional approach is that the same convolution weights can be used for different grid sizes. It thus becomes possible to train \textit{the same} network with point sets of different sizes, and hope that it generalizes. We explore in Sec.~\ref{sec:prop} how it succeeds in generalizing. However, within a single batch, the sample count should remain the same, due to the way batches are processed. For a given batch of size $B$, we thus build a loss that sums contributions for different input grid sizes $\mathcal{S}$ stored in different batches: 
$$\mathcal{L}(\epsilon_\theta, \epsilon_t) = \sum_{j \in \mathcal{S} } \frac1B \sum_{i=1}^B \| \epsilon_\theta(\tilde{x}_{t_i}, t_i) - \epsilon_{t_i} \|^2\,,$$
 for randomly chosen $\{t_i\}$. We typically use $\mathcal{S} = \{8\times 8, 16\times 16, 32\times 32\}$, hence learning from sample sizes $\{64, 256, 1024\}$.
We obtain one trained network, of the same architecture but different training weights, per type of sampler, each able to produce point sets of different sample sizes. 

We train networks to reproduce Sobol' samples with Owen's scrambling~\cite{Sobol67,owen1998scrambling} as a representative LDS matrix-based sampler, LatNetBuilder samples as a representative LDS lattice-based sampler, a Poisson disk sampler (classical dart throwing approach), SOT~\cite{paulin2020sliced} as a representative blue noise sampler using optimal transport, GBN~\cite{GBN22} as a representative kernel-based blue noise sampler, LDBN~\cite{Ahmed:2016:Lowdiscrepancy} as a sampler that combines low discrepancy properties and blue noise spectrum, and Rank-1~\cite{keller2004stratification} as a representative of lattice based sampler.
We train all our models using 64k point sets, except for the SOT sampler trained with only 32 (\textit{not 32k}) point sets to assess robustness to small training datasets. 
We train for a constant time of 3 hours, and synthesis time is typically 35 minutes for 1000 point sets of 1024 samples each using 1000 diffusion steps.

\section{Validation and applications}

\begin{figure*}\setlength\tabcolsep{0pt} 
  \begin{tabular}{lcccccc} 
   & Poisson disk & GBN & SOT & LDBN & Sobol'+Owen & Rank1\\ 
    \multirow{3}{*}{\rotatebox{90}{Original}}&
    \includegraphics[width=2.5cm]{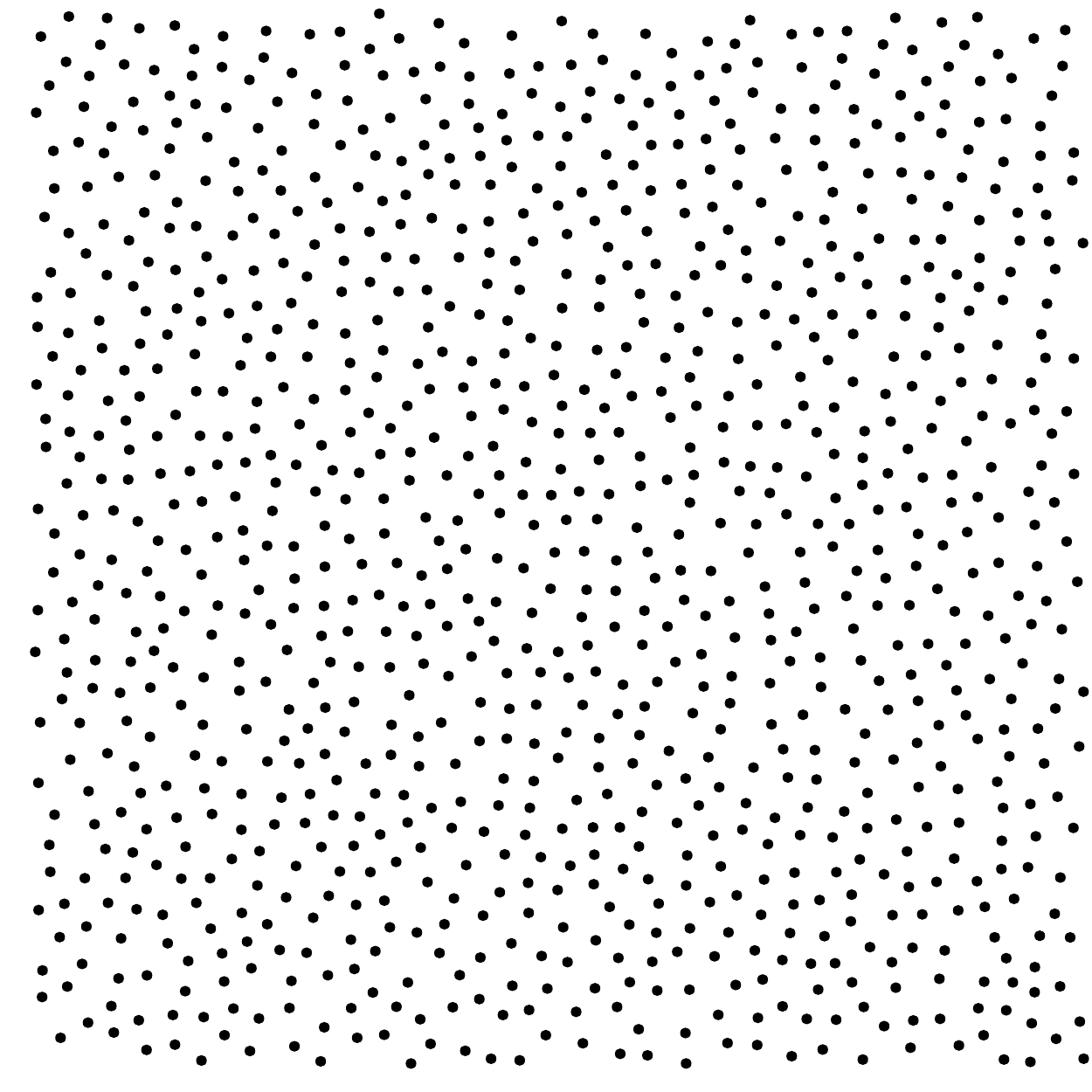}&
    \includegraphics[width=2.5cm]{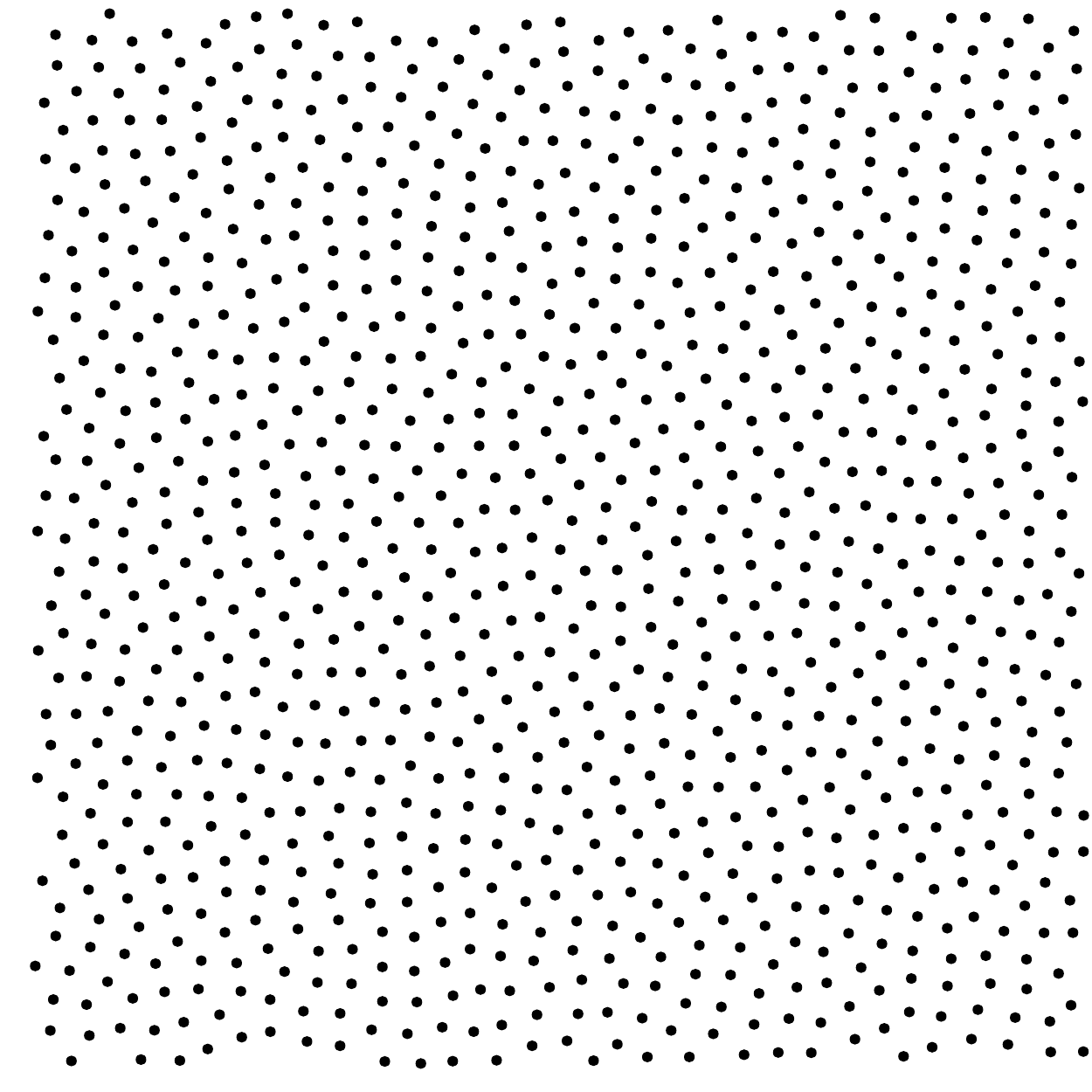}&
    \includegraphics[width=2.5cm]{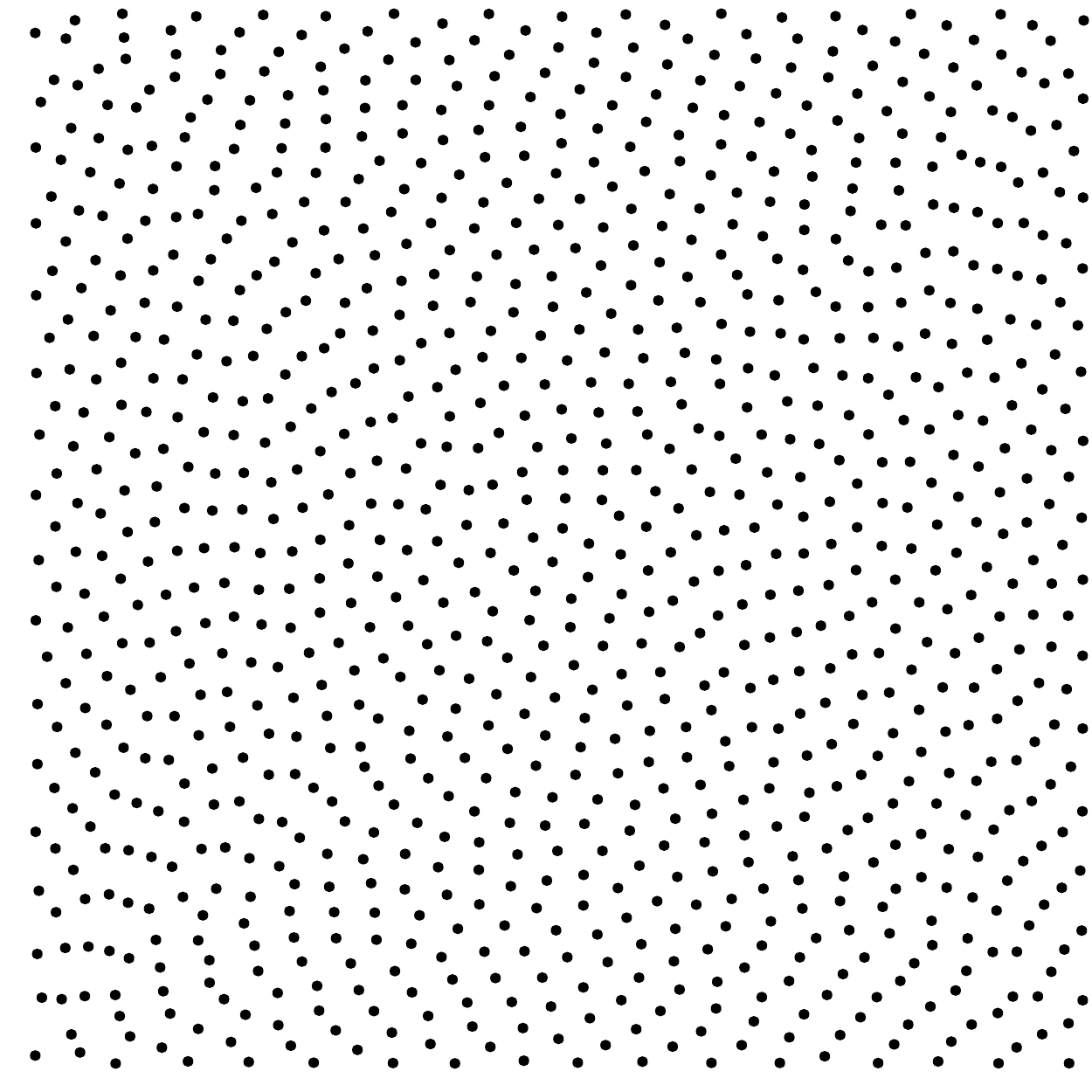}&
    \includegraphics[width=2.5cm]{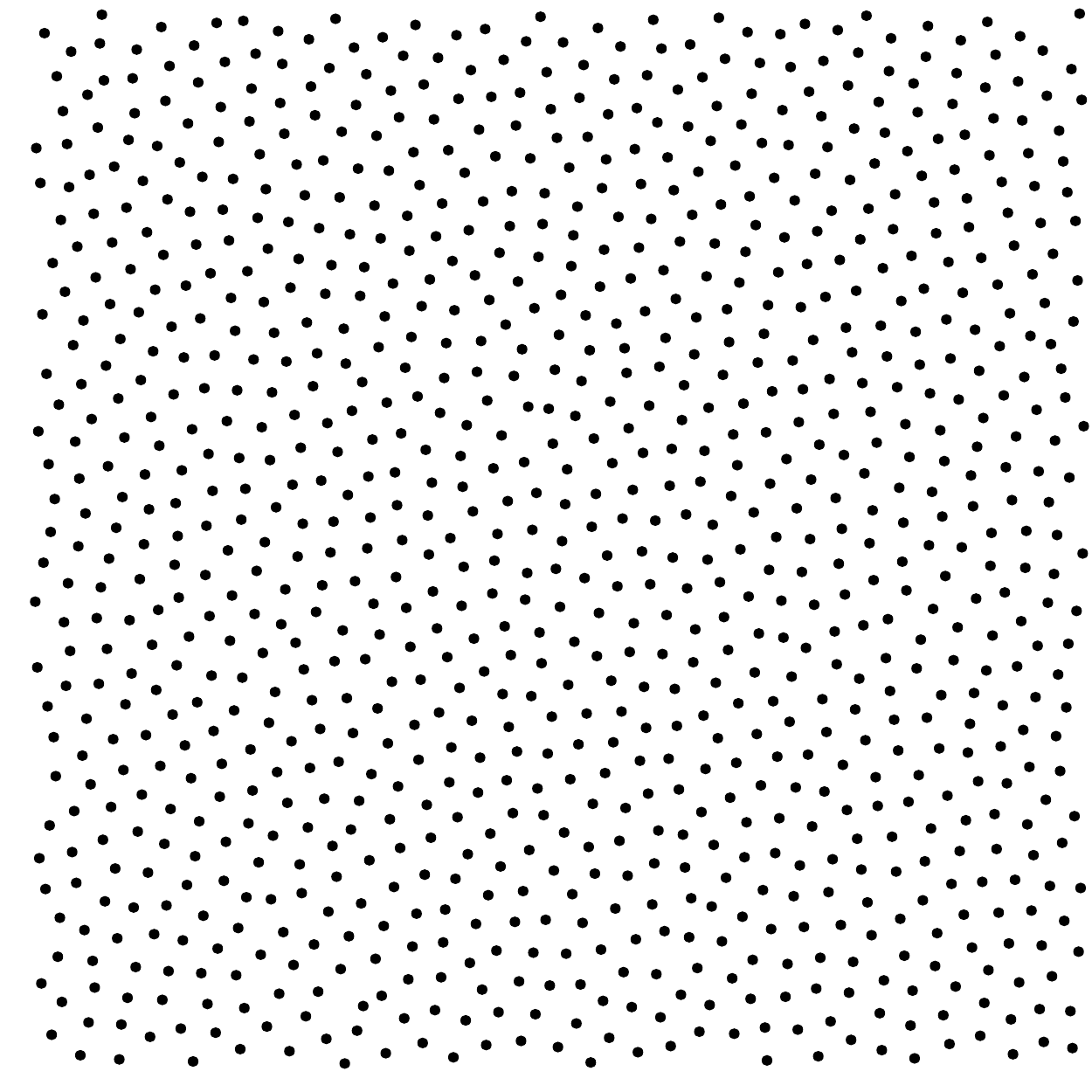}&
    \includegraphics[width=2.5cm]{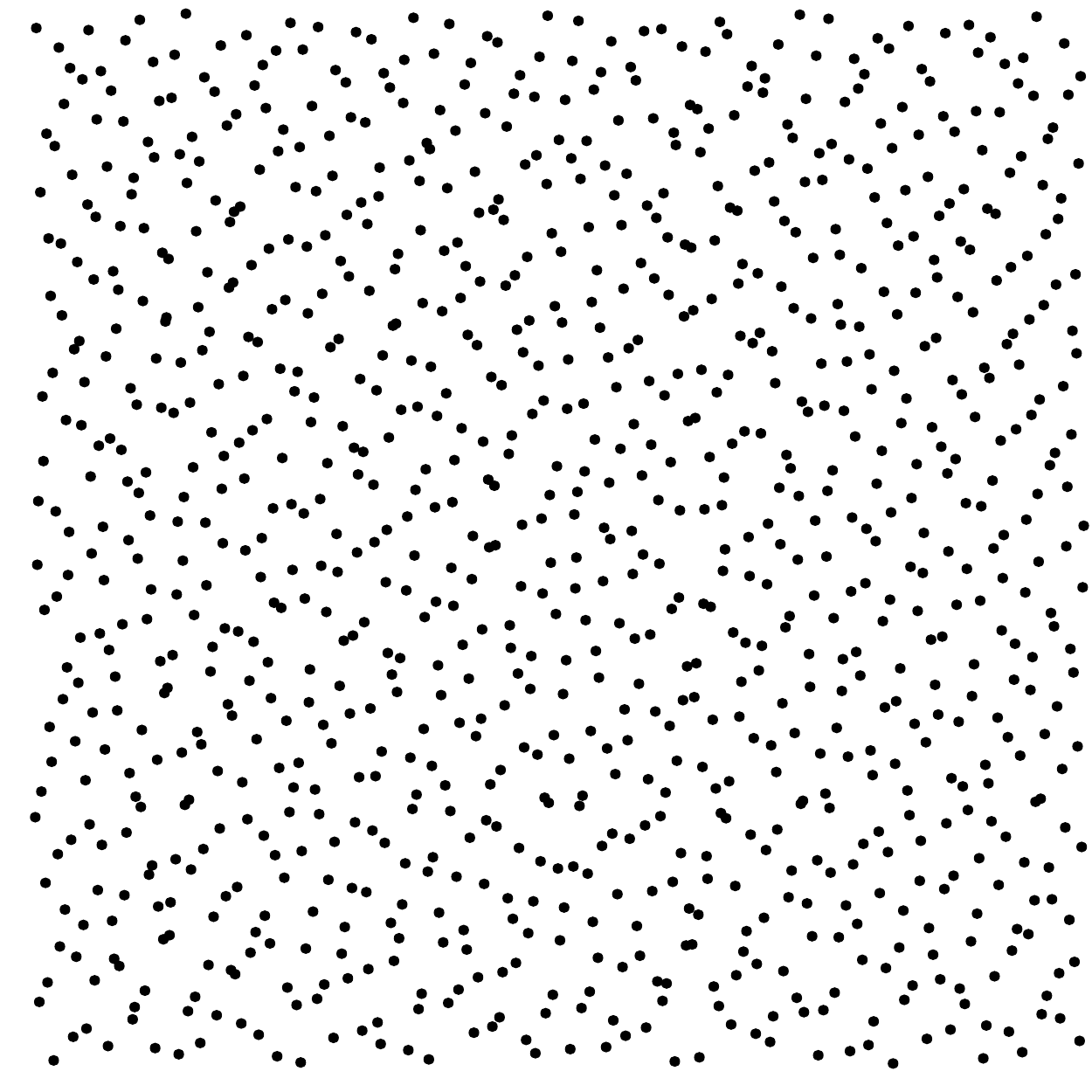}&
    \includegraphics[width=2.5cm]{res/pointsets/Orig/r1-1k}\\
    &
    \includegraphics[width=2.5cm]{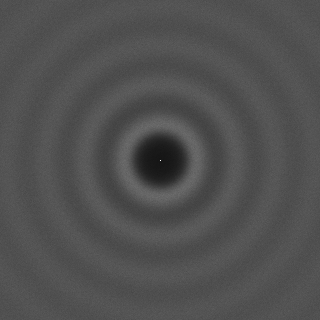}&
    \includegraphics[width=2.5cm]{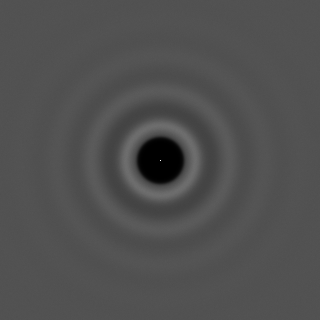}&
    \includegraphics[width=2.5cm]{res/spectrums/Orig/spectrumSOT}&
    \includegraphics[width=2.5cm]{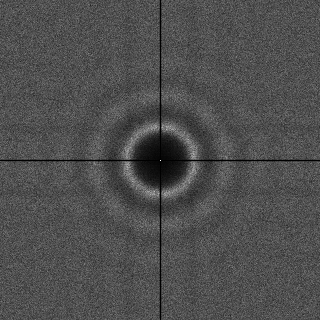}&
    \includegraphics[width=2.5cm]{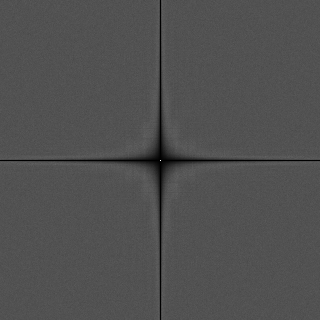}&
    \includegraphics[width=2.5cm]{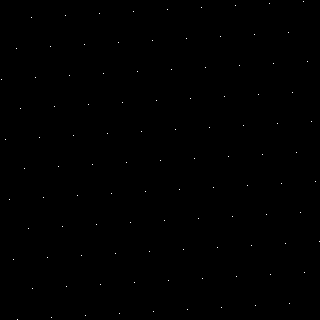}\\
   &
    \scalebox{.3}{\input{res/spectrums/Orig/radialPoisson.tex}}&
    \scalebox{.3}{\input{res/spectrums/Orig/radialGBN.tex}}&
    \scalebox{.3}{\input{res/spectrums/Orig/radialSOT.tex}}&
    \scalebox{.3}{\input{res/spectrums/Orig/radialLDBN.tex}}&
    \scalebox{.3}{\input{res/spectrums/Orig/radialOwen.tex}}&
    \scalebox{.3}{\input{res/spectrums/Orig/radialR1.tex}}\\ 
\multirow{3}{*}{\rotatebox{90}{\cite{leimkuhler2019deep}}}&    \includegraphics[width=2.5cm]{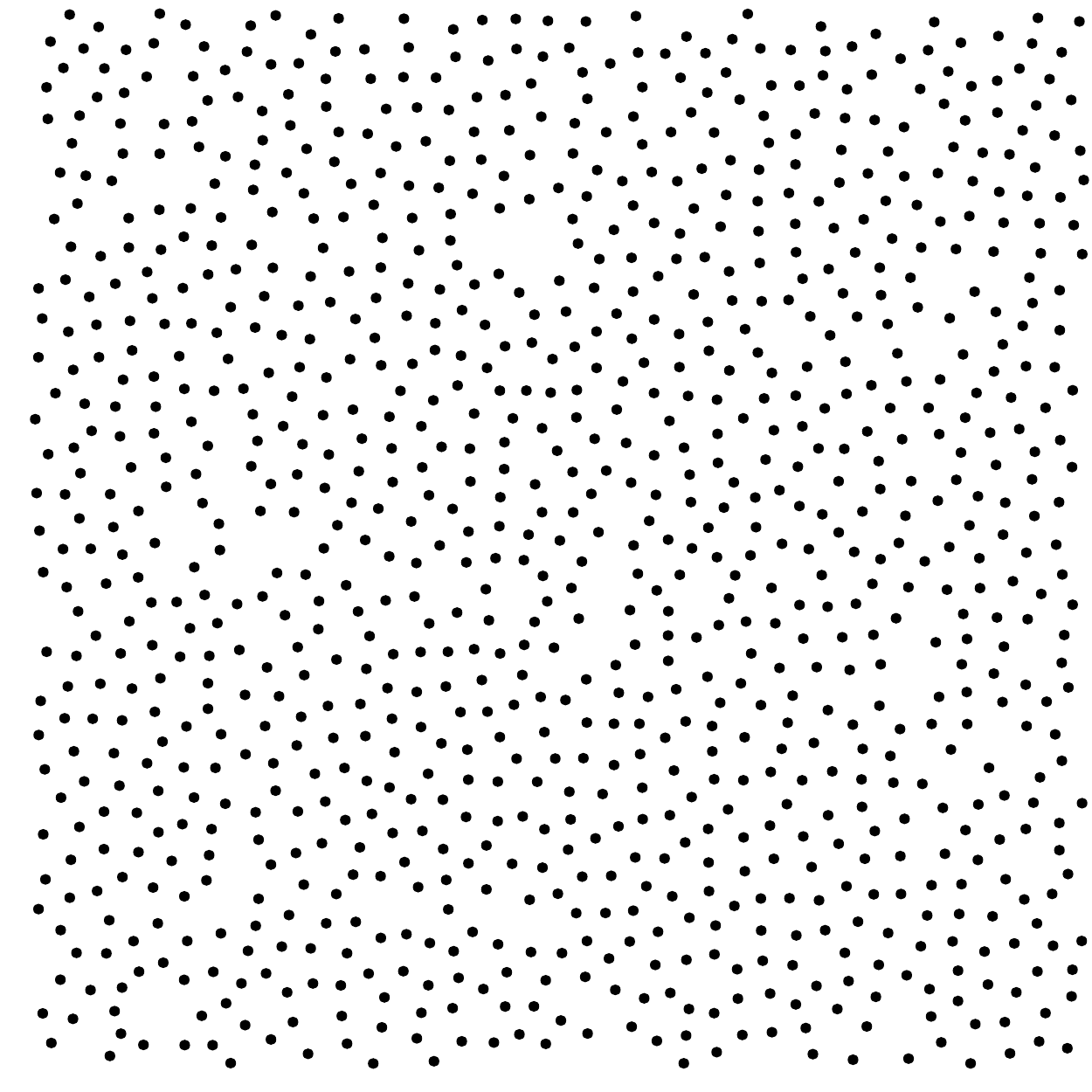}&
    \includegraphics[width=2.5cm]{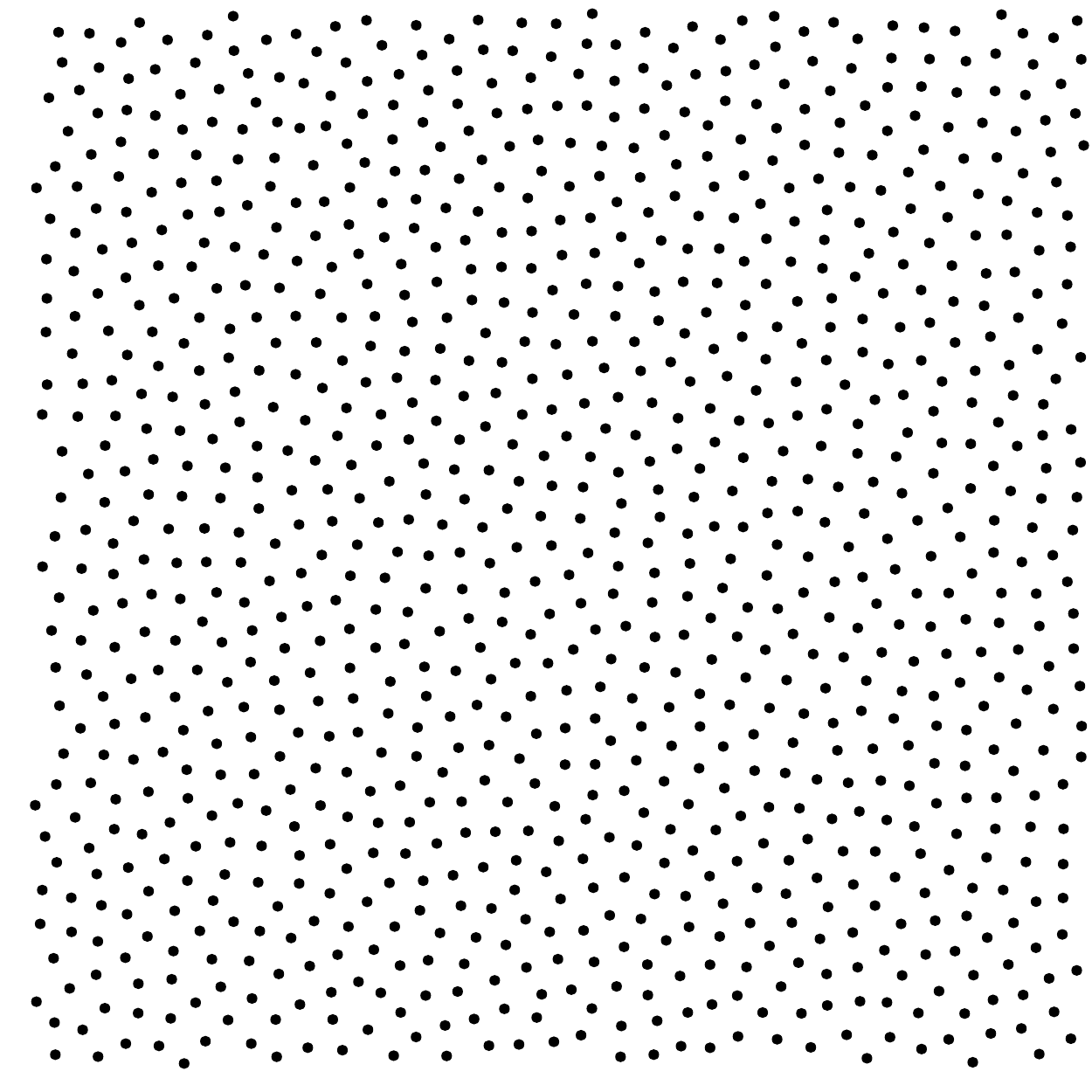}&
    \includegraphics[width=2.5cm]{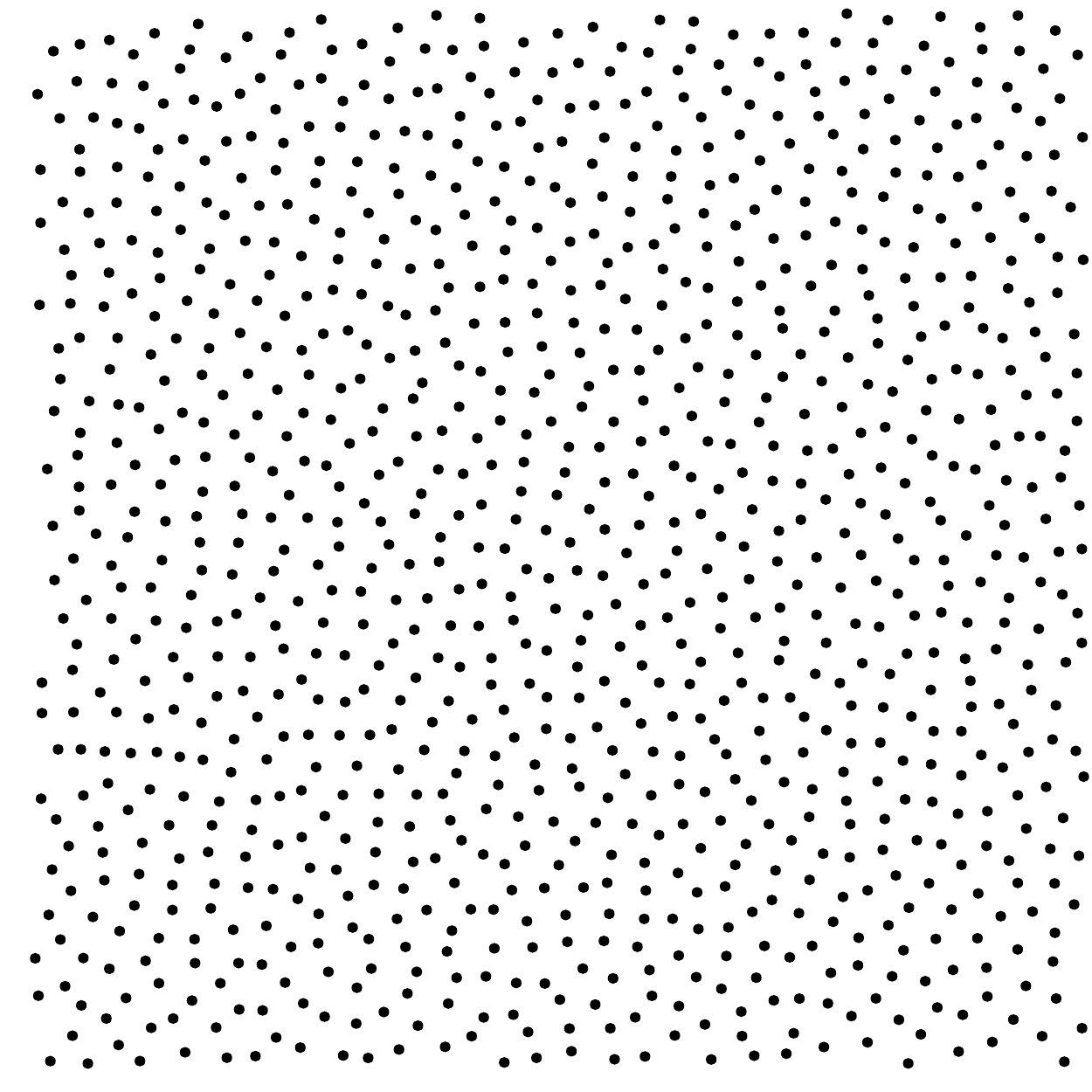}&
    \includegraphics[width=2.5cm]{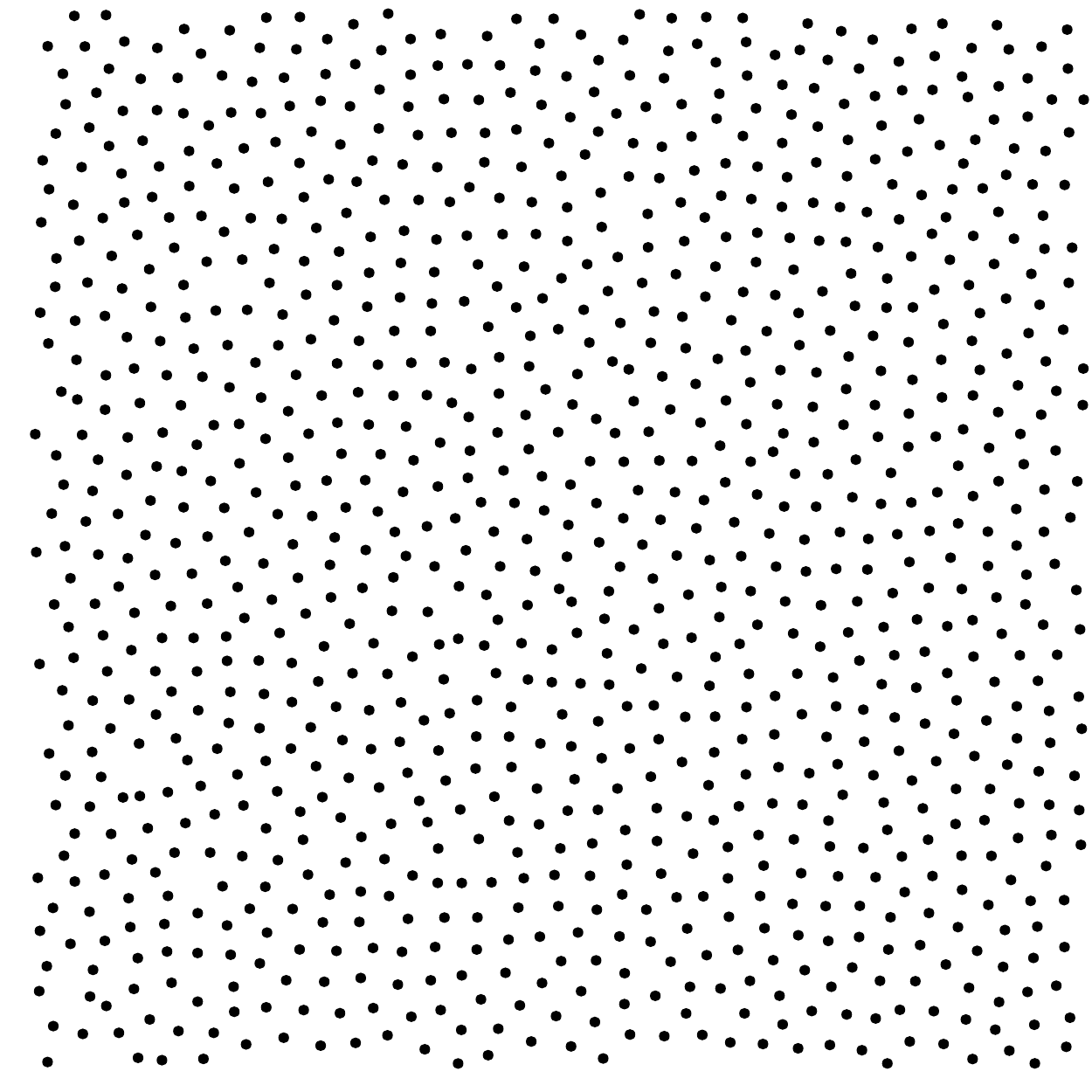}&
    \includegraphics[width=2.5cm]{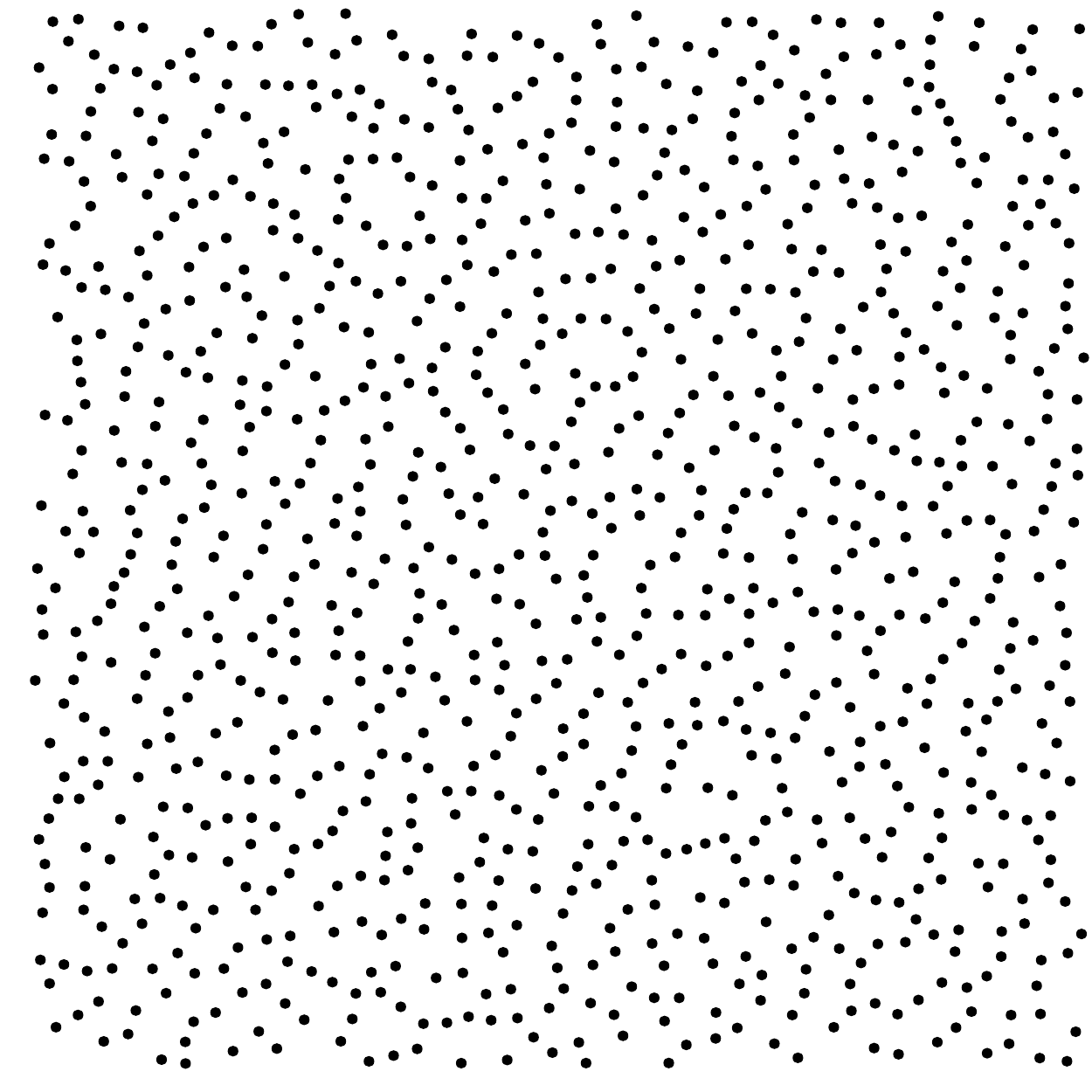}&
    \raisebox{1.cm}{N.A.}\\
    &
    \includegraphics[width=2.5cm]{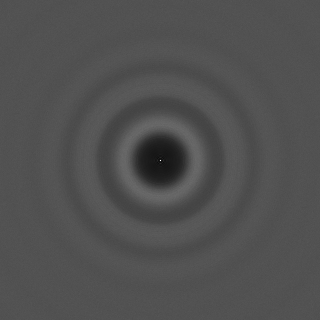}&
    \includegraphics[width=2.5cm]{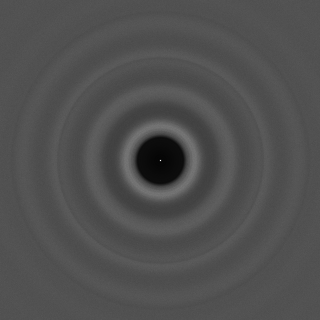}&
    \includegraphics[width=2.5cm]{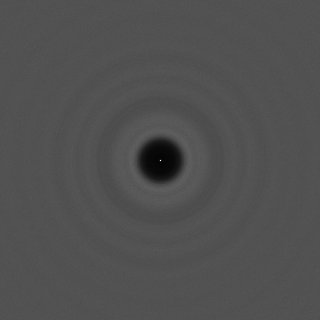}&
    \includegraphics[width=2.5cm]{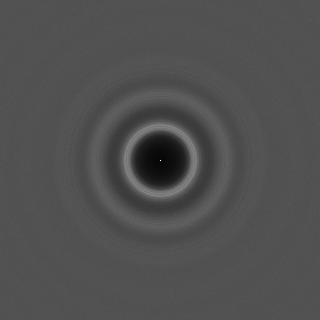}&
    \begin{overpic}[width=2.5cm]{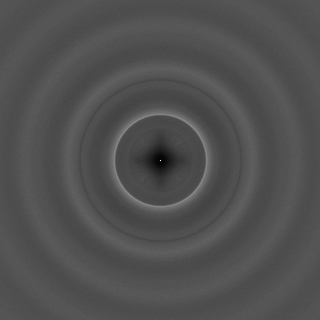}
      \put(37.5,37.5){\begin{tikzpicture}
          \draw[draw=orange,thick] (0,0) rectangle ++(0.6,0.6);
        \end{tikzpicture} }
    \end{overpic}

    &
     \raisebox{1.cm}{N.A.}\\
    &
    \scalebox{.3}{\input{res/spectrums/DeepPointCorrelation/radialPoisson.tex}}&
    \scalebox{.3}{\input{res/spectrums/DeepPointCorrelation/radialGBN.tex}}&
    \scalebox{.3}{\input{res/spectrums/DeepPointCorrelation/radialSOT.tex}}&
    \scalebox{.3}{\input{res/spectrums/DeepPointCorrelation/radialLDBN.tex}}&
    \scalebox{.3}{\input{res/spectrums/DeepPointCorrelation/radialOwenSpectre2d.tex}}&
    \raisebox{.5cm}{N.A.}\\

     \multirow{3}{*}{\rotatebox{90}{Our}}&
     \includegraphics[width=2.5cm]{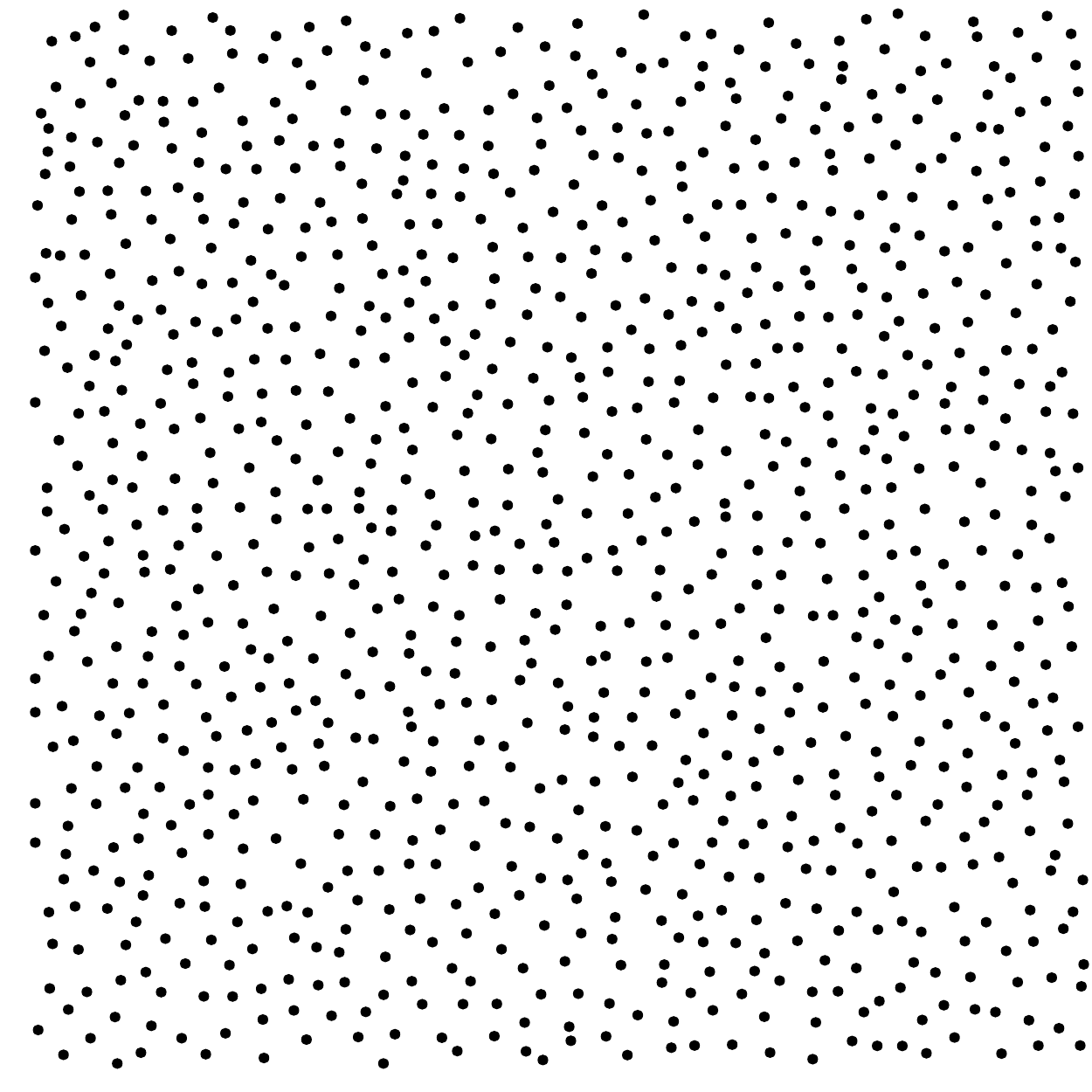}&
    \includegraphics[width=2.5cm]{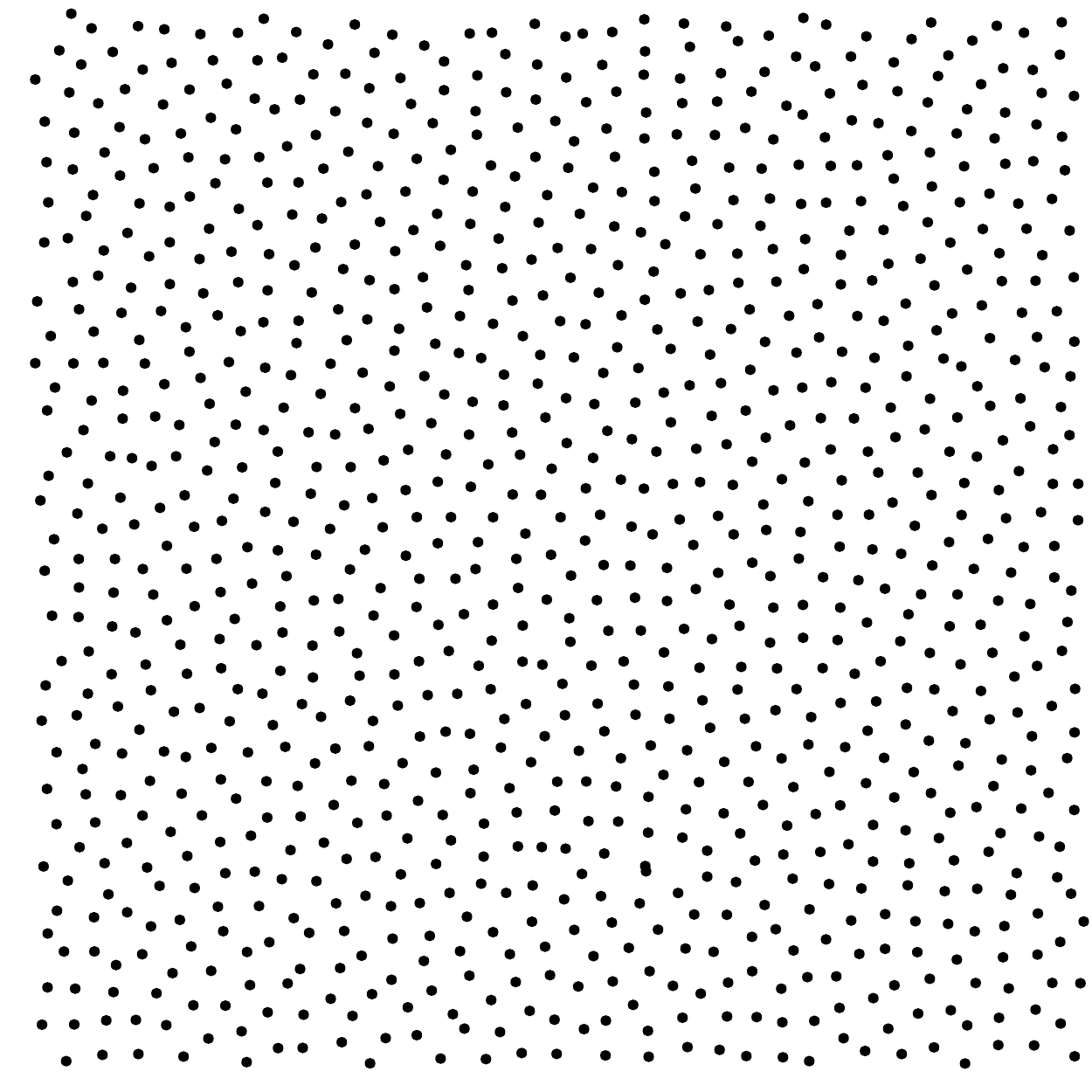}&
    \includegraphics[width=2.5cm]{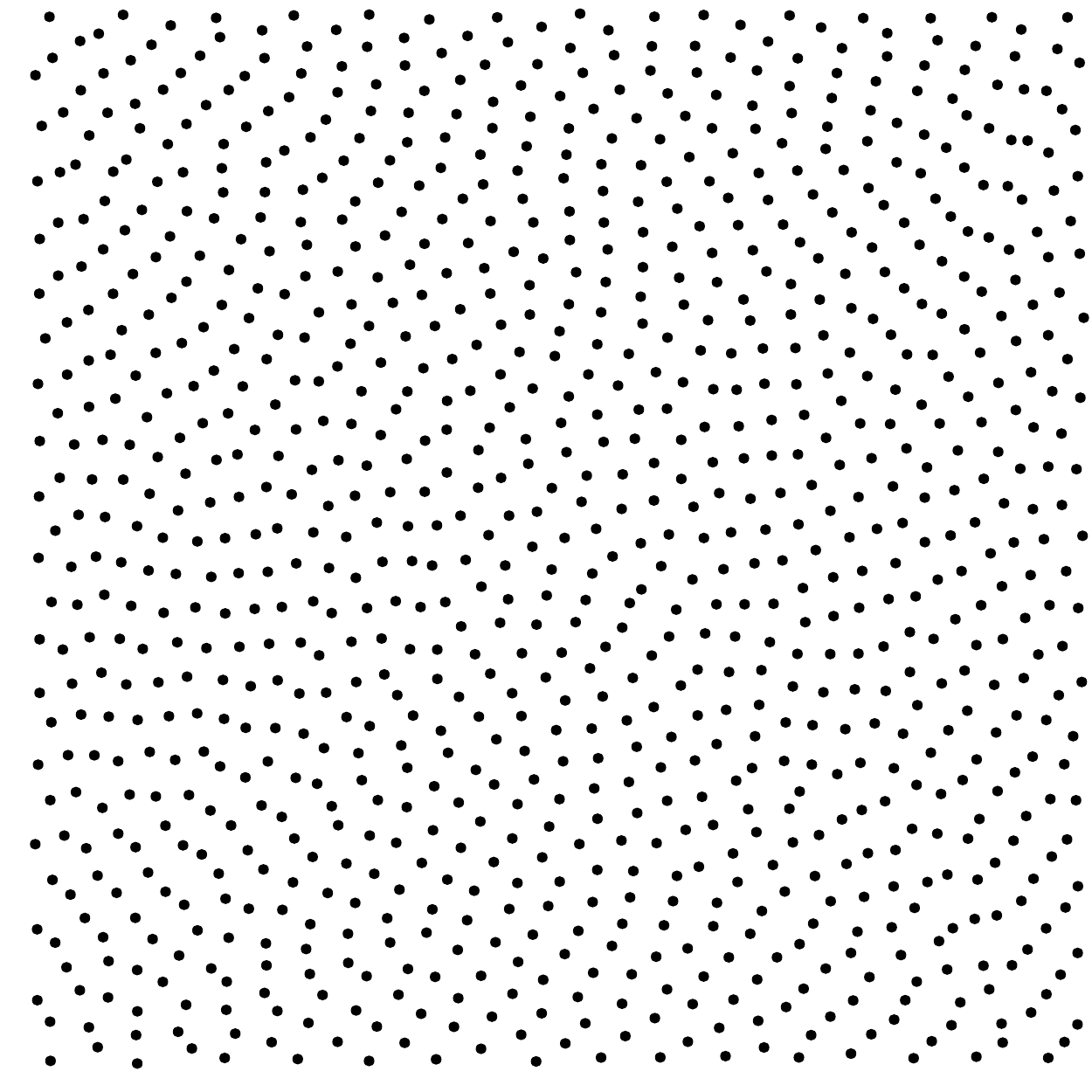}&
    \includegraphics[width=2.5cm]{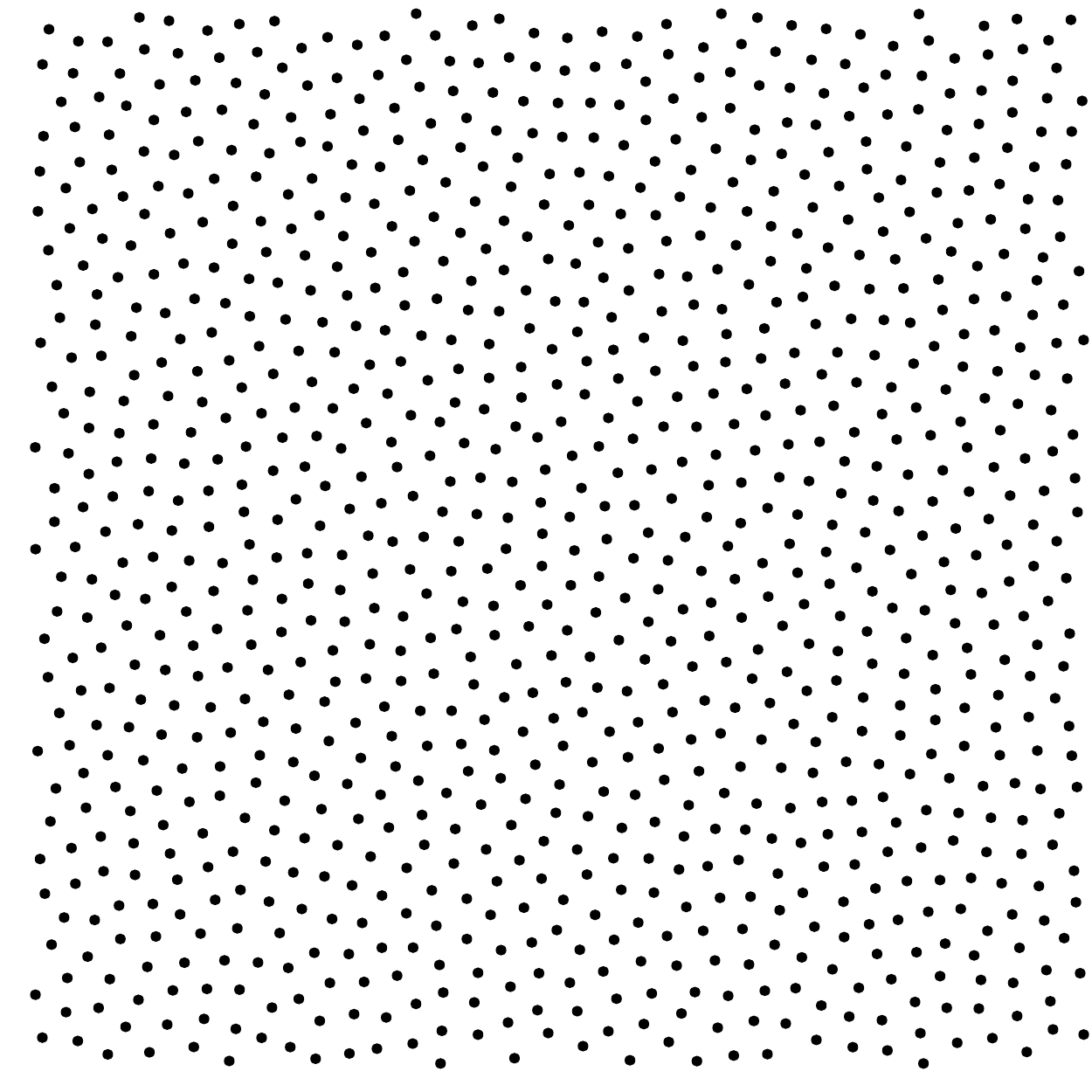}&
    \includegraphics[width=2.5cm]{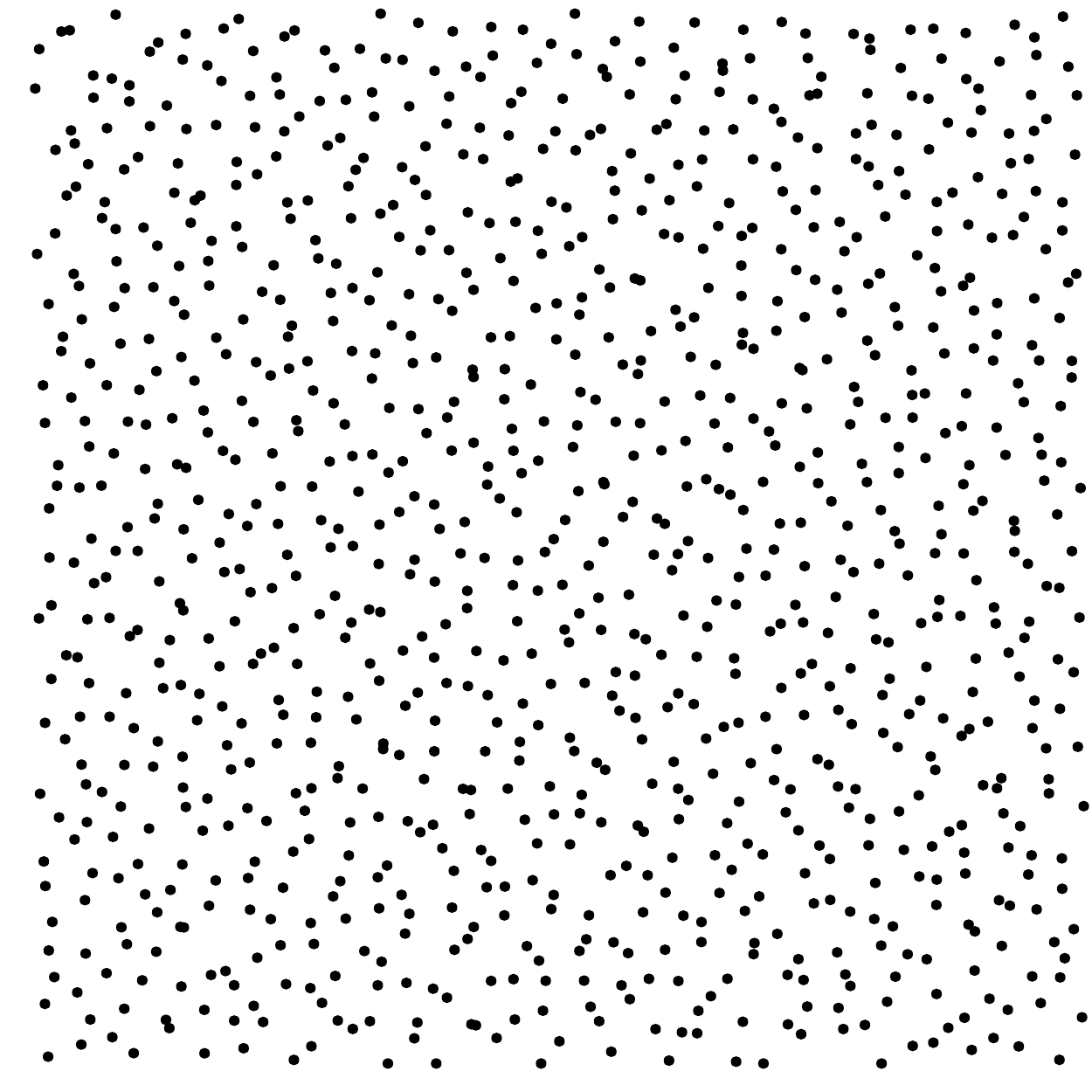}&
    \includegraphics[width=2.5cm]{res/pointsets/NNFinal/r1-1k}\\
    &
    \includegraphics[width=2.5cm]{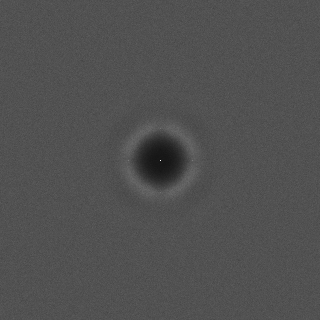}&
    \includegraphics[width=2.5cm]{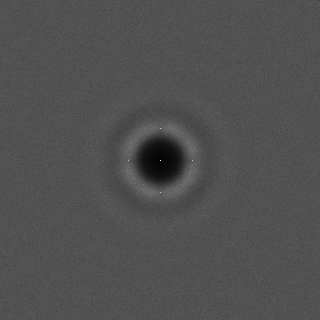}&
    \includegraphics[width=2.5cm]{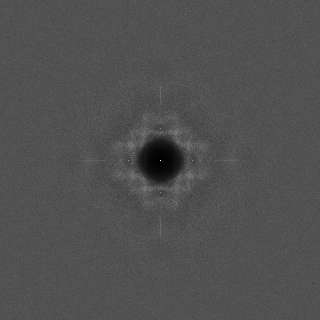}&
    \includegraphics[width=2.5cm]{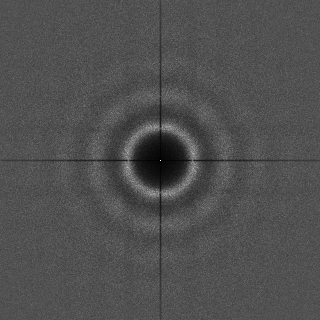}&
    \begin{overpic}[width=2.5cm]{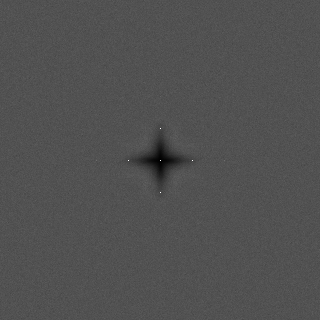}
    \end{overpic}
    &
    \includegraphics[width=2.5cm]{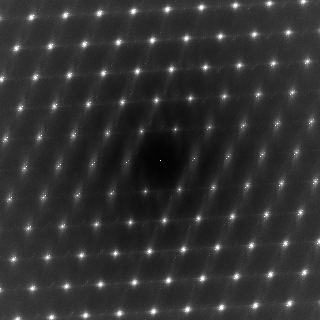}\\
    &
    \scalebox{.3}{\input{res/spectrums/Final/radialPoisson.tex}}&
    \scalebox{.3}{\input{res/spectrums/Final/radialGBN.tex}}&
    \scalebox{.3}{\input{res/spectrums/Final/radialSOT.tex}}&
    \scalebox{.3}{\input{res/spectrums/Final/radialLDBN.tex}}&
    \scalebox{.3}{\input{res/spectrums/Final/radialOwen.tex}}&
    \scalebox{.3}{\input{res/spectrums/Final/radialR1.tex}}\\
  \end{tabular}
  \caption{For various input samplers and their spectral content
    (Fourier power spectrum and radial mean power spectrum), we
    compare our approach (last three rows) with that of 
    \cite{leimkuhler2019deep} (1d radial mean power spectrum loss for
    Poisson disk, GBN, SOT and LDBN; for Sobol'+Owen
    and Rank-1, we used the 2d power spetrum cropped to
    the central part, framed in orange, for the learning to converge).\label{fig:spectra}}
\end{figure*}

\subsection{Properties of generated samples}
\label{sec:prop}

We study power spectra, optimal transport energy, discrepancy,
integration errors and minimum distance statistics of generated point sets,
and verify that they match properties they were trained for. We also
verify how our network generalizes as we increase the number of
samples outside the range it was trained for. For these comparisons, we
compare to the approach of \cite{leimkuhler2019deep}. For stationary and
isotropic point processes or samplers targeting such properties, we
have used their publicly available implementation with a 1d radial mean power
spectrum loss (same learning parameters as the one provided by the
authors for similar experiments). For non-stationary or anisotropic
samplers (\emph{e.g.} Sobol'+Owen and Rank1), we had to design our own
learning experiment following their examples in 1d, 
with losses defined as $l_1$ norm between 2d power
spectra (cropped to the central frequency part). We observe that such
training turns out to be very difficult in 2d and leads to
non-competitive results. In Fig~\ref{fig:spectra}, we only show
 results for Sobol'+Owen in 2d and leave the discussion for Rank-1 in supplementary materials.

\begin{figure}[!tbh]
  \centering\scriptsize
  \setlength\fboxrule{1.5pt}
  \begin{overpic}[width=\linewidth]{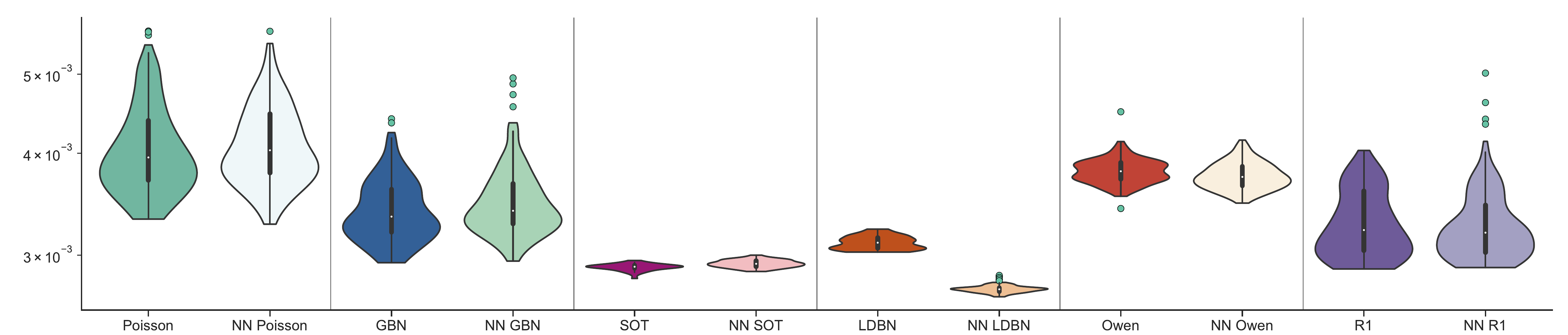}
    \put(-1,5){\rotatebox{90}{\bf OT, 64}}
  \end{overpic}
  \begin{overpic}[width=\linewidth]{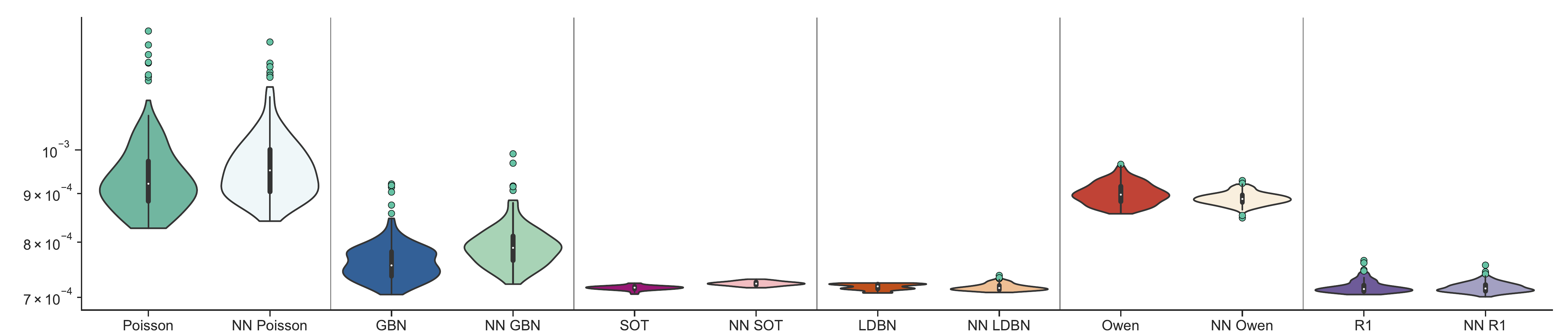}
    \put(-1,5){\rotatebox{90}{\bf OT, 256}}
  \end{overpic}
  \begin{overpic}[width=\linewidth]{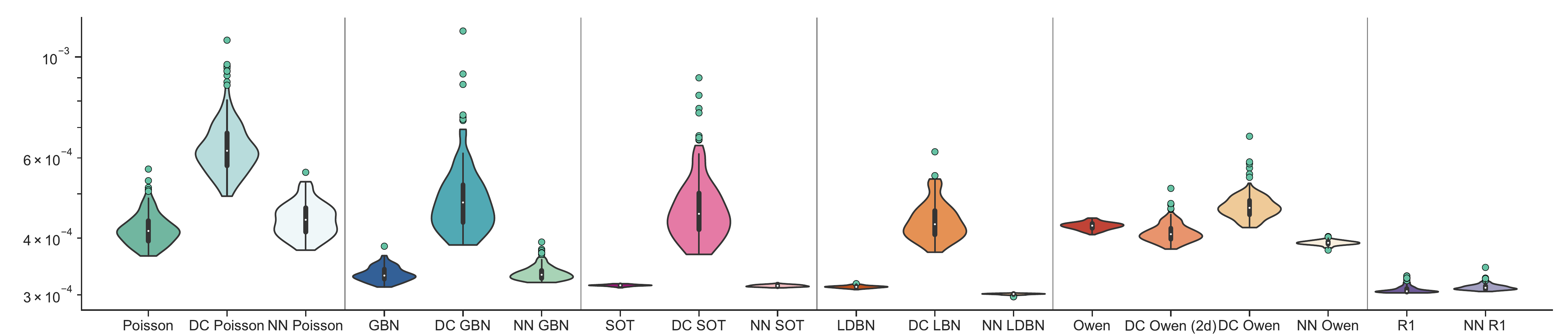}
    \put(-1,5){\rotatebox{90}{\bf OT, 576}}
  \end{overpic}
  \begin{overpic}[width=\linewidth]{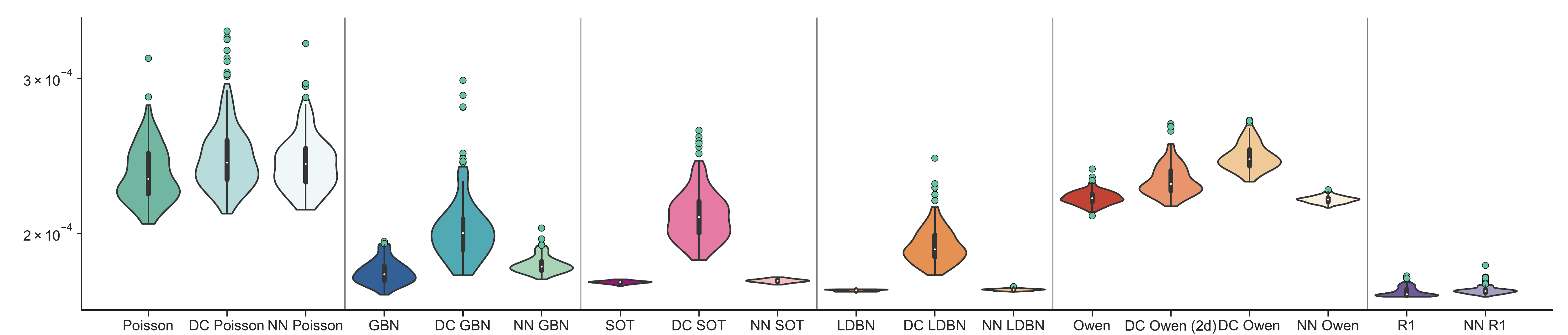}
    \put(-1,5){\rotatebox{90}{\bf OT, 1024}}
  \end{overpic}
   \begin{overpic}[width=\linewidth]{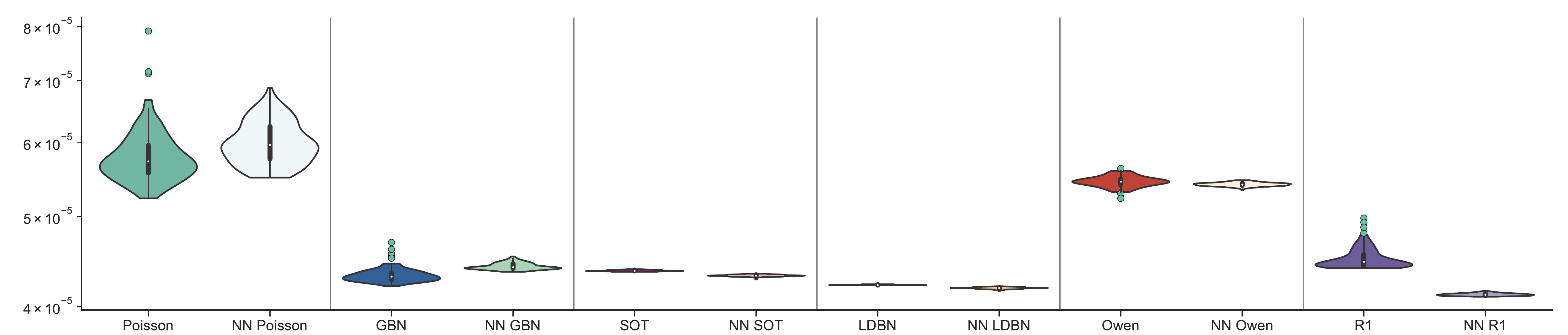}
     \put(-1,5){\rotatebox{90}{\bf OT, 4096}}
     \put(-2.5,9.3){\cfbox{orange}{\begin{minipage}{\linewidth}\vspace{1.65cm}\quad\end{minipage}}}
     \put(-2.5,52){\cfbox{orange}{\begin{minipage}{\linewidth}\vspace{1.65cm}\quad\end{minipage}}}
   \end{overpic}
   \caption{We verify that the point sets predicted by our network match the semi-discrete optimal transport distance to a uniform distribution of the original point sets. These plots show these statistics distributions for 128 point sets from the training set and produced by our network, for sample counts of 64, 256, 576, 1024 and 4096 (top to bottom). The network has only been trained with point sets of 64, 256 and 1024 samples, but successfully predicts point sets of 576 and 4096 samples (results highlighted in an orange frame). Labels prefixed by \textbf{DC} refer to Deep Point Correlation results~\cite{leimkuhler2019deep} (on 1d radial power spectral, unless 2d is specified), while \textbf{NN} refers to results produced by our Neural Network.
   }
   \label{fig:OTenergy}
\end{figure}

\begin{figure}[!tbh]\scriptsize
  \centering
    \setlength\fboxrule{1.5pt}
  \begin{overpic}[width=\linewidth]{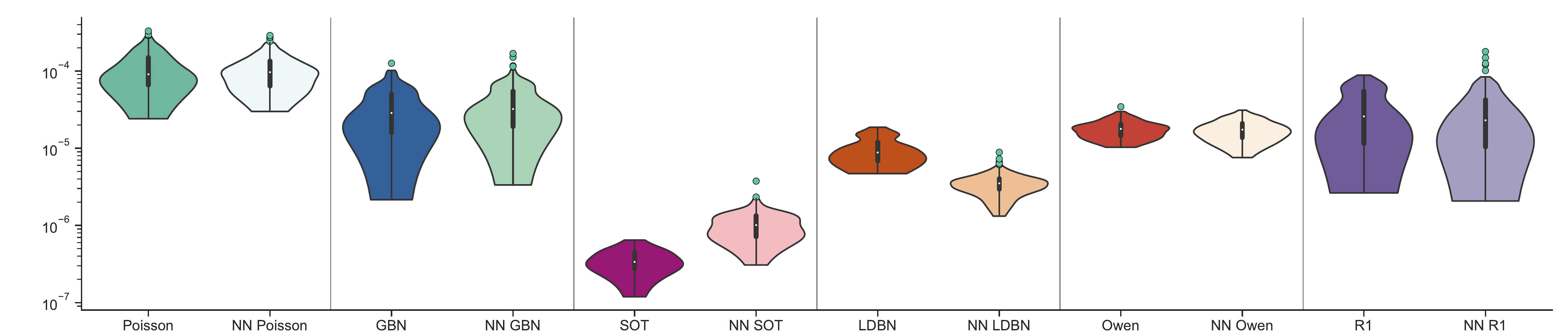}
    \put(-1,4){\rotatebox{90}{\bf Gaussian, 64}}
  \end{overpic}
  \begin{overpic}[width=\linewidth]{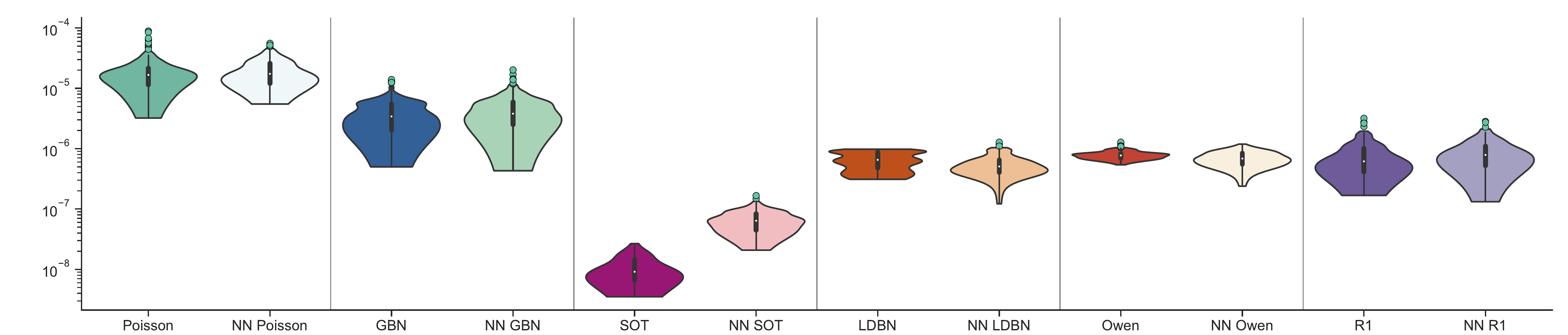}
    \put(-1,4){\rotatebox{90}{\bf Gaussian, 256}}
  \end{overpic}
  \begin{overpic}[width=\linewidth]{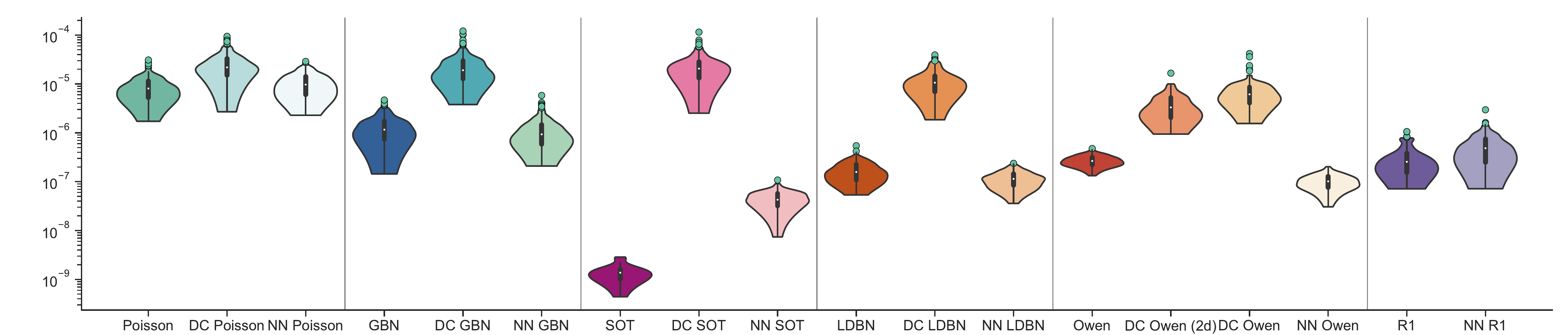}
    \put(-1,4){\rotatebox{90}{\bf Gaussian, 576}}
  \end{overpic}
  \begin{overpic}[width=\linewidth]{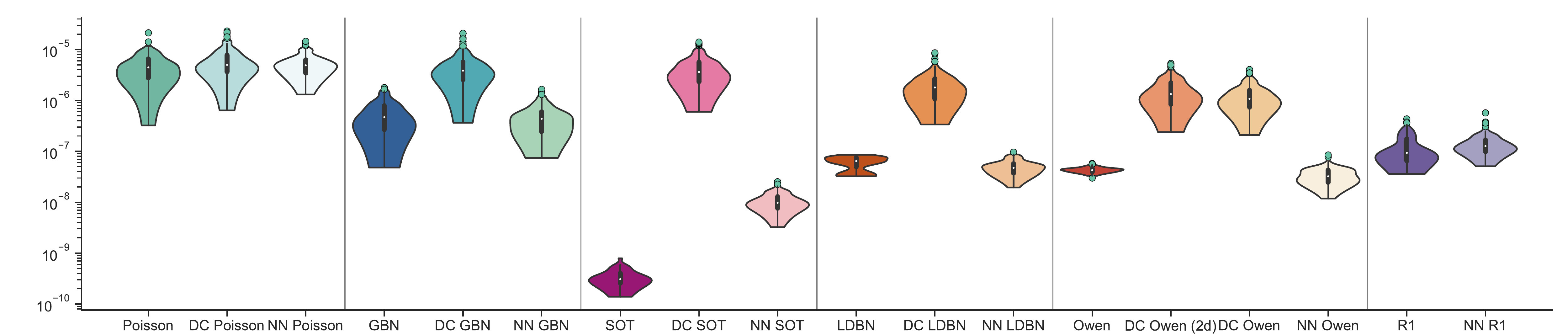}
    \put(-1,4){\rotatebox{90}{\bf Gaussian, 1024}}
  \end{overpic}
  \begin{overpic}[width=\linewidth]{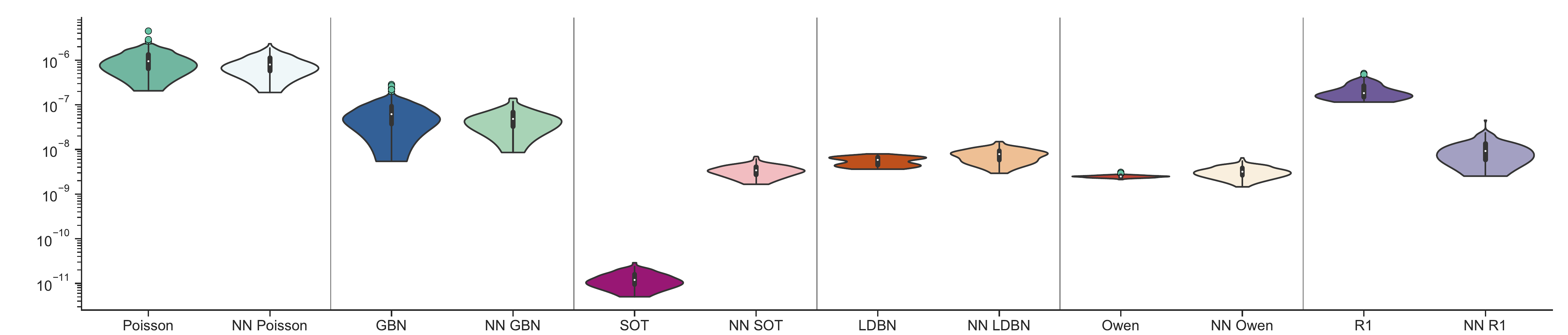}
    \put(-1,3.5){\rotatebox{90}{\bf Gaussian, 4096}}
  \end{overpic}
  \begin{overpic}[width=\linewidth]{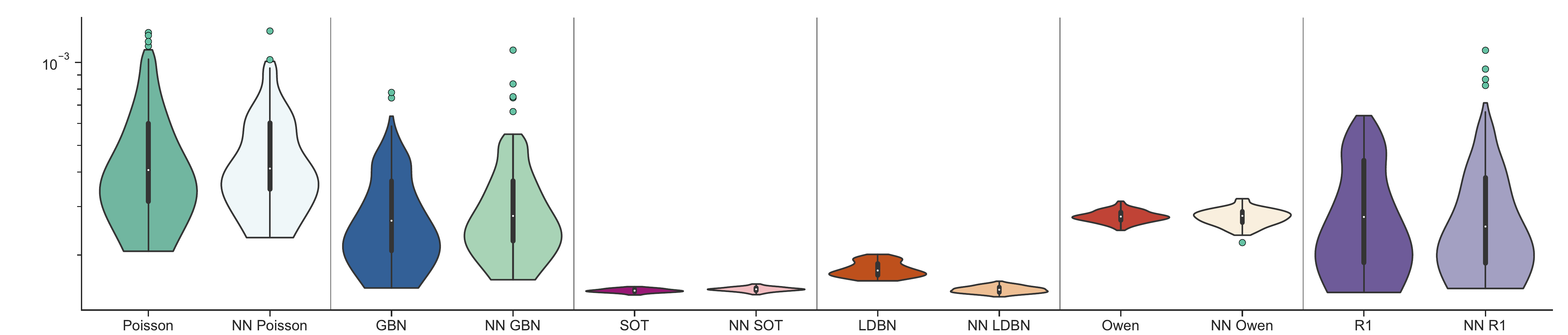}
    \put(-1,3){\rotatebox{90}{\bf Heaviside, 64}}
  \end{overpic}
  \begin{overpic}[width=\linewidth]{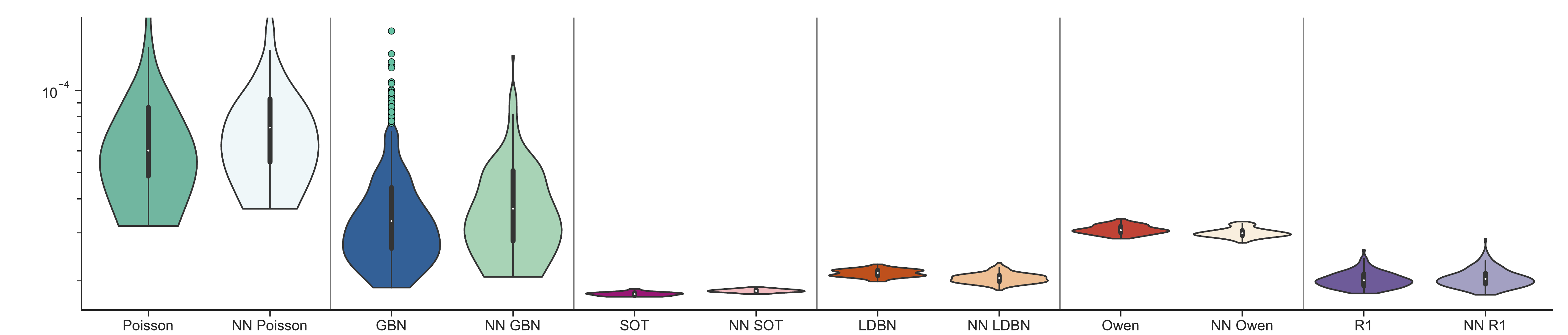}
    \put(-1,3){\rotatebox{90}{\bf Heaviside, 256}}
  \end{overpic}
  \begin{overpic}[width=\linewidth]{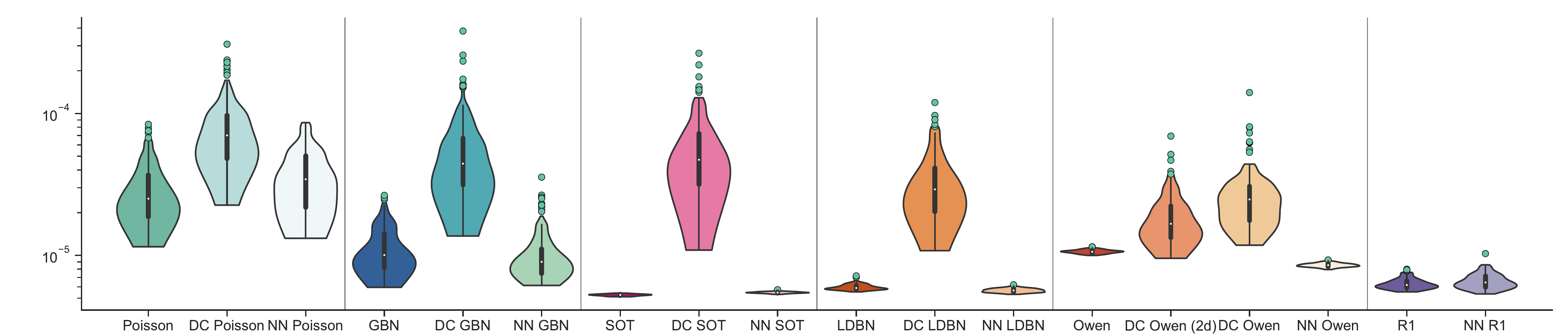}
    \put(-1,3){\rotatebox{90}{\bf Heaviside, 576}}
  \end{overpic}
  \begin{overpic}[width=\linewidth]{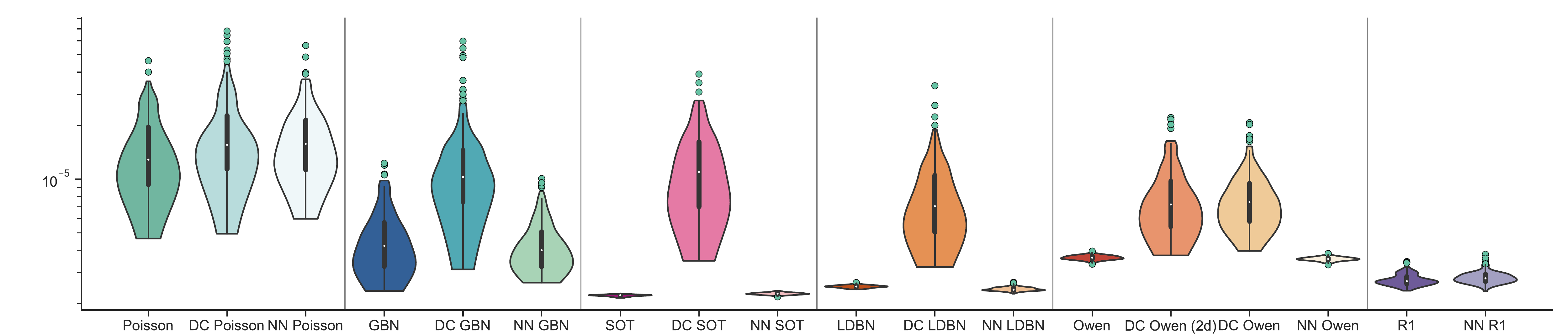}
    \put(-1,3){\rotatebox{90}{\bf Heaviside, 1024}}
   \end{overpic}
  \begin{overpic}[width=\linewidth]{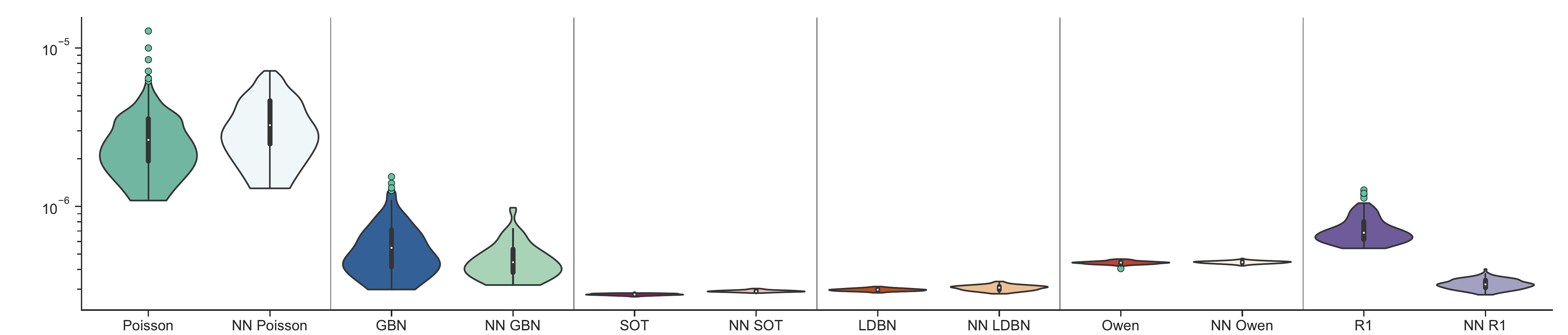}
    \put(-1,3.5){\rotatebox{90}{\bf Heaviside, 4096}}
    \put(-2.5,9.3){\cfbox{orange}{\begin{minipage}{\linewidth}\vspace{1.65cm}\quad\end{minipage}}}
     \put(-2.5,52.5){\cfbox{orange}{\begin{minipage}{\linewidth}\vspace{1.65cm}\quad\end{minipage}}}
     \put(-2.5,116.3){\cfbox{orange}{\begin{minipage}{\linewidth}\vspace{1.65cm}\quad\end{minipage}}}
     \put(-2.5,159.3){\cfbox{orange}{\begin{minipage}{\linewidth}\vspace{1.65cm}\quad\end{minipage}}}
  \end{overpic}
   \caption{Our network matches integration errors on Gaussian integrands (top 4 plots) and Heaviside integrands (bottom 4 plots), even beyond the sample sizes it was trained for ($\{64, 256, 1024\}$). Sample counts are 64, 256, 576, 1024 and 4096 (top to bottom for each integrand). }
   \label{fig:integ}
\end{figure}

\begin{figure}[!tbh]
  \centering\scriptsize    \setlength\fboxrule{1.5pt}
  \begin{overpic}[width=\linewidth]{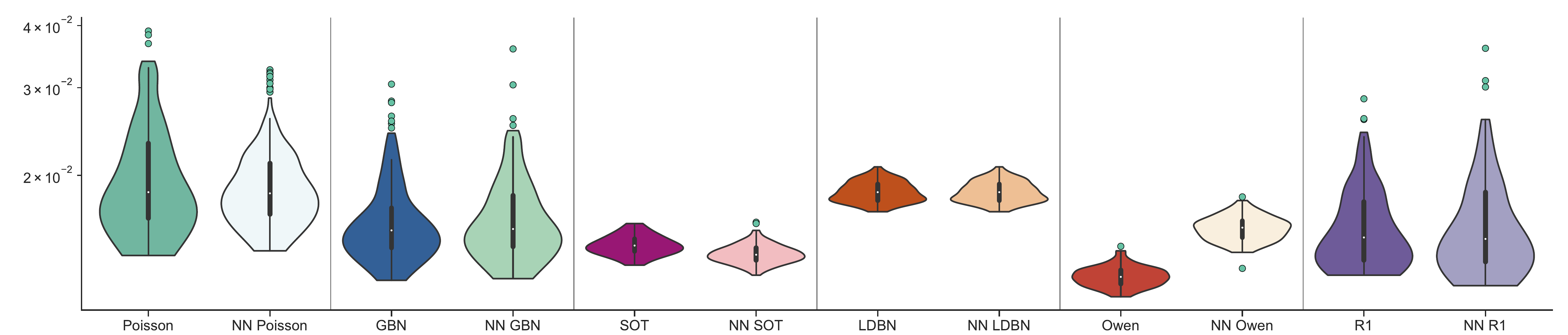}
    \put(-1,1){\rotatebox{90}{\bf Discrepancy, 64}}
  \end{overpic}
  \begin{overpic}[width=\linewidth]{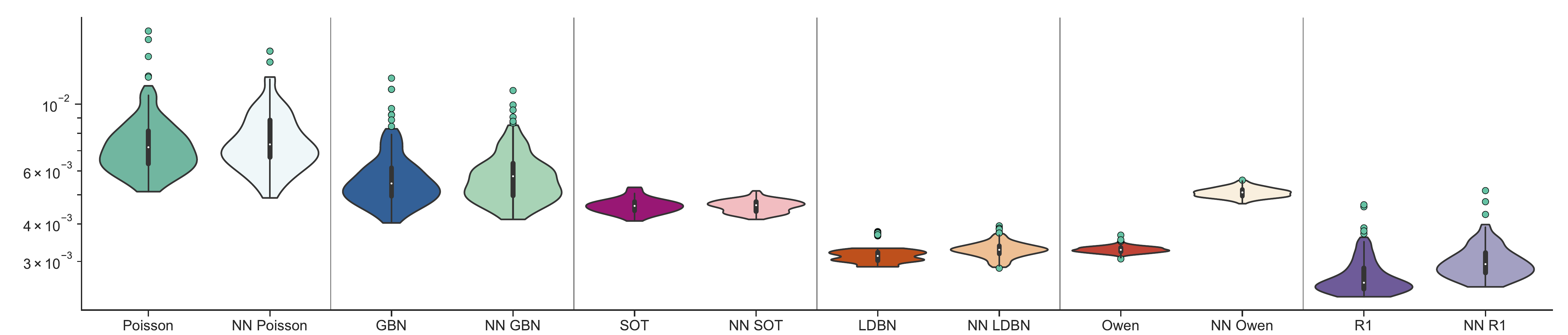}
    \put(-1,1){\rotatebox{90}{\bf Discrepancy, 256}}
  \end{overpic}
  \begin{overpic}[width=\linewidth]{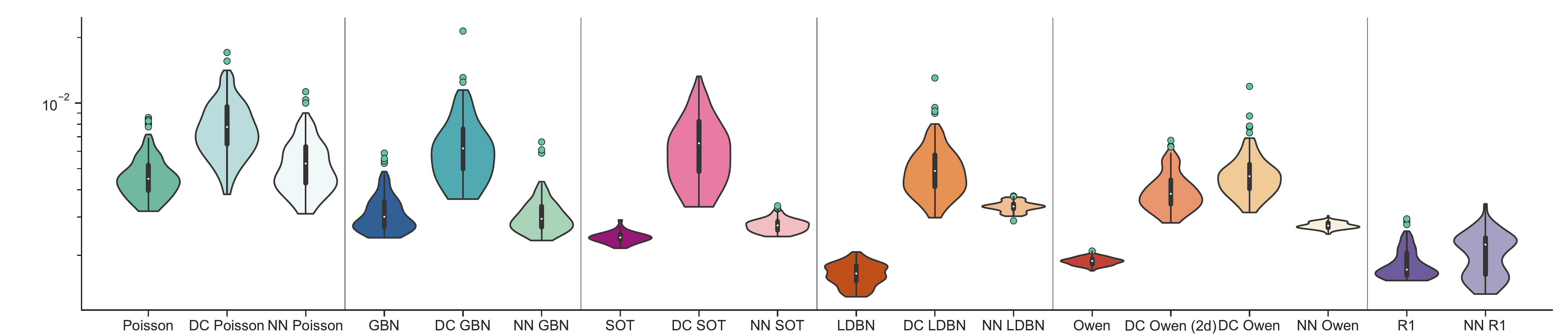}
    \put(-1,1){\rotatebox{90}{\bf Discrepancy, 576}}
  \end{overpic}
  \begin{overpic}[width=\linewidth]{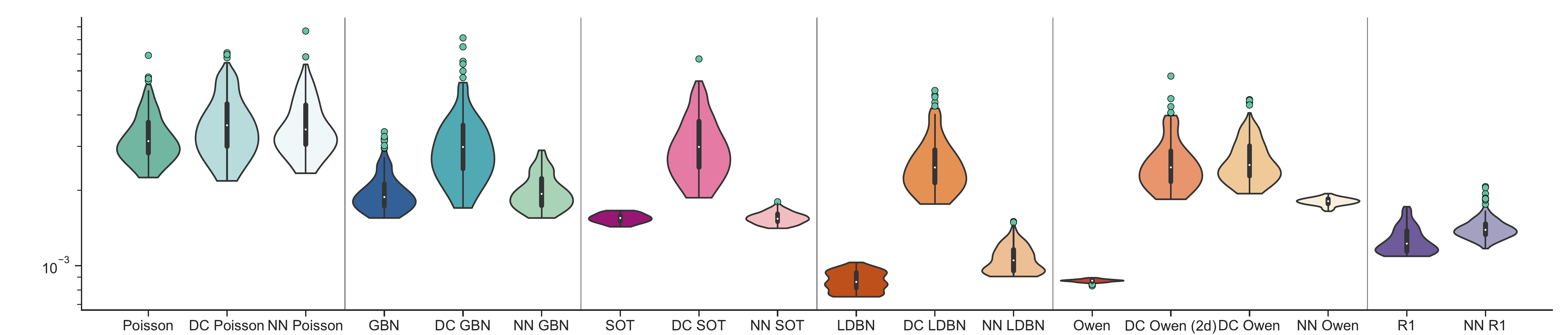}
    \put(-1,1){\rotatebox{90}{\bf Discrepancy, 1024}}
  \end{overpic}
  \begin{overpic}[width=\linewidth]{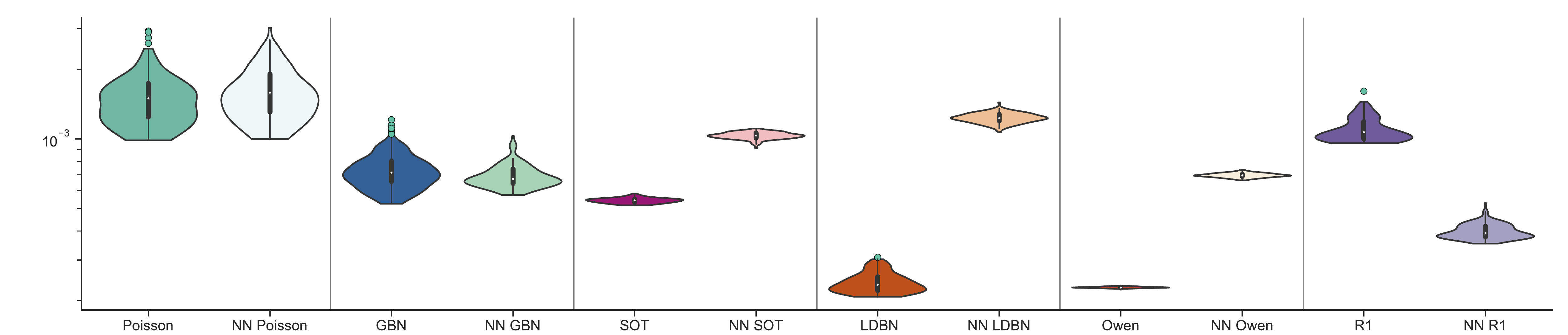}
    \put(-1,1){\rotatebox{90}{\bf Discrepancy, 4096}}
    \put(-2.5,9.3){\cfbox{orange}{\begin{minipage}{\linewidth}\vspace{1.65cm}\quad\end{minipage}}}
    \put(-2.5,52.5){\cfbox{orange}{\begin{minipage}{\linewidth}\vspace{1.65cm}\quad\end{minipage}}}
  \end{overpic}
  \caption{Our network matches the L2 discrepancy of the original point sets. Sample counts are 64, 256, 576, 1024 and 4096 (top to bottom).}
  \label{fig:disc}
\end{figure}

\begin{figure}[!tbh]
  \centering\scriptsize    \setlength\fboxrule{1.5pt}
  \begin{overpic}[width=\linewidth]{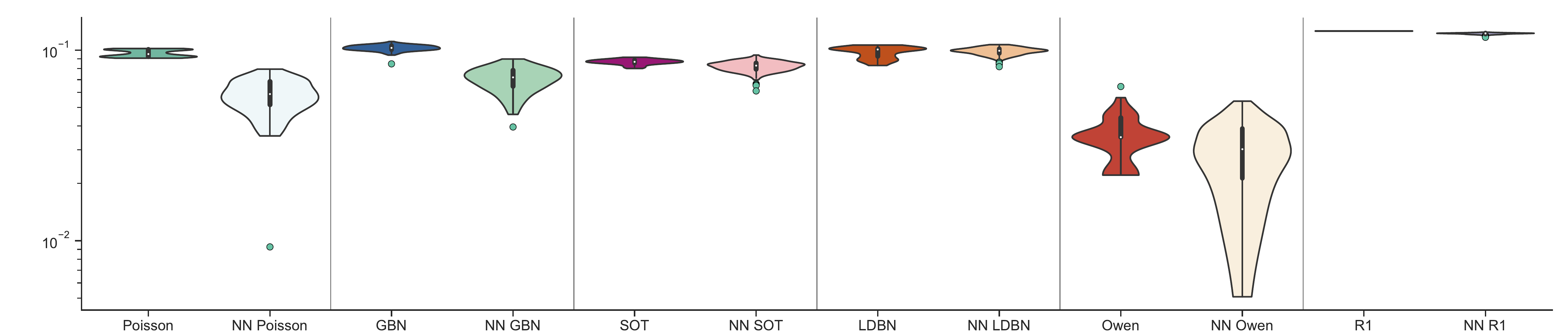}
    \put(-1,4){\rotatebox{90}{\bf MinDist, 64}}
  \end{overpic}
  \begin{overpic}[width=\linewidth]{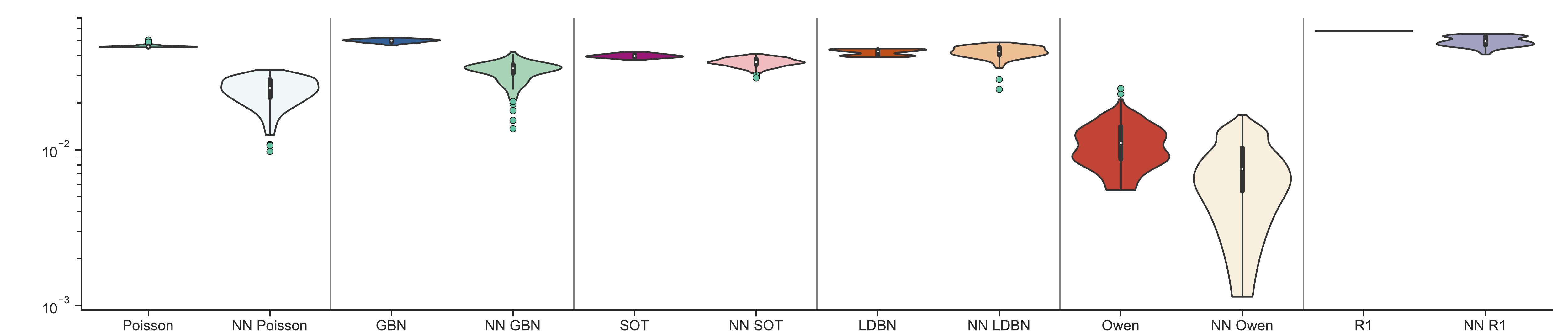}
    \put(-1,4){\rotatebox{90}{\bf MinDist, 256}}
  \end{overpic}
  \begin{overpic}[width=\linewidth]{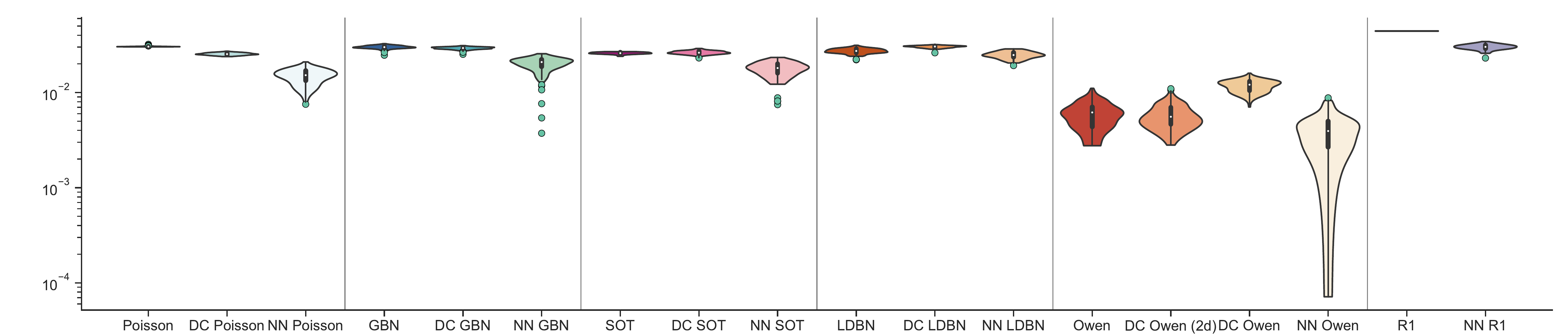}
    \put(-1,4){\rotatebox{90}{\bf MinDist, 576}}
  \end{overpic}
  \begin{overpic}[width=\linewidth]{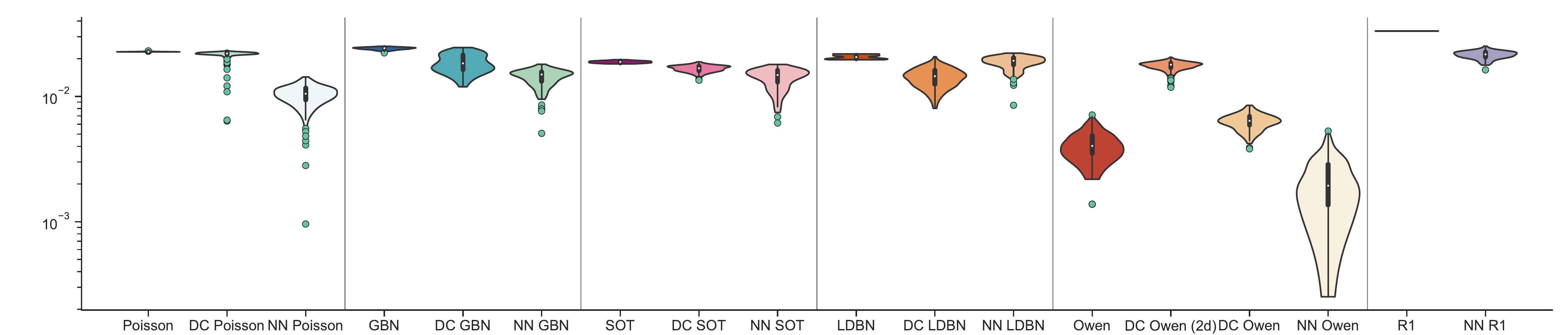}
    \put(-1,4){\rotatebox{90}{\bf MinDist, 1024}}
  \end{overpic}
  \begin{overpic}[width=\linewidth]{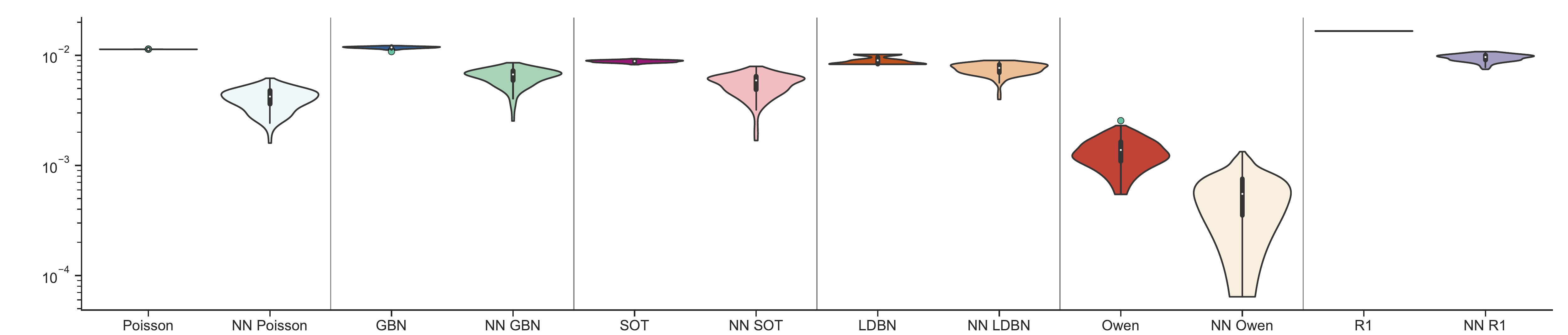}
    \put(-1,4){\rotatebox{90}{\bf MinDist, 4096}}
    \put(-2.5,9){\cfbox{orange}{\begin{minipage}{\linewidth}\vspace{1.65cm}\quad\end{minipage}}}
    \put(-2.5,52){\cfbox{orange}{\begin{minipage}{\linewidth}\vspace{1.65cm}\quad\end{minipage}}}
  \end{overpic}
  \caption{We evaluate the minimum pairwise distance between samples. This property is highly sensitive as it only depends on the location of 2 samples. Our network tends to produce smaller values, while the sample repulsion of \cite{leimkuhler2019deep} better preserve minimum distances. Sample counts are 64, 256, 576, 1024 and 4096 (top to bottom).}
   \label{fig:distances}
\end{figure}

\sloppy While we trained our network on small set of sample sizes ($\{64, 256, 1024\}$), 
we assess the performance of these metrics for other sample sizes ($\{576, 4096\}$).
For most of these properties, we illustrate them with violin plots (Fig.~\ref{fig:OTenergy}, \ref{fig:integ}, \ref{fig:disc}, \ref{fig:distances}), 
that show the distribution of values in the form of vertical histograms  (e.g., similar 
to a population pyramid). We compute them using 128 point sets.

\paragraph{Power spectra.} In Fig.~\ref{fig:spectra}, we first
  show performances of  \cite{leimkuhler2019deep} and our approach to
  recover spectral properties of the training sets (either through
  1d radial mean power spectra for stationay and isotropic point sets,
  or 2d spectra for other ones). As discussed above, capturing
  anisotropic spectra with \cite{leimkuhler2019deep} is very
  challenging using a 2d spectra loss function. Our approach fully
  captures such characteristics.

\paragraph{Optimal transport energy.} Optimal transport (OT) provides
a way to characterize the uniformity of a point set by computing the
(squared) semi-discrete optimal transport distance between the point
set and a uniform distribution~\cite{merigot2011multiscale}. 
Fig.~\ref{fig:OTenergy} illustrates how we match the OT energy.

\paragraph{Discrepancy and integration error.} Fig.~\ref{fig:integ} and \ref{fig:disc} show how our network matches integration errors and discrepancy of point sets. For discrepancy, we used the L2 discrepancy~\cite{Niederreiter1992,heinrich1996efficient}. For integration error, we compute the average MSE on the integration of wide anisotropic  Gaussians (anisotropic ratio between 1:1 and 1:9, and Gaussian sizes ranging from 0.1 to 0.333 for its largest axis) or Heaviside distributions randomly linearly dividing the unit square. We randomly chose 64k integrands among 1 million, whose integral has been estimated with maximum precision as reference.  These statistics also often match for sample sizes not seen during training ($\{576, 4096\})$.

\paragraph{Minimum distance.} For distributions such as Poisson Disk, the minimum distance between any pair of samples can be important. We assess this statistics in Fig.~\ref{fig:distances}. This property is highly sensitive as it only depends on the location of 2 points within the entire point set. For this property, the approach of \cite{leimkuhler2019deep} performs remarkably well, due to the repulsion of points introduced during learning. In our approach, we tend to produce points with lower minimum distance value.

\subsection{Non-uniform distributions}
\label{sec:nonuniform}
\begin{figure}[!tbh]
  \centering
  \begin{tabular}{cc}
  \includegraphics[width=3cm]{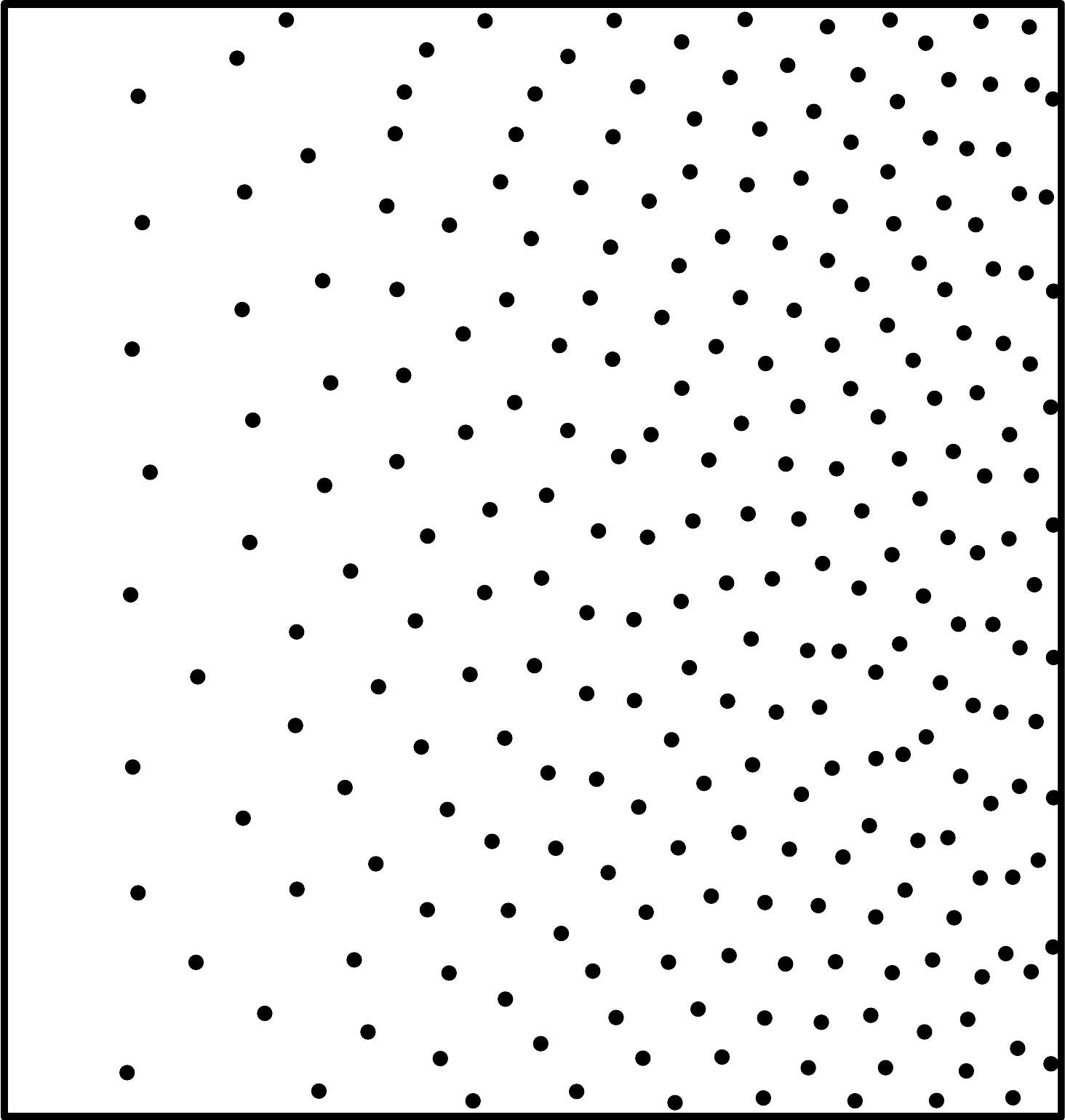} & \includegraphics[width=3cm]{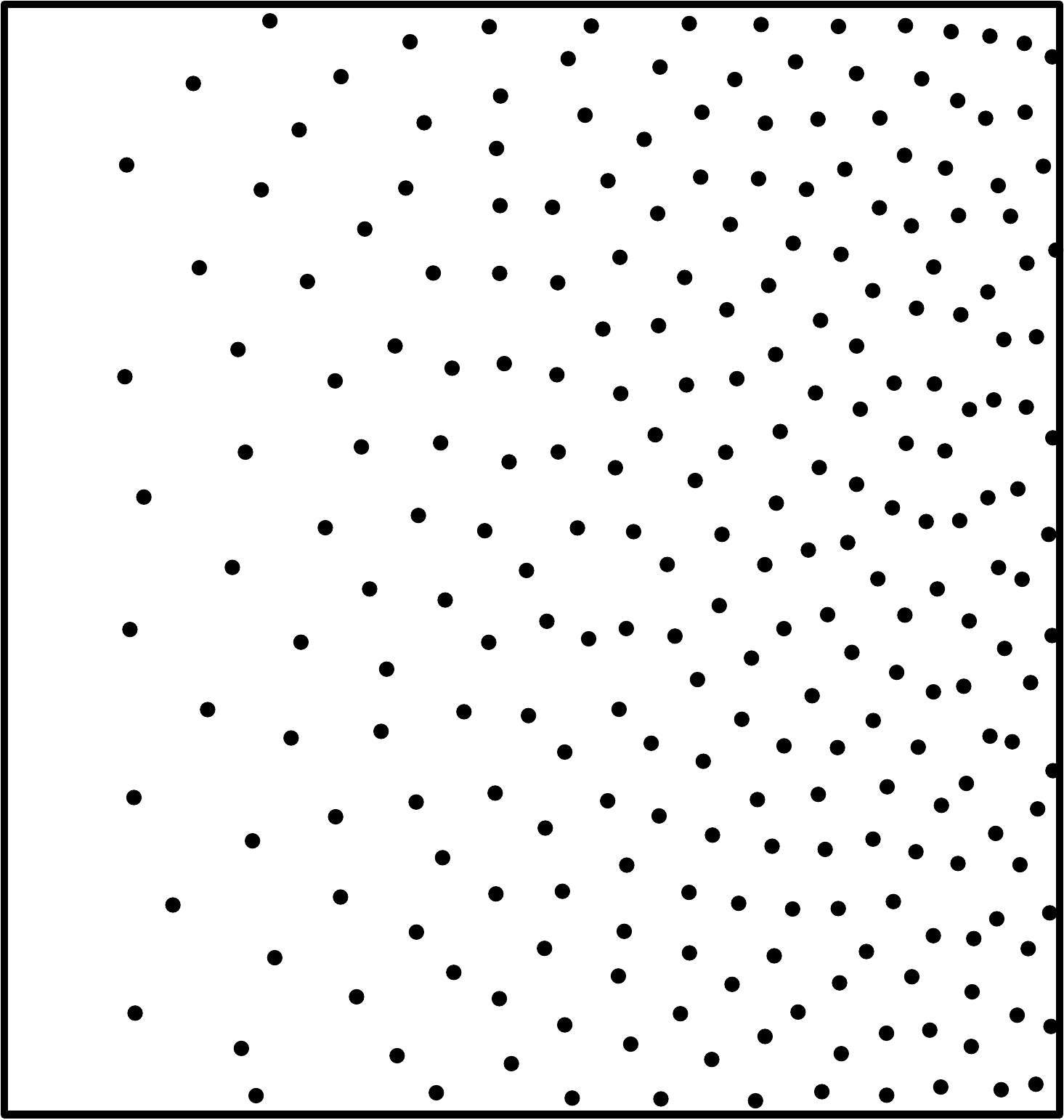} \\
   \includegraphics[width=3cm]{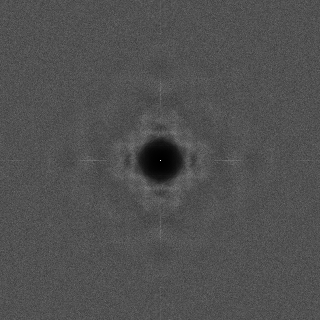} & \includegraphics[width=3cm]{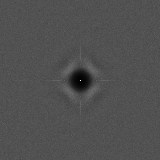} 
   \end{tabular}
   \caption{We sample from a learned sliced OT linear ramp. \textbf{Top row, left.} One example point set used for training (among 66,035). \textbf{Top row, right.} One synthesized point set. \textbf{Bottom row.} Unwarping example and synthesized point sets to recover a uniform distribution shows that their spectra match. The uniformity of the unwarped samples can also be measured: the semi-discrete optimal transport energy averaged for 128 realizations of 256 samples is $7.24 . 10^{-4}$ for the neural network output, compared with $7.16 . 10^{-4}$ for the original sliced OT uniform samples. }
   \label{fig:nonuniform}
\end{figure}

\begin{figure}[!tbh]
  \centering
  \begin{overpic}[width=1.5cm]{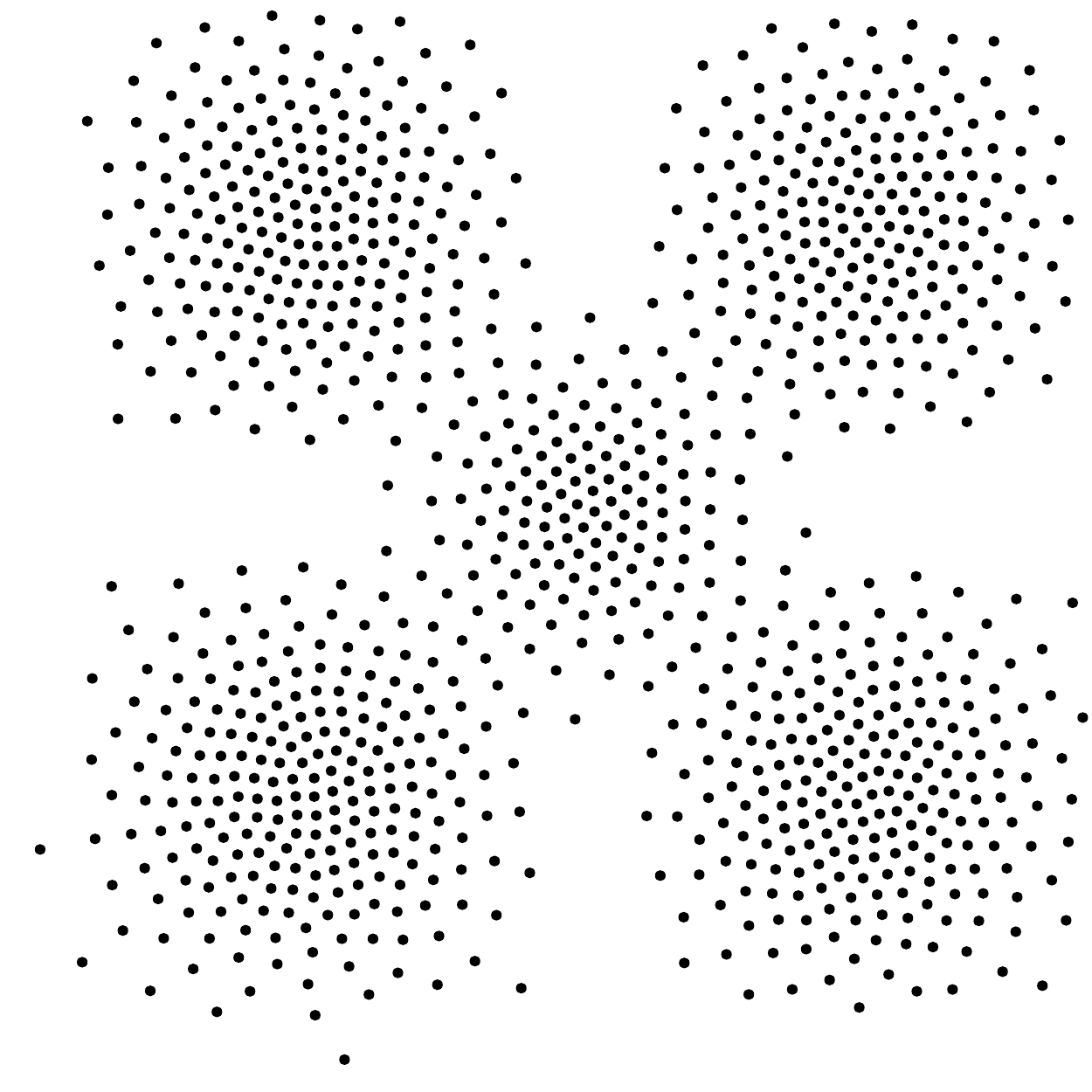}
    \put(-20,0){{\includegraphics[width=1cm]{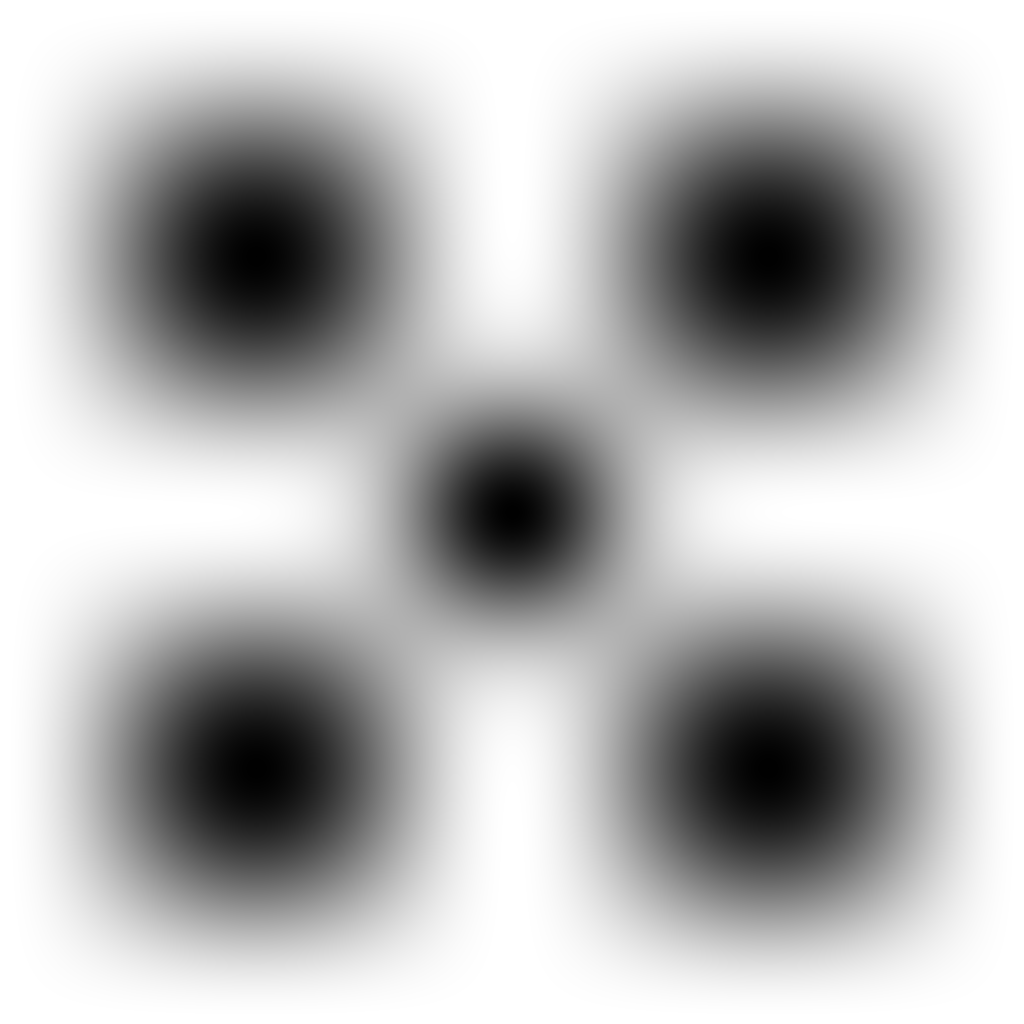}}}
  \end{overpic}
  \includegraphics[width=1.5cm]{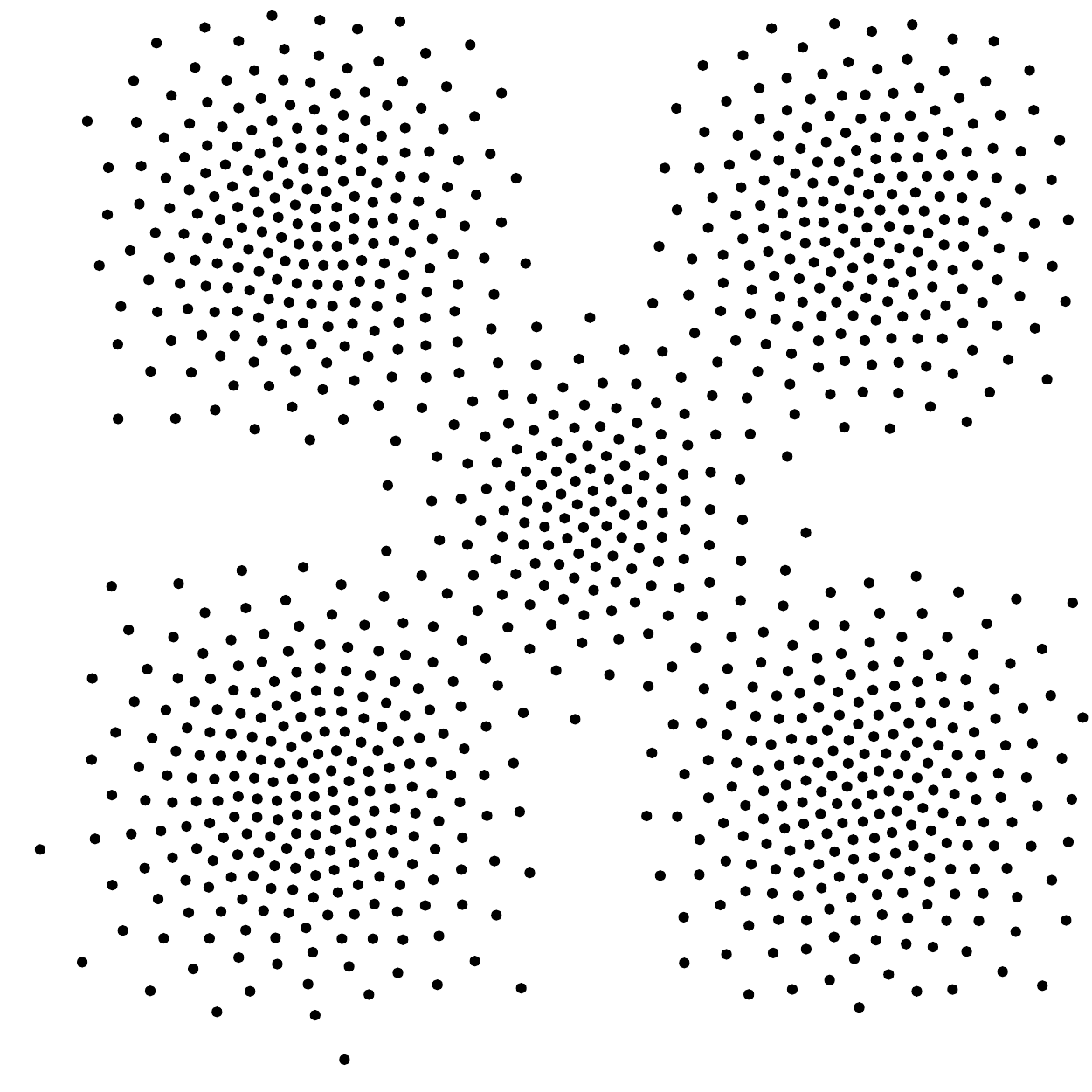}
  \includegraphics[width=1.5cm]{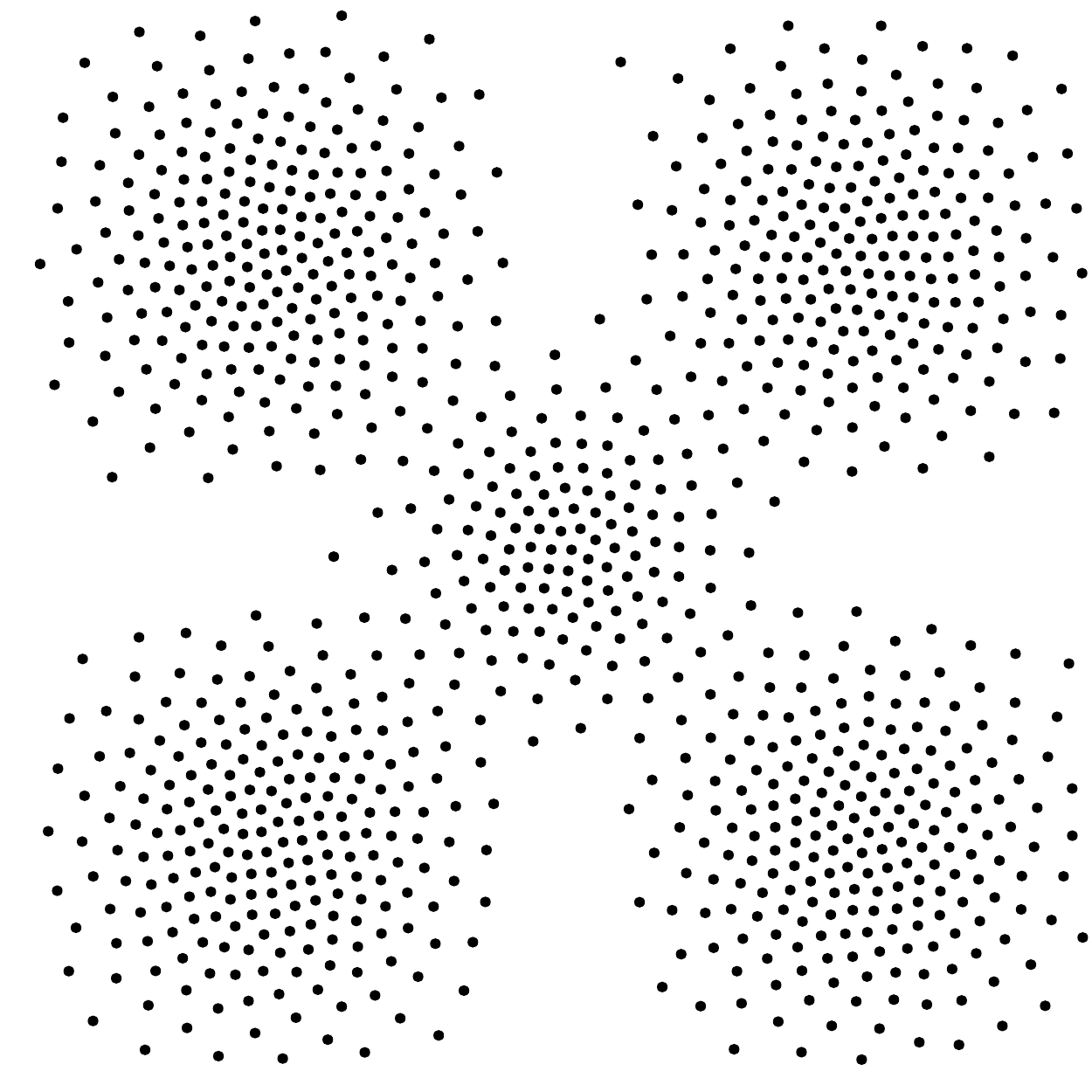}
  \includegraphics[width=1.5cm]{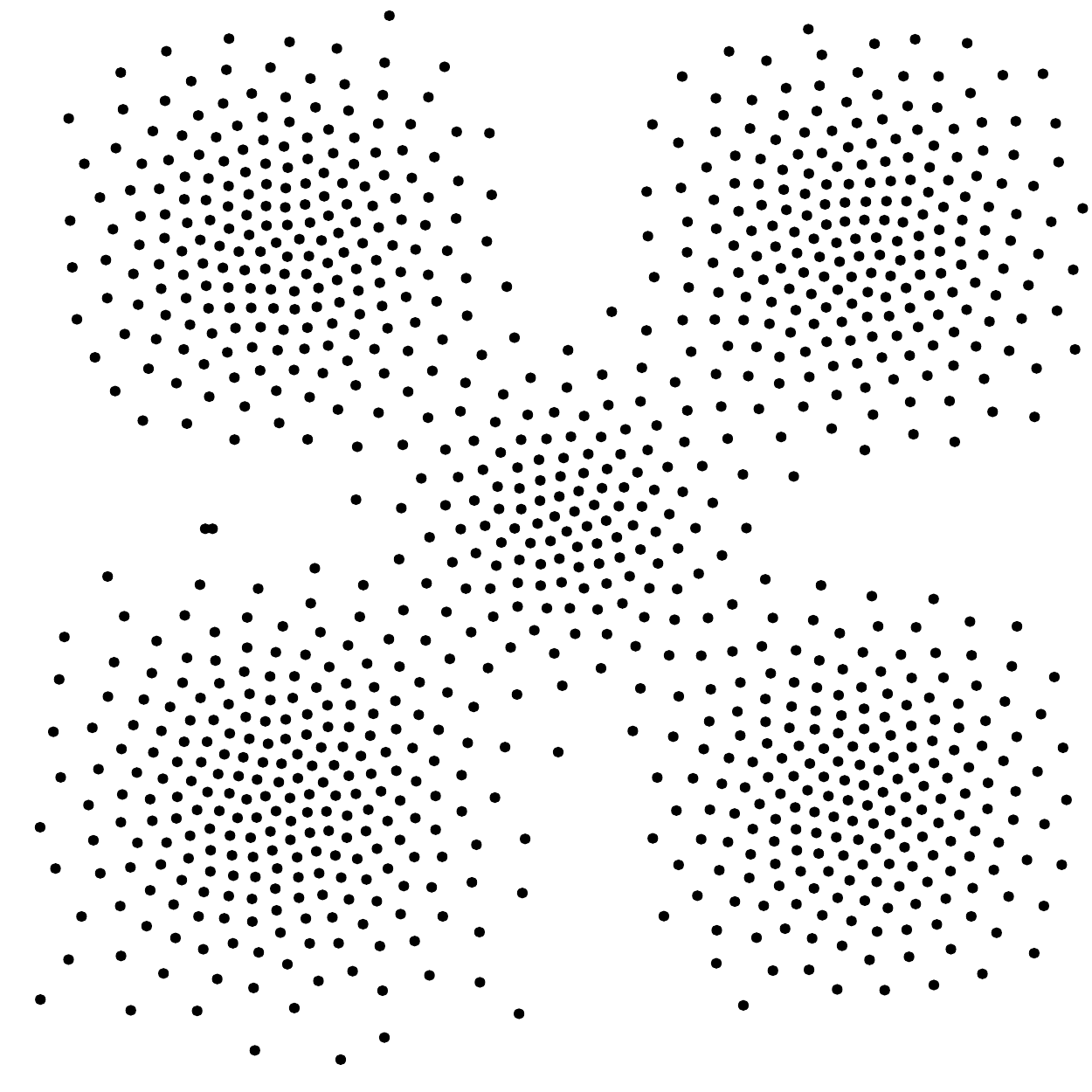}
  \includegraphics[width=1.5cm]{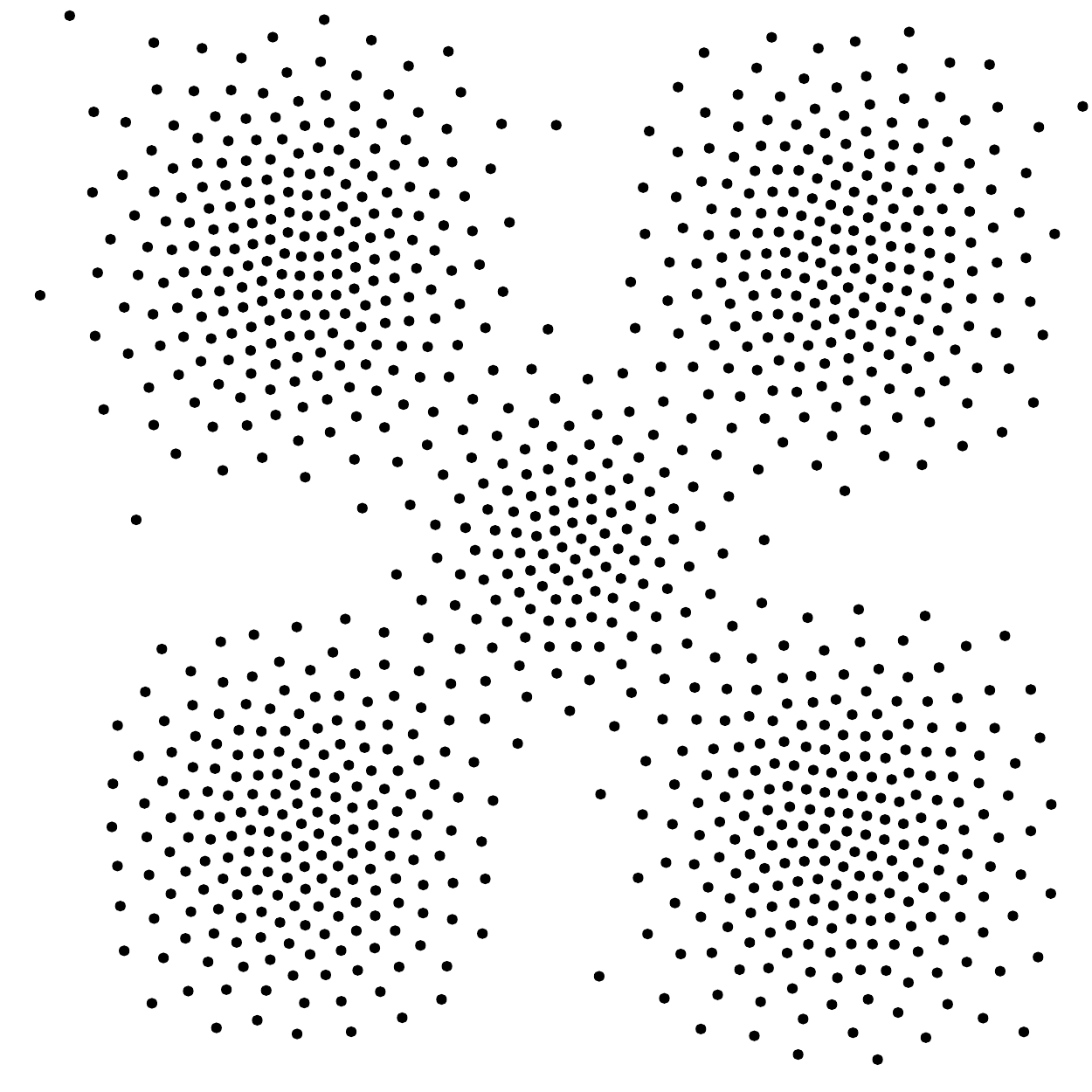} \raisebox{.7cm}{~~~~~\ldots}\\
  ~~\\
  \includegraphics[width=1.5cm]{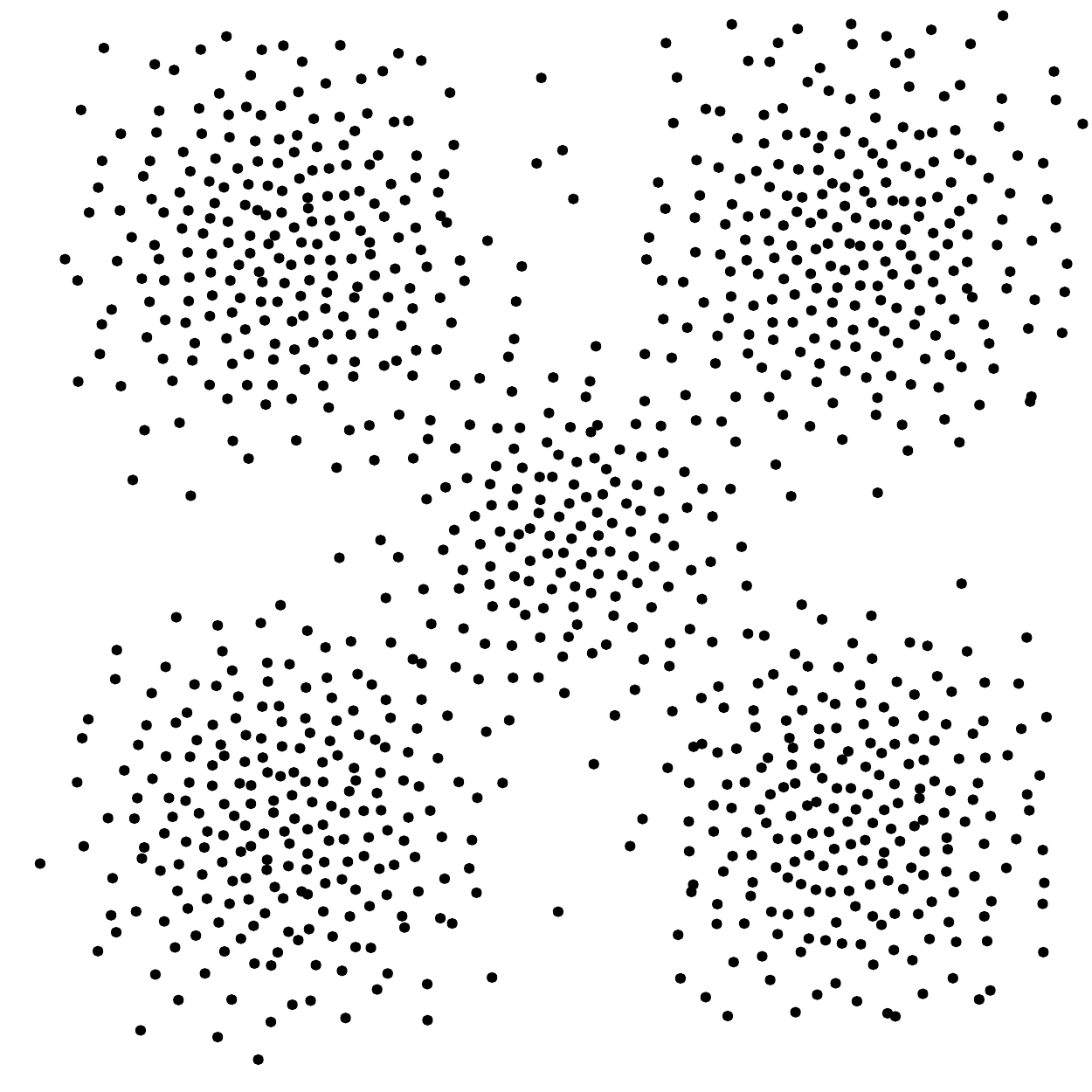}
  \includegraphics[width=1.5cm]{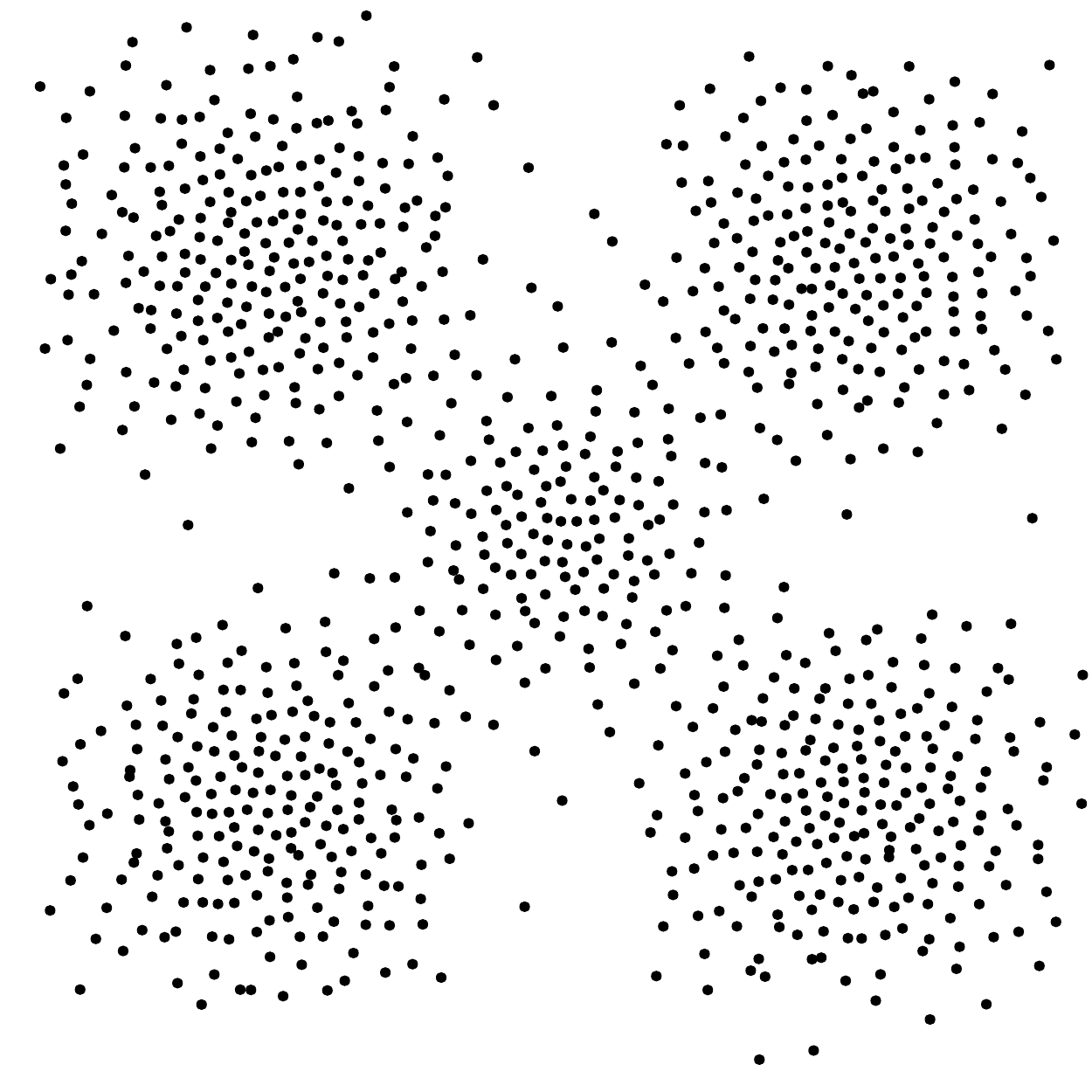}
  \includegraphics[width=1.5cm]{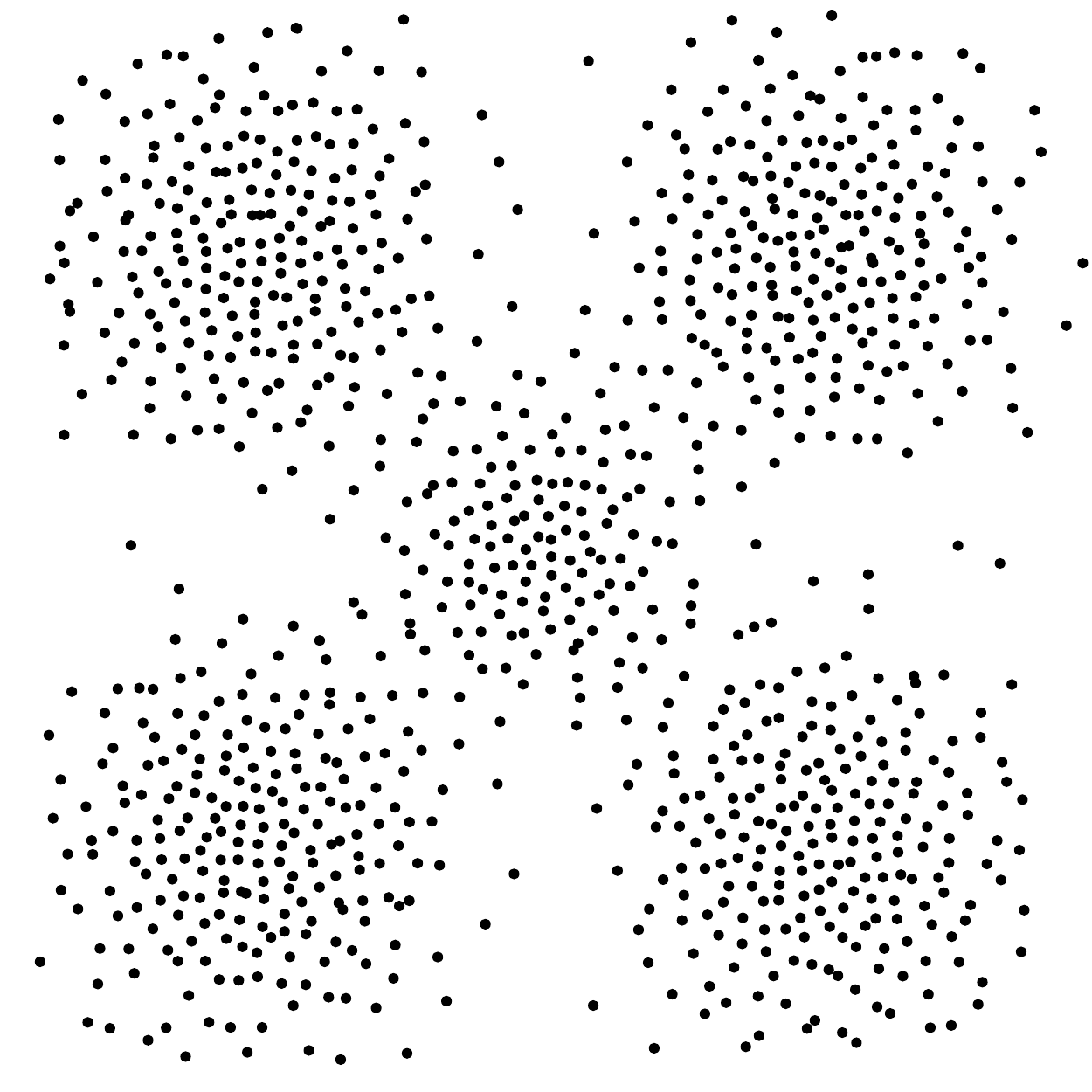}
  \includegraphics[width=1.5cm]{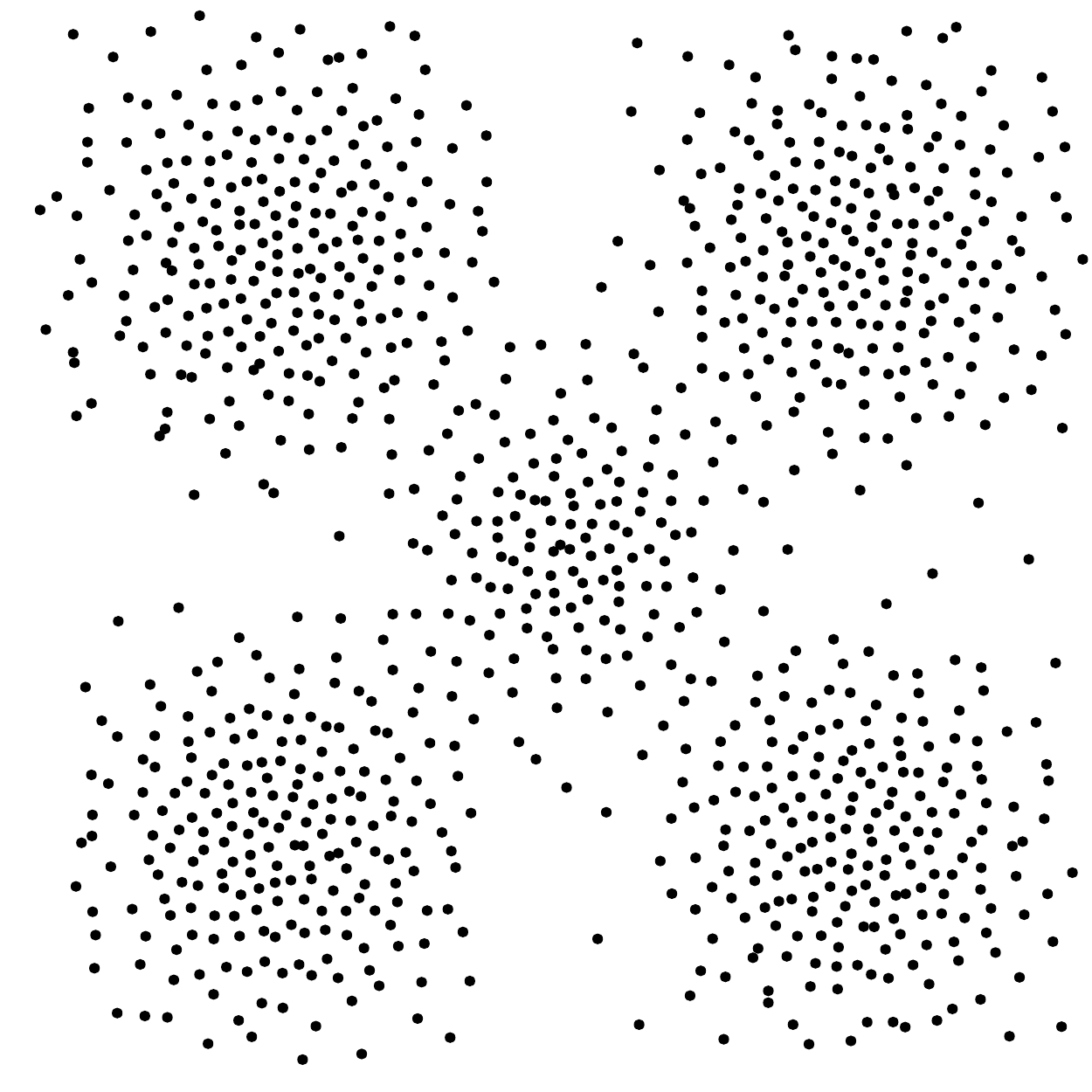}
  \includegraphics[width=1.5cm]{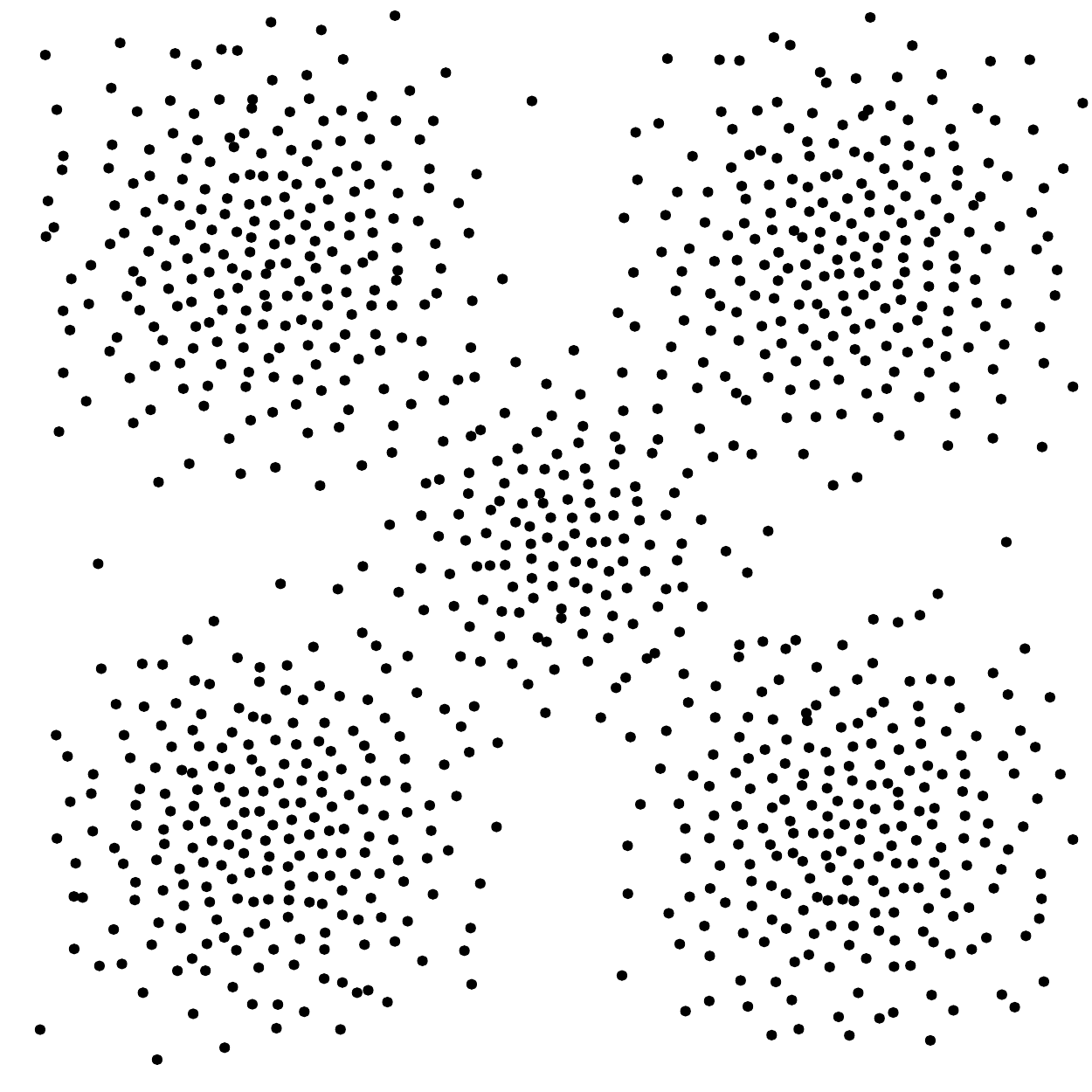} \raisebox{.7cm}{~~~~~\ldots}\\
  
   \caption{As a stress test, we sample from the density $0.2 e^{-20(x^2 + y^2)} + 0.2 \sin(\pi x)^2 \sin(\pi y)^2$ by importance sampling using GBN as a training set (first row). 
   Our sampler reproduces the density well and mostly preserves important characteristics of the sampler (second row).}
   \label{fig:nonuniform2}
\end{figure}

The goal of our optimal transport matching to a uniform grid is to infer neighborhood information on the point sets from neighborhood information on the grid, that is, neighboring points on the grid are expected to correspond to neighboring samples. In Fig.~\ref{fig:nonuniform}, using a non-uniform linear ramp sliced optimal transport sampling, we show that, even for non-uniform sampling, our network successfully learns from examples and preserve spectral noise characteristics of the sampler. As a stress test, we also learn to sample a blobby function shown in Fig.~\ref{fig:nonuniform2}. In this example, we learn from importance sampled GBN point sets obtained by rejection sampling. Our network reproduces the sampling density well, and mostly preserves important characteristics of the GBN sampler despite inaccuracies in neighborhood information due to the grid embedding. 
Non-uniform sampling is not possible with the approach of~\cite{leimkuhler2019deep}.

\subsection{Applications}

Aside from the fast generation of point sets, we also benefit from the differentiability of our network to further optimize point sets within their class.

\begin{figure}[!tbh]
  \centering
  \begin{overpic}[width=\linewidth]{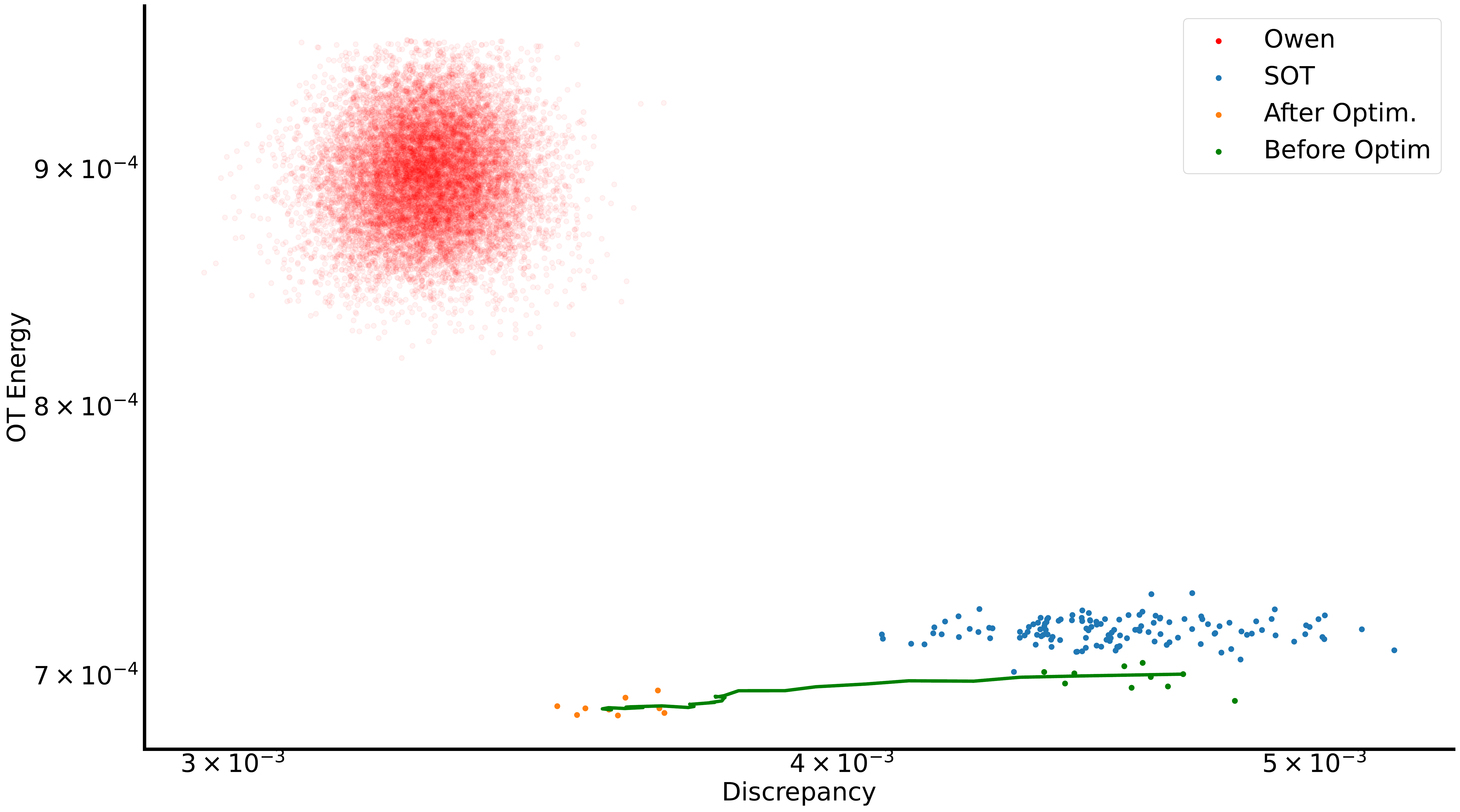}
    \put(65,20){\includegraphics[width=1.5cm]{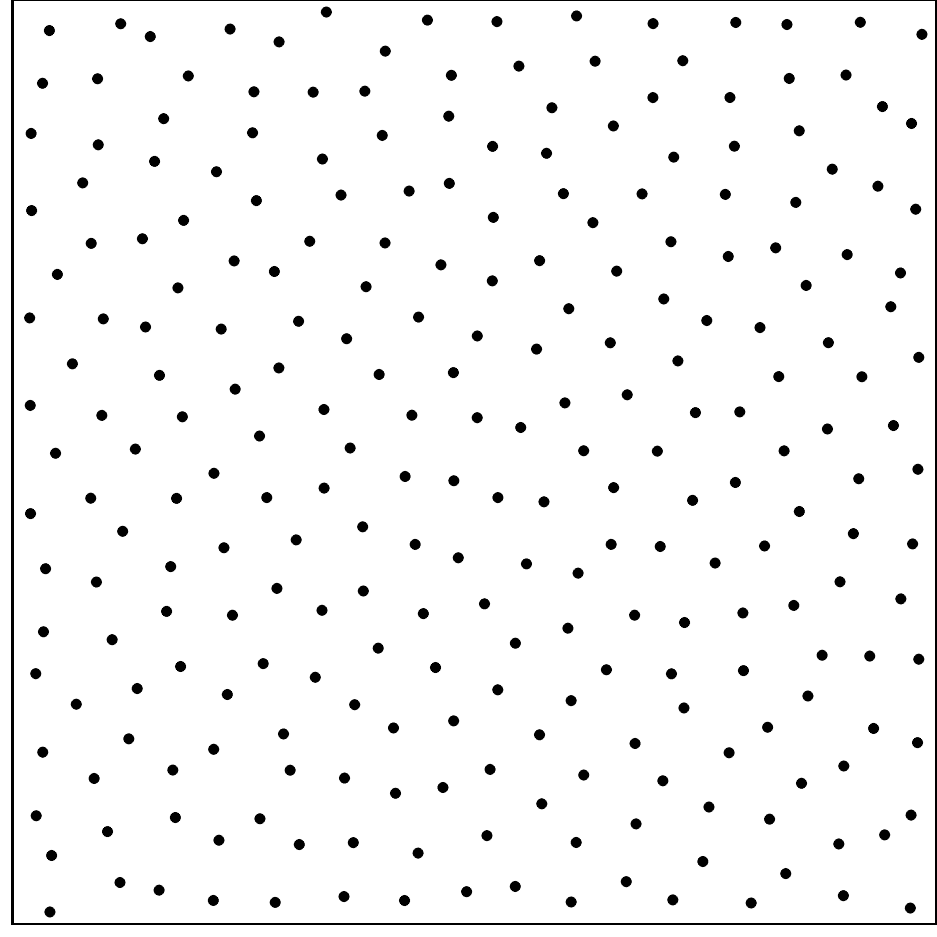}}
    \put(20,10){\includegraphics[width=1.5cm]{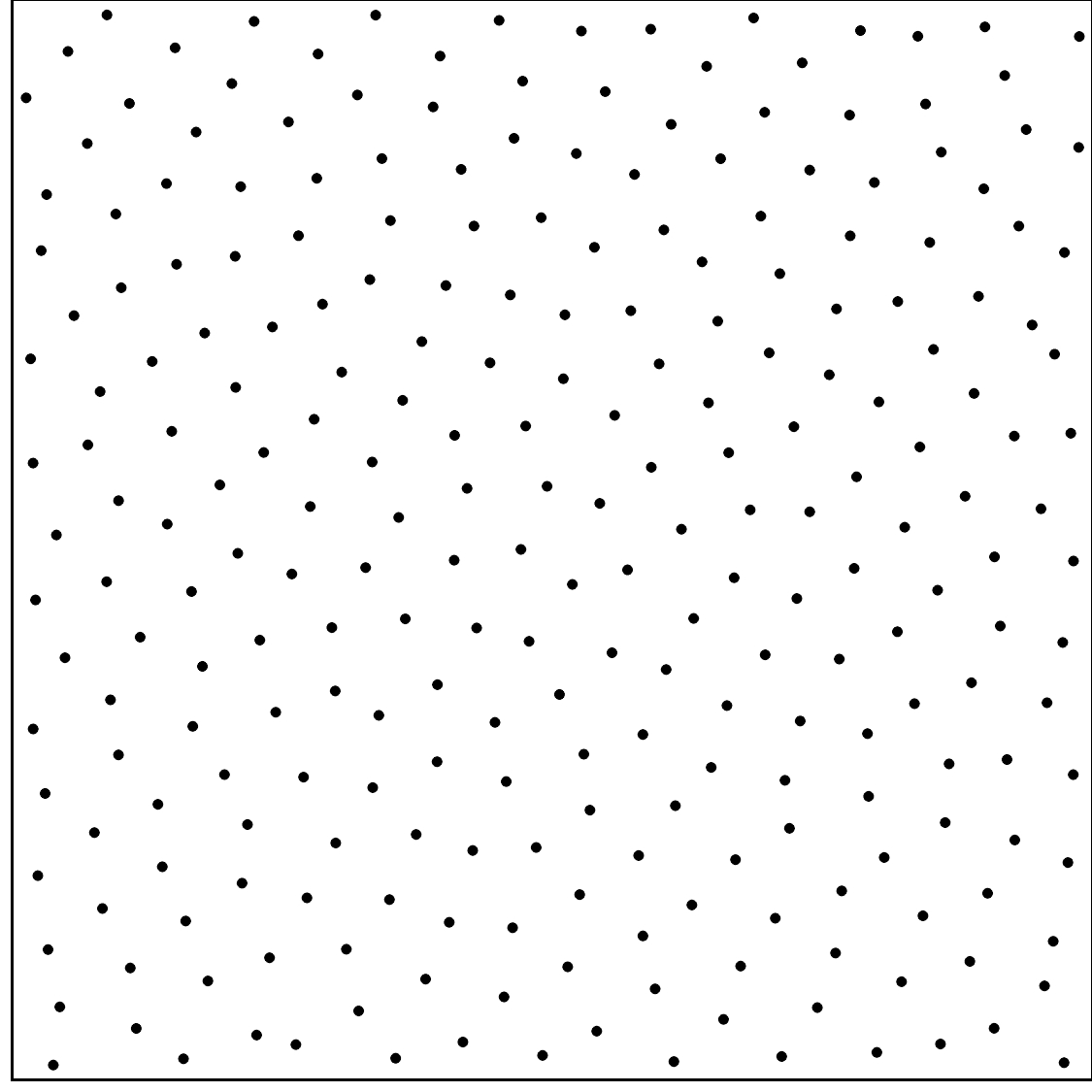}}
  \end{overpic}
   \caption{We used a trained SOT sampling network to optimize the discrepancy of the generated point sets among the class of SOT point sets. For 128 SOT (blue) and some Sobol'+Owen (red) point sets as representative of blue noise and LDS samplers, we show their distribution of OT and discrepancy statistics. In orange, we illustrate the SOT and discrepancy value for 10 optimized point sets as well as a representative trajectory during the optimization process. We also show a representative point set before (right) and after (left) optimization.}
   \label{fig:optimization}
\end{figure}

We illustrate how the differentiability of our network can be used to add properties to generated point sets. Here, we wish to add low discrepancy properties to a sliced optimal transport sampler~\cite{paulin2020sliced}, to benefit from both low discrepancy and low optimal transport energy. We train the network on SOT and then fix the trained weights of the network. Then we optimize the initial white noise samples with an objective function aimed at minimizing the L2 discrepancy measure. As backpropagation requires significant memory overhead, we reduce the number of diffusion steps to 100 (instead of 1000) in the diffusion model. 
In Fig.~\ref{fig:optimization}, we illustrate the result of our optimization in terms of discrepancy and optimal transport energy, and illustrate with an example generated point set.

\section{Discussions \& Perspectives}

We showed that diffusion models provide a powerful tool for learning how to generate point sets directly from examples across a wide range of samplers and they generalize well with sample size. Generalization hints at the fact that the network is correctly learning the general principles that make each point set so particular. An interesting future work would involve conditioning the network with respect to the particular sampler, sampler type or more general desired properties. This would allow for a single trained network to produce point sets of types. Preliminary experiments showed subpar results, but more complex architectures could alleviate this issue. The capacity of our network to produce possibly non-uniform example-based point sets may open the door to syntheses where sampling data are only available through a small number of measurements (e.g., distribution of trees, cells, etc.) and optimizing only for summarized statistics (power spectrum or PCF) is not desired. This is a promising direction as we have successfully trained our network with 32 examples of the SOT sampler.

While in principle our method would work in arbitrary dimension, the efficiency gained through our convolutions on grids would be lost as storing higher dimensional grids becomes impractical, both in terms of storage (that exponentially grows with dimension) and supported sample size (in the form $k^d$ for some $k$, similarly to stratified samplers). To date, higher dimensional data would be better supported by the approach of~\cite{leimkuhler2019deep} that does not rely on grids.
To remove this grid-dependency in the Monte Carlo sampling, one could adapt recent diffusion models for 3D point cloud shape synthesis~\cite{luo2021diffusion,zeng2022lion}. While our network is reasonably efficient, other recent architectures have been proposed to accelerate diffusion models and could be explored as well~\cite{song2020denoising}.

However, in the settings we focus on, in most cases our samples preserve characteristics of major samplers well, including their power spectrum, Monte Carlo integration quality, distance statistics, optimal transport energy and discrepancy. Our diffusion-based sampler allows to generate point sets much faster than some optimization-based samplers by learning from their output. Aside for the fast generation of diverse point sets, we have shown use for our network's differentiability by adding a low discrepancy property to an optimal transport-based sampler. Rendering applications could benefit from our samplers, e.g., through differentiable rendering pipelines~\cite{Jakob2022} or for generating point sets nicely distributing Monte Carlo error in a blue noise fashion in screen space~\cite{salaun2022scalable}.

\section*{Acknowledgments}
This work was funded in part by the french Agence Nationale de la Recherche, grant ANR-20-CE45-0025.

\bibliographystyle{ACM-Reference-Format}
\bibliography{bibliography}

\newpage

\twocolumn[
\begin{center}
\textbf{\Large{Supplementary document}}
\end{center}
]
\setcounter{section}{0}

\vspace{2em}

\section{Diffusion model}

Diffusion models date back to the work of \citet{Sohl15} but were popularized by \citet{Ho20} for image synthesis.
This section recalls the details for completeness.

Probabilistic Denoising Diffusion models involve a \emph{forward} process, where noise is gradually added to the signal (here an image) and a \emph{reverse} process where noise is removed through a learnable network.
The forward diffusion process is a Markov Chain, where each transition adds Gaussian Noise to the image, following:
\begin{equation}
 q(x_{t}|x_{t-1}) = \mathcal N(x_{t}; \sqrt{1-\beta_t}x_{t-1},\beta_t I)\,,
\end{equation}
where $(\beta_t)_{t=0}^T$ are the noise variances for each time $t$. The variance schedule is chosen such that nothing distinguishes $x_T$ from a white noise.
In our model, we set the variances $\beta_t$ to be constant $\beta_t=\beta$

One has:
\begin{equation}
 q_{x_1:T|x_0}=\prod_{t=1\cdots T} q(x_t|x_{t-1})\,.
\end{equation}

The reverse (denoising) process is also a Markov Chain, with transitions:
\begin{equation}
 p_\theta(x_{t-1}|x_t) = \mathcal N(x_{t-1};\mu_\theta(x_t,t), \Sigma_\theta(x_t,t))\,,
\end{equation}
$\mu_\theta$ and $\Sigma_\theta$ are learned by examples. To simplify, following the work of \citet{Ho20}, we consider that $\Sigma_\theta = \sigma_t I$, with $\sigma_t=\beta_t=\beta$.
The forward process allows to sample $x_t$ with arbitrary $t$ from $x_0$, following:
\begin{equation}
q(x_t|x_0)=\mathcal N(x_t;\sqrt{\bar\alpha_t} x_0, (1 -\bar\alpha_t) I)\,,
\end{equation}
with $\alpha_t = 1 - \beta_t $ and $\bar \alpha_t = \prod_{s=1}^t \alpha_s$.

During training, and image $x_0$ is drawn from the set of examples, along with a random time $t\in{1\cdots T}$, a random noise image $\epsi$ is drawn following $\mathcal N(0,I)$ and the algorithm tries to minimize:
\begin{equation}
 \|\epsi - \epsi_\theta(\sqrt{\bar \alpha_t}x_0+\sqrt{1-\alpha_t}\epsi,t)\|^2\,,
\end{equation}
by gradient descent.

During sampling a random noise image $z\sim \mathcal N(0,I)$ is drawn and iteratively denoised by applying:
\begin{equation}
 x_{t-1} = \frac1{\sqrt{\bar \alpha_t}}(x_t - \frac{1-\alpha_t}{\sqrt{1-\bar \alpha_t}}\epsi_\theta(x_t,t))+\sigma_t z\,,
\end{equation}
where $z$ is a random noise and in our case, we take $\sigma_t=\beta_t$
The key ingredient of diffusion models is the approximator $\epsi_\theta$, which is modeled by a neural network.

\section{Network}

\begin{figure}[htb]
 \begin{center}
    \includegraphics[width=\linewidth]{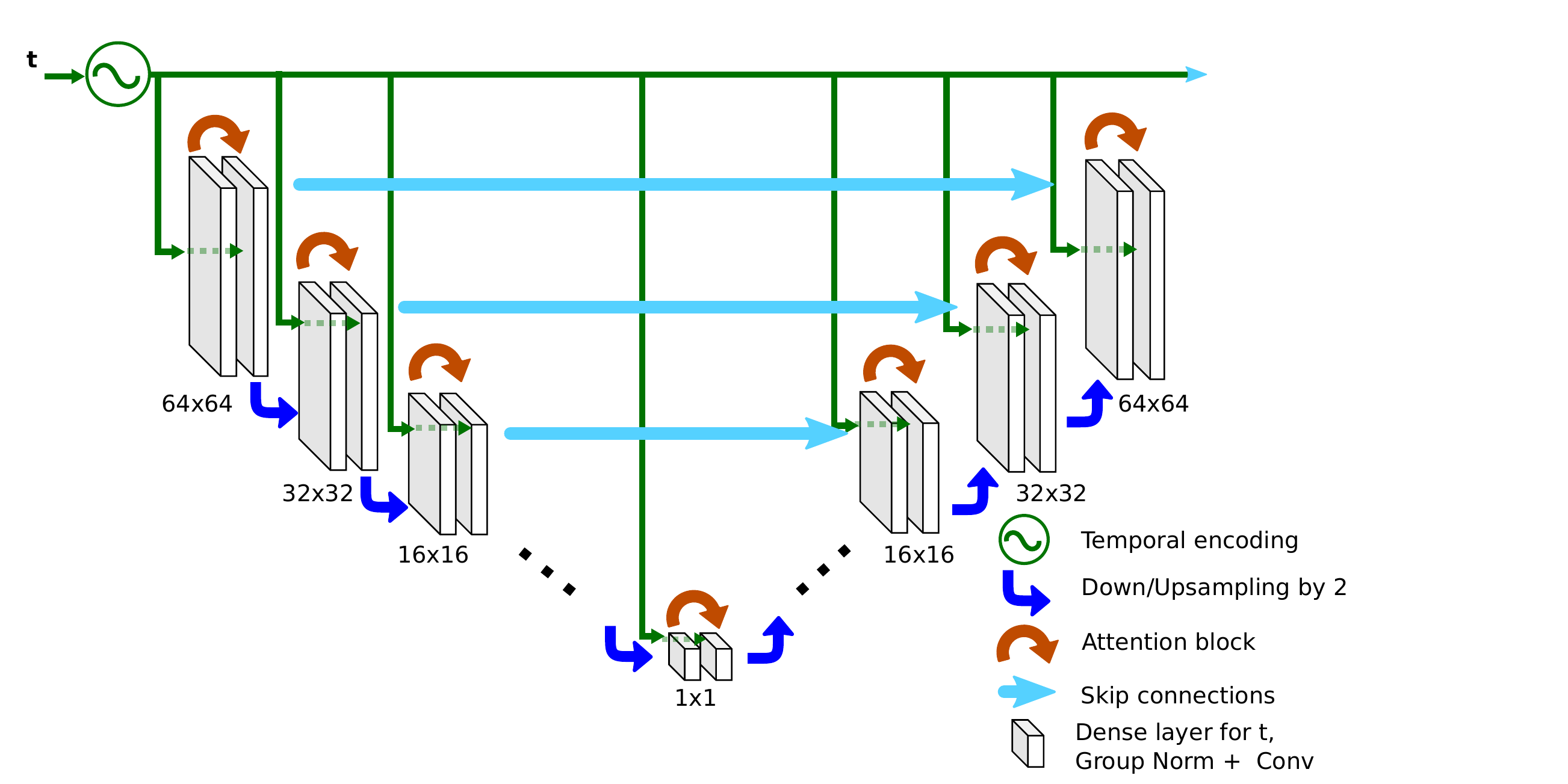}
 \end{center}
\caption{Diffusion network architecture}
\label{fig:network}
\end{figure}

Our network is a slightly modified version of the denoising network of \citet{Ho20} and is summarized on Figure \ref{fig:network}.

\section{Learning Rank-1 realizations with  \cite{leimkuhler2019deep}}

\begin{figure}
 \raisebox{.5cm}{\rotatebox{90}{Original}} \includegraphics[width=2cm]{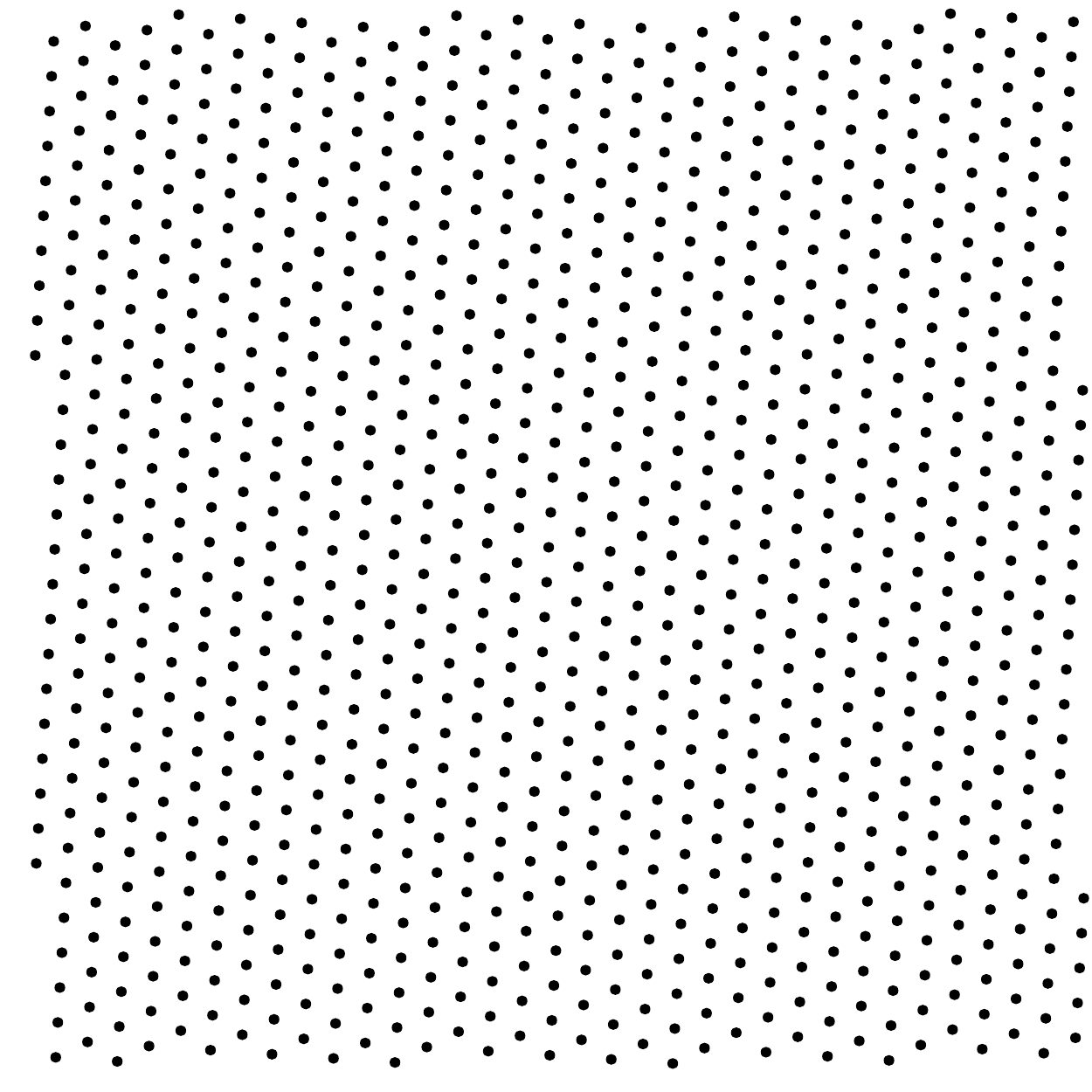}
  \includegraphics[width=2cm]{res/spectrums/Orig/spectrumR1} \raisebox{.3cm}{\scalebox{0.3}{\input{res/spectrums/Orig/radialR1.tex}}}\\
 \raisebox{.5cm}{\rotatebox{90}{Our}} \includegraphics[width=2cm]{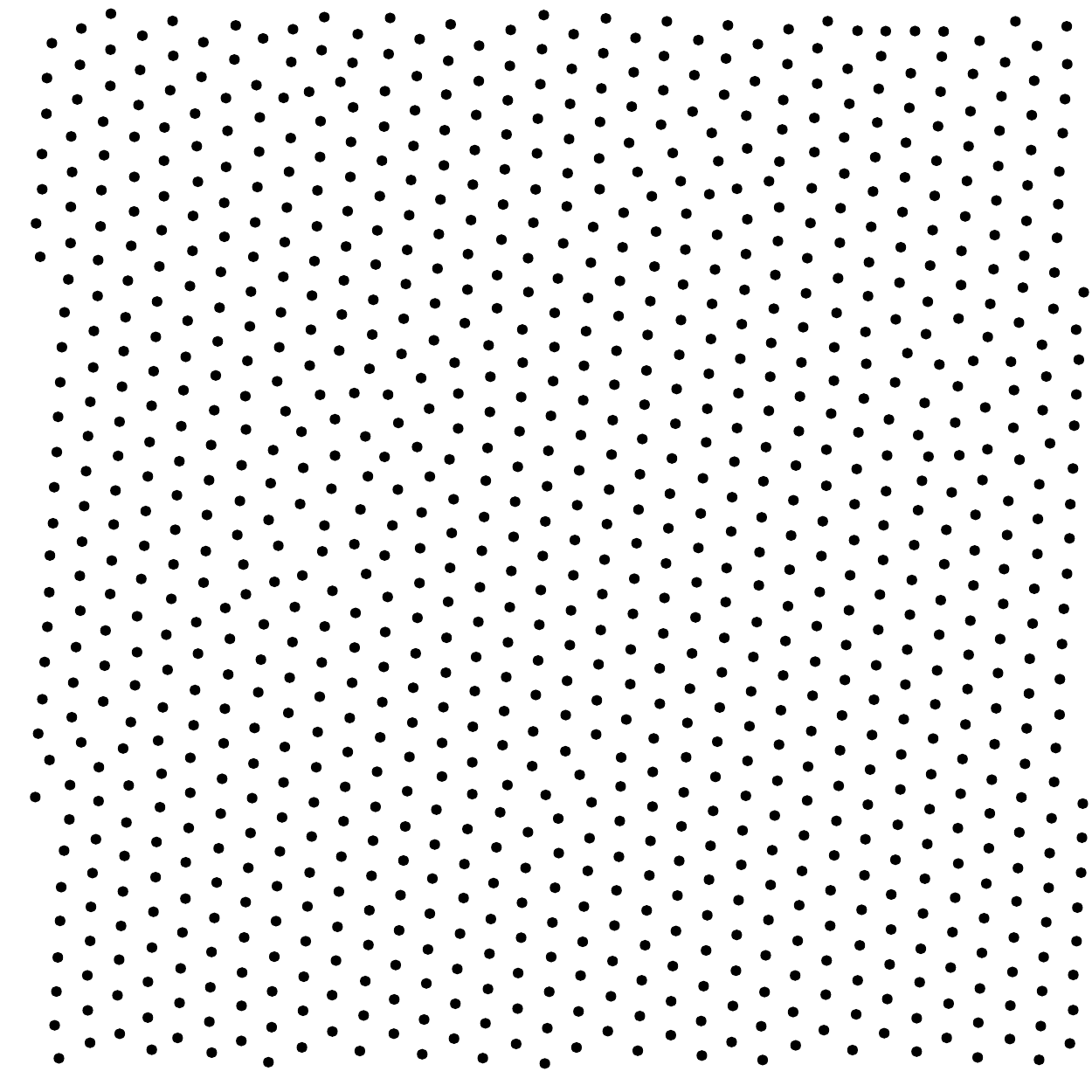}
  \includegraphics[width=2cm]{res/spectrums/Final/spectrumR1} \raisebox{.3cm}{\scalebox{0.3}{\input{res/spectrums/Final/radialR1.tex}}}\\
  \raisebox{.5cm}{\rotatebox{90}{2d target}} \includegraphics[width=2cm]{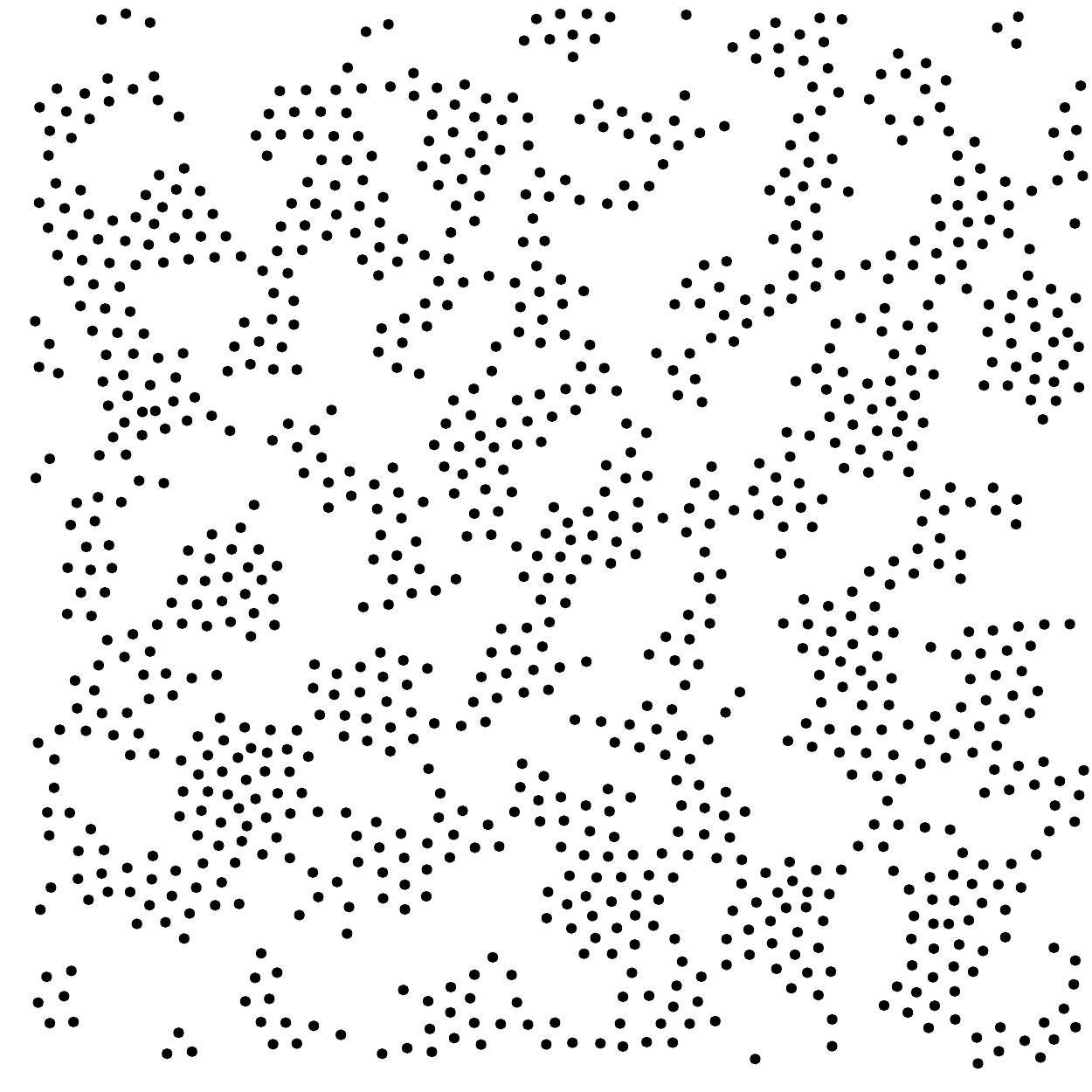}
  \includegraphics[width=2cm]{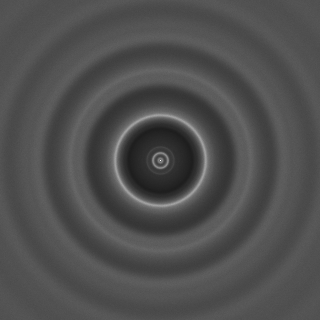} \raisebox{.3cm}{\scalebox{0.3}{\input{res/spectrums/DeepPointCorrelation/radialr1-2d.tex}}}\\
 \raisebox{.5cm}{\rotatebox{90}{1d target}} \includegraphics[width=2cm]{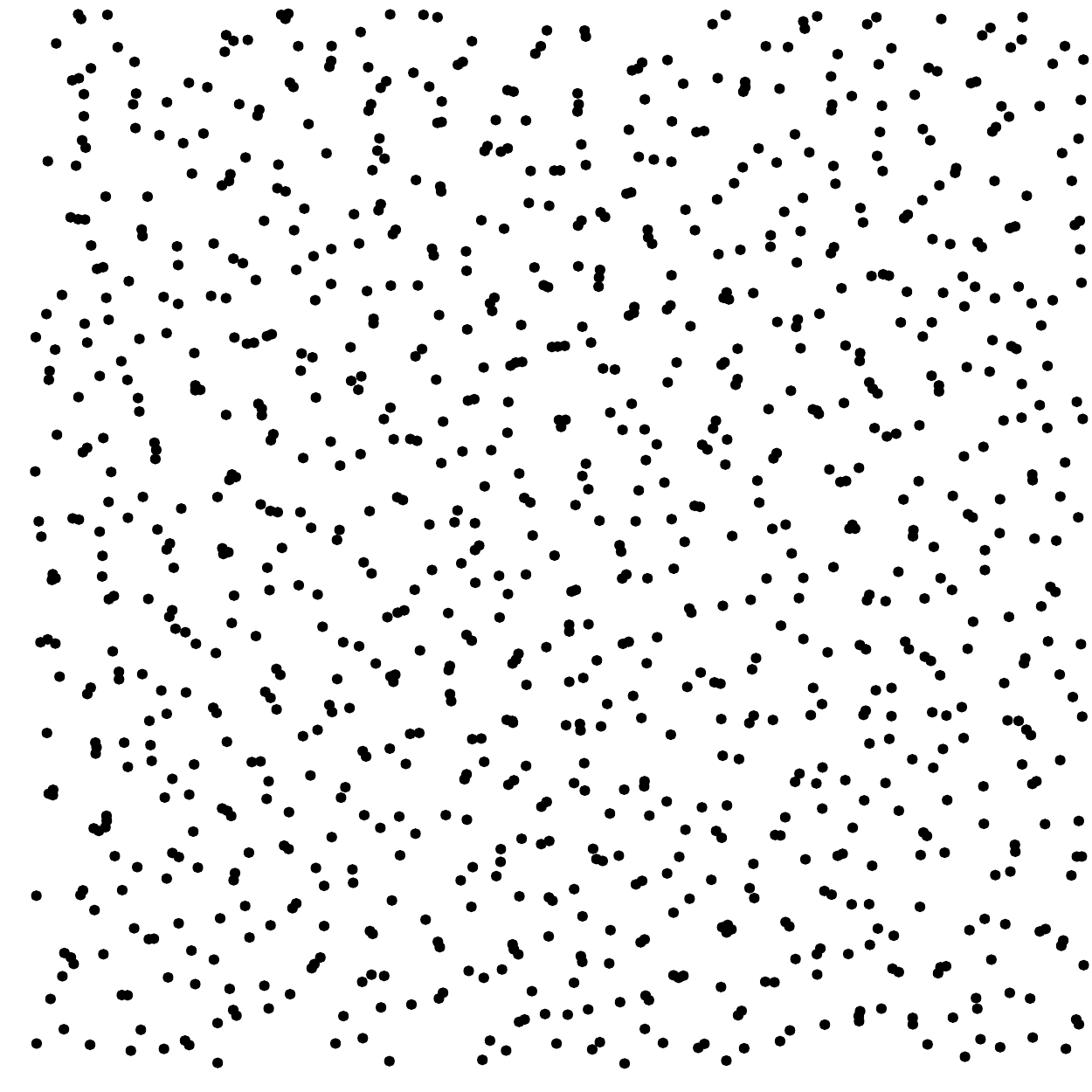}
  \includegraphics[width=2cm]{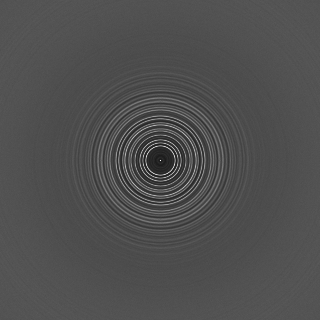} \raisebox{.3cm}{\scalebox{0.3}{\input{res/spectrums/DeepPointCorrelation/radialr1-1d.tex}}}\\
  \caption{Learning Rank-1 realizations using
    \cite{leimkuhler2019deep} using a 2d power spectra loss, and a 1d
    power spectra loss. We recall the original properties and our
    results for completeness.\label{fig:rank1b}}
\end{figure}

\citet{leimkuhler2019deep} proposed a neural network based point
process design using losses defined from spectral or pair-correlation
information.  In most examples provided by the authors, 1d losses (or
combination of 1d losses) are considered using 1d radial mean power
spectra or 1d pair correlation functions (allowing complex designs
such as a high-dimensional point process with some specific spectral
properties for given 1d or 2d projections).  When targeting isotropic
samplers, the authors provided their experimental settings in
\url{https://github.com/sinbag/deepsampling}. We use the same
parameters for Poisson disk, GBN, SOT, LDBN, targeting their
respective radial power spectra. For Sobol'+Owen, we keep the same
settings but updated the loss function to target a 2d power
spectrum. Cropping the spectra to the central part of the domain allowed
us to obtain a convergence of the learning step (in our experiments,
increasing the cropping domain does not help the convergence). For Rank-1, no
satisfactory results have been obtained, for the sake of
completeness, we illustrate in Fig.\ref{fig:rank1b} the point set and
spectra we have obtained. Note that Rank-1 is a very specific
anisotropic sampler which is far from the context of
\cite{leimkuhler2019deep}. Although, additional investigation would be
interesting to continue.

\end{document}

%% file: macros.tex
\definecolor{codegreen}{rgb}{0,0.6,0}
\definecolor{codegray}{rgb}{0.5,0.5,0.5}
\definecolor{codepurple}{rgb}{0.58,0,0.82}
\definecolor{backcolour}{rgb}{0.95,0.95,0.92}
\definecolor{red}{rgb}{1.0,0.0,0.0}

\lstdefinestyle{mystyle}{
    commentstyle=\color{codegreen},
    keywordstyle=\color{magenta},
    numberstyle=\tiny\color{codegray},
    stringstyle=\color{codepurple},
    basicstyle=\ttfamily\footnotesize,
    captionpos=b,                    
    keepspaces=true,                 
    showspaces=false,                
    showstringspaces=false,
    showtabs=false,                  
    tabsize=2
}
\newcommand{\cfbox}[2]{%
    \colorlet{currentcolor}{.}%
    {\color{#1}%
    \fbox{\color{currentcolor}#2}}%
}

%% file: res/spectrums/Orig/radialPoisson.tex
\begin{tikzpicture}
  \begin{axis}[
    width=10cm,
    height=5cm,
    ymin = 0.000000e+00,
    ymax = 4.200000e+00,
    xmin = 0.000000e+00,
    xmax=4.937500e+00,
    xtick align = outside,
  ]
  \addplot[red, thick, name path=A] table {
    0.06250000000000000   0.06377653825999113
    0.12500000000000000   0.06707315950325266
    0.18750000000000000   0.07070029556466928
    0.25000000000000000   0.07483615500263455
    0.31250000000000000   0.07937433627444124
    0.37500000000000000   0.08854943419651216
    0.43750000000000000   0.10165579756119111
    0.50000000000000000   0.11907732409774793
    0.56250000000000000   0.14985676446951271
    0.62500000000000000   0.20132564706124964
    0.68750000000000000   0.28421140896829039
    0.75000000000000000   0.42507007633266164
    0.81250000000000000   0.63579428674753236
    0.87500000000000000   0.91367982281076487
    0.93750000000000000   1.19750449321964925
    1.00000000000000000   1.46246478615292341
    1.06250000000000000   1.62022284121075355
    1.12500000000000000   1.65486687170131774
    1.18750000000000000   1.58541355256833105
    1.25000000000000000   1.44889162600772514
    1.31250000000000000   1.29685964194505376
    1.37500000000000000   1.12979262248371692
    1.43750000000000000   0.99001021184840121
    1.50000000000000000   0.87428643440785447
    1.56250000000000000   0.79040302522081873
    1.62500000000000000   0.73035025791535035
    1.68750000000000000   0.70559063420977830
    1.75000000000000000   0.70370858791148849
    1.81250000000000000   0.72265053088288655
    1.87500000000000000   0.76702789765897572
    1.93750000000000000   0.82482775547189258
    2.00000000000000000   0.89226861502757193
    2.06250000000000000   0.98198652486233973
    2.12500000000000000   1.07161760149623642
    2.18750000000000000   1.13642125425131590
    2.25000000000000000   1.20125151279045705
    2.31250000000000000   1.23152475702997299
    2.37500000000000000   1.22991930436410679
    2.43750000000000000   1.20523676735814678
    2.50000000000000000   1.15515212839556702
    2.56250000000000000   1.09856934285690944
    2.62500000000000000   1.03828193206026587
    2.68750000000000000   0.97370150263252464
    2.75000000000000000   0.92370620791830382
    2.81250000000000000   0.88582809203025414
    2.87500000000000000   0.86360379949217869
    2.93750000000000000   0.84590946934568412
    3.00000000000000000   0.84856422007313703
    3.06250000000000000   0.86618998064157482
    3.12500000000000000   0.89795125760085437
    3.18750000000000000   0.92964378639373846
    3.25000000000000000   0.97567742590254425
    3.31250000000000000   1.01272445197186589
    3.37500000000000000   1.05357558116886896
    3.43750000000000000   1.09159692059070879
    3.50000000000000000   1.11530001113909272
    3.56250000000000000   1.12133122840791888
    3.62500000000000000   1.11372003456542079
    3.68750000000000000   1.10431830106492668
    3.75000000000000000   1.08001154831331747
    3.81250000000000000   1.04463789417798236
    3.87500000000000000   1.01113272688426981
    3.93750000000000000   0.97420208693877131
    4.00000000000000000   0.94548216432103627
    4.06250000000000000   0.92906068634824868
    4.12500000000000000   0.91261626646050853
    4.18750000000000000   0.90798395986886016
    4.25000000000000000   0.90900395742047102
    4.31250000000000000   0.92448830098309021
    4.37500000000000000   0.94336063357201572
    4.43750000000000000   0.96528356150054728
    4.50000000000000000   0.99483639400552015
    4.56250000000000000   1.01939642450213119
    4.62500000000000000   1.04364534051946634
    4.68750000000000000   1.06226556989396514
    4.75000000000000000   1.07390109509136766
    4.81250000000000000   1.07565834552142925
    4.87500000000000000   1.07801694972011508
    4.93750000000000000   1.06400707636804981
  };
  \addplot[draw=none, domain=0.062500:4.937500, name path=B] {0};
  \addplot[red!10] fill between[of=A and B];
  \addplot[help lines,dashed, domain=0:4.937500] {1};

\end{axis};
\end{tikzpicture}

%% file: res/spectrums/Orig/radialGBN.tex
‰\begin{tikzpicture}
  \begin{axis}[
    width=10cm,
    height=5cm,
    ymin = 0.000000e+00,
    ymax = 4.200000e+00,
    xmin = 0.000000e+00,
    xmax=4.937500e+00,
    xtick align = outside,
  ]
  \addplot[red, thick, name path=A] table {
    0.06250000000000000   0.00000000101112869
    0.12500000000000000   0.00000000141518402
    0.18750000000000000   0.00000000296818520
    0.25000000000000000   0.00000000959705938
    0.31250000000000000   0.00000004307616534
    0.37500000000000000   0.00000029941464551
    0.43750000000000000   0.00000249997843462
    0.50000000000000000   0.00002619488196669
    0.56250000000000000   0.00035219713723960
    0.62500000000000000   0.00680559167238622
    0.68750000000000000   0.11710246697452233
    0.75000000000000000   0.55765496133917203
    0.81250000000000000   0.96475112324108137
    0.87500000000000000   1.24281731886475932
    0.93750000000000000   1.53417646548083986
    1.00000000000000000   1.79843750576686223
    1.06250000000000000   1.90068254219375166
    1.12500000000000000   1.76127034312383479
    1.18750000000000000   1.46312605980633426
    1.25000000000000000   1.16620012650569804
    1.31250000000000000   0.93779480156181338
    1.37500000000000000   0.79399531609560925
    1.43750000000000000   0.71392058442792428
    1.50000000000000000   0.67668345560250531
    1.56250000000000000   0.66847204427164009
    1.62500000000000000   0.68209908732002256
    1.68750000000000000   0.72277095337900144
    1.75000000000000000   0.78805387753832934
    1.81250000000000000   0.87235444367635440
    1.87500000000000000   0.96623978939173105
    1.93750000000000000   1.06108060880954080
    2.00000000000000000   1.14571063105838178
    2.06250000000000000   1.21224476352489785
    2.12500000000000000   1.24583782061722670
    2.18750000000000000   1.23426273859733548
    2.25000000000000000   1.18440364858767588
    2.31250000000000000   1.11263677391961857
    2.37500000000000000   1.03636211571942760
    2.43750000000000000   0.97162078514354011
    2.50000000000000000   0.92243706172238726
    2.56250000000000000   0.89465050770499266
    2.62500000000000000   0.88155093458266853
    2.68750000000000000   0.88334269898626550
    2.75000000000000000   0.89553755804049451
    2.81250000000000000   0.91712892544376590
    2.87500000000000000   0.94520413058209640
    2.93750000000000000   0.97402952298583034
    3.00000000000000000   1.00466065876797694
    3.06250000000000000   1.03206097152675769
    3.12500000000000000   1.05348746999684439
    3.18750000000000000   1.06704570779045205
    3.25000000000000000   1.07088014589296110
    3.31250000000000000   1.06568284305698890
    3.37500000000000000   1.05105876343041005
    3.43750000000000000   1.03243864714136024
    3.50000000000000000   1.01276120237446698
    3.56250000000000000   0.99366876663459869
    3.62500000000000000   0.97790440034613579
    3.68750000000000000   0.96734252695966705
    3.75000000000000000   0.96314430075557933
    3.81250000000000000   0.96339652751924598
    3.87500000000000000   0.96786824876643940
    3.93750000000000000   0.97433974521282452
    4.00000000000000000   0.98366114535007354
    4.06250000000000000   0.99335288913008413
    4.12500000000000000   1.00116637433901867
    4.18750000000000000   1.00851247588941373
    4.25000000000000000   1.01487180615086792
    4.31250000000000000   1.01856857357615405
    4.37500000000000000   1.01945810290151129
    4.43750000000000000   1.01763764870254336
    4.50000000000000000   1.01523339587665551
    4.56250000000000000   1.01051282145441013
    4.62500000000000000   1.00512562903929203
    4.68750000000000000   1.00037768268191751
    4.75000000000000000   0.99442338021623411
    4.81250000000000000   0.99232240976305897
    4.87500000000000000   0.99044879478232406
    4.93750000000000000   0.98919682016841448
  };
  \addplot[draw=none, domain=0.062500:4.937500, name path=B] {0};
  \addplot[red!10] fill between[of=A and B];
  \addplot[help lines,dashed, domain=0:4.937500] {1};

\end{axis};
\end{tikzpicture}

%% file: res/spectrums/Orig/radialSOT.tex
\begin{tikzpicture}
  \begin{axis}[
    width=10cm,
    height=5cm,
    ymin = 0.000000e+00,
    ymax = 4.200000e+00,
    xmin = 0.000000e+00,
    xmax=4.937500e+00,
    xtick align = outside,
  ]
  \addplot[red, thick, name path=A] table {
    0.06250000000000000   0.00008820461366917
    0.12500000000000000   0.00029637237113491
    0.18750000000000000   0.00072850656170143
    0.25000000000000000   0.00167717641766048
    0.31250000000000000   0.00417611973256693
    0.37500000000000000   0.01044020698543466
    0.43750000000000000   0.02289798807456273
    0.50000000000000000   0.04674879170238843
    0.56250000000000000   0.09496441549590809
    0.62500000000000000   0.25224074603129004
    0.68750000000000000   0.51905593683196505
    0.75000000000000000   0.85880617451252539
    0.81250000000000000   1.13639547087874004
    0.87500000000000000   1.30510101636208820
    0.93750000000000000   1.29225534364555128
    1.00000000000000000   1.18722383652236396
    1.06250000000000000   1.14625735911146487
    1.12500000000000000   1.16537581355481112
    1.18750000000000000   1.19359636241948786
    1.25000000000000000   1.18520037492988961
    1.31250000000000000   1.16416109328353445
    1.37500000000000000   1.11661418677150648
    1.43750000000000000   1.06221728344866895
    1.50000000000000000   0.99840839113395319
    1.56250000000000000   0.95372129832131392
    1.62500000000000000   0.91575055444719433
    1.68750000000000000   0.90860874399218750
    1.75000000000000000   0.88288443993389054
    1.81250000000000000   0.91280929914327358
    1.87500000000000000   0.89742284656071269
    1.93750000000000000   0.90018633163709327
    2.00000000000000000   0.94765027005240321
    2.06250000000000000   0.96603573853771629
    2.12500000000000000   1.00015488985845225
    2.18750000000000000   1.01977422545892638
    2.25000000000000000   1.03243101555820749
    2.31250000000000000   1.04862362698254286
    2.37500000000000000   1.04135873468966000
    2.43750000000000000   1.05065650043199921
    2.50000000000000000   1.04262400610392447
    2.56250000000000000   1.03327342657723475
    2.62500000000000000   1.03633378992969094
    2.68750000000000000   1.02120401382829962
    2.75000000000000000   1.00813986625994589
    2.81250000000000000   1.00576577815942203
    2.87500000000000000   0.99008651554915972
    2.93750000000000000   0.97902718384267817
    3.00000000000000000   0.97919601620100727
    3.06250000000000000   0.98902714270178926
    3.12500000000000000   0.99296227739260912
    3.18750000000000000   0.97004440107519729
    3.25000000000000000   0.98730136797513701
    3.31250000000000000   0.98121851616188327
    3.37500000000000000   0.99366455928193831
    3.43750000000000000   1.00948219262419103
    3.50000000000000000   1.00986074980523211
    3.56250000000000000   1.00894842282536779
    3.62500000000000000   1.00402991730400348
    3.68750000000000000   0.99976120410097979
    3.75000000000000000   0.99496287532575256
    3.81250000000000000   0.98619349290459657
    3.87500000000000000   1.00130251534220149
    3.93750000000000000   1.00391912981794174
    4.00000000000000000   0.99739954206594261
    4.06250000000000000   0.99940992711028842
    4.12500000000000000   0.99646839915511665
    4.18750000000000000   1.00480092633956586
    4.25000000000000000   1.00777731306288820
    4.31250000000000000   0.99994641236778836
    4.37500000000000000   1.01358318615627185
    4.43750000000000000   1.00153968598925691
    4.50000000000000000   1.00477030324428762
    4.56250000000000000   1.00762498811826373
    4.62500000000000000   1.00148849474091550
    4.68750000000000000   0.99657556811320724
    4.75000000000000000   1.00559911109183675
    4.81250000000000000   0.98816245967584959
    4.87500000000000000   0.99853433799319602
    4.93750000000000000   0.99624581145507851
  };
  \addplot[draw=none, domain=0.062500:4.937500, name path=B] {0};
  \addplot[red!10] fill between[of=A and B];
  \addplot[help lines,dashed, domain=0:4.937500] {1};

\end{axis};
\end{tikzpicture}

%% file: res/spectrums/Orig/radialLDBN.tex
\begin{tikzpicture}
  \begin{axis}[
    width=10cm,
    height=5cm,
    ymin = 0.000000e+00,
    ymax = 4.200000e+00,
    xmin = 0.000000e+00,
    xmax=4.937500e+00,
    xtick align = outside,
  ]
  \addplot[red, thick, name path=A] table {
    0.06250000000000000   0.00410681299572806
    0.12500000000000000   0.00750500520570725
    0.18750000000000000   0.00929281792353444
    0.25000000000000000   0.01441110939608423
    0.31250000000000000   0.01904438017491587
    0.37500000000000000   0.02396513100151785
    0.43750000000000000   0.03444581626343149
    0.50000000000000000   0.04408406489141255
    0.56250000000000000   0.06173128015609518
    0.62500000000000000   0.08935924799909163
    0.68750000000000000   0.13375979066506874
    0.75000000000000000   0.22523276144831952
    0.81250000000000000   0.45111082192510848
    0.87500000000000000   0.98927374483850294
    0.93750000000000000   1.75733789372642324
    1.00000000000000000   2.61058032023542719
    1.06250000000000000   2.56179951584455567
    1.12500000000000000   1.90049385187645581
    1.18750000000000000   1.29018972273181909
    1.25000000000000000   0.86633869504630567
    1.31250000000000000   0.67920039513172425
    1.37500000000000000   0.56536254801339259
    1.43750000000000000   0.50323169216516239
    1.50000000000000000   0.50889719242438736
    1.56250000000000000   0.57336229077139178
    1.62500000000000000   0.65554473114610234
    1.68750000000000000   0.75872873531410068
    1.75000000000000000   0.88486428772315129
    1.81250000000000000   1.05402235734521743
    1.87500000000000000   1.17753795574332343
    1.93750000000000000   1.24236915399107017
    2.00000000000000000   1.35820535241217466
    2.06250000000000000   1.32956422917894046
    2.12500000000000000   1.23294297231881989
    2.18750000000000000   1.15270224214437333
    2.25000000000000000   1.04560815613377667
    2.31250000000000000   0.95245066598087913
    2.37500000000000000   0.88343497827909423
    2.43750000000000000   0.82479930170275095
    2.50000000000000000   0.84866353231110225
    2.56250000000000000   0.84708090407457237
    2.62500000000000000   0.86354465184701978
    2.68750000000000000   0.90559490758517336
    2.75000000000000000   0.97618676598245169
    2.81250000000000000   1.01461689484012241
    2.87500000000000000   1.02688208163067185
    2.93750000000000000   1.07612391423732912
    3.00000000000000000   1.07030287253814937
    3.06250000000000000   1.09241430065199951
    3.12500000000000000   1.07623457043033444
    3.18750000000000000   1.07681538572645197
    3.25000000000000000   1.04436409917977491
    3.31250000000000000   1.00324894458766400
    3.37500000000000000   0.96623432592759040
    3.43750000000000000   0.97875722547401345
    3.50000000000000000   0.95185623431429967
    3.56250000000000000   0.94811544963546768
    3.62500000000000000   0.93897338352130022
    3.68750000000000000   0.95728345584881092
    3.75000000000000000   0.98662325684479812
    3.81250000000000000   0.97570504497095090
    3.87500000000000000   1.03862969714215758
    3.93750000000000000   1.00276589190117282
    4.00000000000000000   1.00955755834178551
    4.06250000000000000   1.04004098523296906
    4.12500000000000000   1.01366958490511450
    4.18750000000000000   1.01581851378776888
    4.25000000000000000   1.02625278764075034
    4.31250000000000000   1.03022933776363246
    4.37500000000000000   0.99970150667495550
    4.43750000000000000   0.98766108826127452
    4.50000000000000000   0.97574242403440747
    4.56250000000000000   0.98633966046572696
    4.62500000000000000   0.98550977646511917
    4.68750000000000000   0.97652438300384692
    4.75000000000000000   0.99778779371333814
    4.81250000000000000   0.98238746747874117
    4.87500000000000000   1.01566931603528521
    4.93750000000000000   1.00820918684730909
  };
  \addplot[draw=none, domain=0.062500:4.937500, name path=B] {0};
  \addplot[red!10] fill between[of=A and B];
  \addplot[help lines,dashed, domain=0:4.937500] {1};

\end{axis};
\end{tikzpicture}

%% file: res/spectrums/Orig/radialOwen.tex
\begin{tikzpicture}
  \begin{axis}[
    width=10cm,
    height=5cm,
    ymin = 0.000000e+00,
    ymax = 4.200000e+00,
    xmin = 0.000000e+00,
    xmax=4.937500e+00,
    xtick align = outside,
  ]
  \addplot[red, thick, name path=A] table {
    0.06250000000000000   0.00371932642589829
    0.12500000000000000   0.01939090161191327
    0.18750000000000000   0.05313319976287044
    0.25000000000000000   0.11287007105525375
    0.31250000000000000   0.19487525948590720
    0.37500000000000000   0.31076160534517866
    0.43750000000000000   0.42088590119511399
    0.50000000000000000   0.54755397115258486
    0.56250000000000000   0.66541754653626595
    0.62500000000000000   0.77785443920952224
    0.68750000000000000   0.83854177420993226
    0.75000000000000000   0.91314378111988181
    0.81250000000000000   0.95511650478548238
    0.87500000000000000   0.96506002345106623
    0.93750000000000000   0.99305107728564435
    1.00000000000000000   0.99137485480410159
    1.06250000000000000   0.99992426639975340
    1.12500000000000000   1.00272611028750469
    1.18750000000000000   0.99826548052137820
    1.25000000000000000   0.98945142253948204
    1.31250000000000000   0.98628176710851767
    1.37500000000000000   0.97835307079109779
    1.43750000000000000   0.98496157144191188
    1.50000000000000000   0.99532976102761861
    1.56250000000000000   0.98856041442811304
    1.62500000000000000   0.99388883639522418
    1.68750000000000000   1.00409934083840224
    1.75000000000000000   0.99140783227420226
    1.81250000000000000   1.00356071484979803
    1.87500000000000000   1.00489518044595805
    1.93750000000000000   1.00338384280824000
    2.00000000000000000   1.00708184622485741
    2.06250000000000000   1.00503176214450485
    2.12500000000000000   1.00305597010072245
    2.18750000000000000   0.99791729266443141
    2.25000000000000000   0.99417009026876113
    2.31250000000000000   0.99761011505374209
    2.37500000000000000   0.99892910949961344
    2.43750000000000000   0.99892897720784735
    2.50000000000000000   1.00167203599645060
    2.56250000000000000   0.99712644247176285
    2.62500000000000000   0.99595729779743480
    2.68750000000000000   0.99589126454883314
    2.75000000000000000   0.99969987055600518
    2.81250000000000000   0.99697617067491751
    2.87500000000000000   1.00210459547858810
    2.93750000000000000   0.99638240469655426
    3.00000000000000000   1.00025252870138237
    3.06250000000000000   1.00091888298608445
    3.12500000000000000   0.99860232122273840
    3.18750000000000000   1.00025044546445763
    3.25000000000000000   0.99895482647902656
    3.31250000000000000   1.00233232900341385
    3.37500000000000000   0.99811561998943299
    3.43750000000000000   1.00121256326637331
    3.50000000000000000   0.99990623995917716
    3.56250000000000000   1.00132301980445515
    3.62500000000000000   0.99995317579596565
    3.68750000000000000   0.99667647559713923
    3.75000000000000000   0.99668522295143658
    3.81250000000000000   1.00058068326045269
    3.87500000000000000   0.99767327128284666
    3.93750000000000000   0.99995420358485820
    4.00000000000000000   1.00274564000181088
    4.06250000000000000   1.00234128227713160
    4.12500000000000000   0.99881542048981131
    4.18750000000000000   1.00580310734568612
    4.25000000000000000   0.99897730850202726
    4.31250000000000000   1.00076238761285219
    4.37500000000000000   1.00060118690596833
    4.43750000000000000   1.00045550164859454
    4.50000000000000000   0.99865469525048789
    4.56250000000000000   0.99793654163625078
    4.62500000000000000   0.99497697591393586
    4.68750000000000000   1.00340922685833633
    4.75000000000000000   0.99762815157691398
    4.81250000000000000   0.99642453451265123
    4.87500000000000000   1.00114138432716193
    4.93750000000000000   0.99863495969448501
  };
  \addplot[draw=none, domain=0.062500:4.937500, name path=B] {0};
  \addplot[red!10] fill between[of=A and B];
  \addplot[help lines,dashed, domain=0:4.937500] {1};

\end{axis};
\end{tikzpicture}

%% file: res/spectrums/Orig/radialR1.tex
\begin{tikzpicture}
  \begin{axis}[
    width=10cm,
    height=5cm,
    ymin = 0.000000e+00,
    ymax = 4.200000e+00,
    xmin = 0.000000e+00,
    xmax=4.937500e+00,
    xtick align = outside,
  ]
  \addplot[red, thick, name path=A] table {
    0.06250000000000000   0.00000000000000000
    0.12500000000000000   0.00000000000000000
    0.18750000000000000   0.00000000000000000
    0.25000000000000000   0.00000000000000000
    0.31250000000000000   0.00000000000000000
    0.37500000000000000   0.00000000000000000
    0.43750000000000000   0.00000000000000000
    0.50000000000000000   0.00000000000000000
    0.56250000000000000   0.00000000000000000
    0.62500000000000000   0.00000000000000000
    0.68750000000000000   0.00000000000000000
    0.75000000000000000   0.00000000000000000
    0.81250000000000000   0.00000000000000000
    0.87500000000000000   0.00000000000000000
    0.93750000000000000   0.00000000000000000
    1.00000000000000000   0.00000000000000000
    1.06250000000000000  14.09174311926604339
    1.12500000000000000   0.00000000000000000
    1.18750000000000000   0.00000000000000000
    1.25000000000000000   0.00000000000000000
    1.31250000000000000   0.00000000000000000
    1.37500000000000000   0.00000000000000000
    1.43750000000000000   0.00000000000000000
    1.50000000000000000   0.00000000000000000
    1.56250000000000000   0.00000000000000000
    1.62500000000000000   0.00000000000000000
    1.68750000000000000   0.00000000000000000
    1.75000000000000000   0.00000000000000000
    1.81250000000000000   5.59562841530055355
    1.87500000000000000   2.62564102564102875
    1.93750000000000000   0.00000000000000000
    2.00000000000000000   0.00000000000000000
    2.06250000000000000   0.00000000000000000
    2.12500000000000000   7.07834101382489500
    2.18750000000000000   0.00000000000000000
    2.25000000000000000   0.00000000000000000
    2.31250000000000000   0.00000000000000000
    2.37500000000000000   0.00000000000000000
    2.43750000000000000   0.00000000000000000
    2.50000000000000000   0.00000000000000000
    2.56250000000000000   0.00000000000000000
    2.62500000000000000   0.00000000000000000
    2.68750000000000000   0.00000000000000000
    2.75000000000000000   0.00000000000000000
    2.81250000000000000  10.89361702127660081
    2.87500000000000000   0.00000000000000000
    2.93750000000000000   0.00000000000000000
    3.00000000000000000   0.00000000000000000
    3.06250000000000000   0.00000000000000000
    3.12500000000000000   0.00000000000000000
    3.18750000000000000   3.23028391167192730
    3.25000000000000000   1.55623100303951323
    3.31250000000000000   0.00000000000000000
    3.37500000000000000   0.00000000000000000
    3.43750000000000000   0.00000000000000000
    3.50000000000000000   0.00000000000000000
    3.56250000000000000   0.00000000000000000
    3.62500000000000000   0.00000000000000000
    3.68750000000000000   2.71618037135278767
    3.75000000000000000   1.37634408602150193
    3.81250000000000000   3.84962406015038061
    3.87500000000000000   3.93846153846153424
    3.93750000000000000   0.00000000000000000
    4.00000000000000000   0.00000000000000000
    4.06250000000000000   0.00000000000000000
    4.12500000000000000   0.00000000000000000
    4.18750000000000000   0.00000000000000000
    4.25000000000000000   2.37587006960556790
    4.31250000000000000   1.19347319347319347
    4.37500000000000000   0.00000000000000000
    4.43750000000000000   0.00000000000000000
    4.50000000000000000   0.00000000000000000
    4.56250000000000000   0.00000000000000000
    4.62500000000000000   3.24050632911392666
    4.68750000000000000   3.28907922912205564
    4.75000000000000000   0.00000000000000000
    4.81250000000000000   1.04489795918367423
    4.87500000000000000   2.05210420841683527
    4.93750000000000000   3.13469387755101714
  };
  \addplot[draw=none, domain=0.062500:4.937500, name path=B] {0};
  \addplot[red!10] fill between[of=A and B];
  \addplot[help lines,dashed, domain=0:4.937500] {1};

\end{axis};
\end{tikzpicture}

%% file: res/spectrums/DeepPointCorrelation/radialPoisson.tex
\begin{tikzpicture}
  \begin{axis}[
    width=10cm,
    height=5cm,
    ymin = 0.000000e+00,
    ymax = 4.200000e+00,
    xmin = 0.000000e+00,
    xmax=4.937500e+00,
    xtick align = outside,
  ]
  \addplot[teal, thick, name path=A] table {
    0.06250000000000000   0.08167468344393311
    0.12500000000000000   0.06744000089144962
    0.18750000000000000   0.07575329856969018
    0.25000000000000000   0.07468260025350272
    0.31250000000000000   0.08265650779506913
    0.37500000000000000   0.09134046786775583
    0.43750000000000000   0.10607994048629965
    0.50000000000000000   0.12438247935639879
    0.56250000000000000   0.15449945015157177
    0.62500000000000000   0.20387723743805949
    0.68750000000000000   0.28950468939038521
    0.75000000000000000   0.42295929116805303
    0.81250000000000000   0.63679506575375766
    0.87500000000000000   0.91272410103518276
    0.93750000000000000   1.19534308543110379
    1.00000000000000000   1.43277084163477975
    1.06250000000000000   1.61861643297656999
    1.12500000000000000   1.64567125891173704
    1.18750000000000000   1.60786400233547533
    1.25000000000000000   1.46493454301312420
    1.31250000000000000   1.28655713950228878
    1.37500000000000000   1.12547553699361802
    1.43750000000000000   0.97411430102455354
    1.50000000000000000   0.87448807440738641
    1.56250000000000000   0.78076160658518257
    1.62500000000000000   0.72363548490184926
    1.68750000000000000   0.69683492389152890
    1.75000000000000000   0.69094562401195114
    1.81250000000000000   0.70476668484826177
    1.87500000000000000   0.74560181380078361
    1.93750000000000000   0.82879365977034614
    2.00000000000000000   0.97766884257855136
    2.06250000000000000   1.07763365674759748
    2.12500000000000000   1.13848610211290402
    2.18750000000000000   1.15579919901746719
    2.25000000000000000   1.18460847896961230
    2.31250000000000000   1.18840705969517413
    2.37500000000000000   1.14613851490716057
    2.43750000000000000   1.14145669775583558
    2.50000000000000000   1.12218206234003248
    2.56250000000000000   1.08370547468163347
    2.62500000000000000   1.02594626441363612
    2.68750000000000000   0.96655800220765231
    2.75000000000000000   0.90480401447104930
    2.81250000000000000   0.86828640441749294
    2.87500000000000000   0.84854141585433218
    2.93750000000000000   0.85879042694098462
    3.00000000000000000   0.88644753601533233
    3.06250000000000000   0.92147082274654746
    3.12500000000000000   0.96532570802390982
    3.18750000000000000   1.00164283139622468
    3.25000000000000000   1.03845282000944628
    3.31250000000000000   1.06157078312844777
    3.37500000000000000   1.07247561671491920
    3.43750000000000000   1.07217479849501807
    3.50000000000000000   1.06648564887731245
    3.56250000000000000   1.05457619496521326
    3.62500000000000000   1.04166179088658040
    3.68750000000000000   1.02314048368102029
    3.75000000000000000   1.00329452239427197
    3.81250000000000000   0.98714556156295730
    3.87500000000000000   0.97356385907046694
    3.93750000000000000   0.96377170828668124
    4.00000000000000000   0.96201270481083767
    4.06250000000000000   0.96858953870661602
    4.12500000000000000   0.97660148816038661
    4.18750000000000000   0.98390608574035487
    4.25000000000000000   0.99598247353461000
    4.31250000000000000   1.00539265153190516
    4.37500000000000000   1.00921326025014135
    4.43750000000000000   1.00943855927386306
    4.50000000000000000   1.00916919266972260
    4.56250000000000000   1.00666169835100883
    4.62500000000000000   1.00333551008902999
    4.68750000000000000   1.00112674220635589
    4.75000000000000000   0.99495409903650678
    4.81250000000000000   0.99298370283492299
    4.87500000000000000   0.99142095631618155
    4.93750000000000000   0.99921422672389815
  };
  \addplot[draw=none, domain=0.062500:4.937500, name path=B] {0};
  \addplot[teal!10] fill between[of=A and B];
  \addplot[help lines,dashed, domain=0:4.937500] {1};

\end{axis};
\end{tikzpicture}

%% file: res/spectrums/DeepPointCorrelation/radialGBN.tex
\begin{tikzpicture}
  \begin{axis}[
    width=10cm,
    height=5cm,
    ymin = 0.000000e+00,
    ymax = 4.200000e+00,
    xmin = 0.000000e+00,
    xmax=4.937500e+00,
    xtick align = outside,
  ]
  \addplot[teal, thick, name path=A] table {
    0.06250000000000000   0.00407119012337028
    0.12500000000000000   0.00442799595227187
    0.18750000000000000   0.00622621028618871
    0.25000000000000000   0.00731886643898973
    0.31250000000000000   0.00932195637816877
    0.37500000000000000   0.01146558081035113
    0.43750000000000000   0.01422008973931820
    0.50000000000000000   0.01606555800088429
    0.56250000000000000   0.01794202438994254
    0.62500000000000000   0.01944389543336893
    0.68750000000000000   0.03687371736120282
    0.75000000000000000   0.30638825638315759
    0.81250000000000000   0.82063900617187779
    0.87500000000000000   1.22619104804536994
    0.93750000000000000   1.68848024433640598
    1.00000000000000000   2.07951289897687586
    1.06250000000000000   2.12435506681738939
    1.12500000000000000   1.83504082113089262
    1.18750000000000000   1.43471498987269186
    1.25000000000000000   1.09212655435062977
    1.31250000000000000   0.85247193810563771
    1.37500000000000000   0.72451249982855181
    1.43750000000000000   0.65469455757322081
    1.50000000000000000   0.62969064265469676
    1.56250000000000000   0.63374833506509376
    1.62500000000000000   0.63429477676426282
    1.68750000000000000   0.67781733031494562
    1.75000000000000000   0.74518815648922887
    1.81250000000000000   0.86053198397875375
    1.87500000000000000   0.99864545589445408
    1.93750000000000000   1.11728665794172644
    2.00000000000000000   1.20662404515415855
    2.06250000000000000   1.26255071698386456
    2.12500000000000000   1.31867831736971031
    2.18750000000000000   1.31872772525009396
    2.25000000000000000   1.25844620987676636
    2.31250000000000000   1.12641477234747356
    2.37500000000000000   1.00180785534396666
    2.43750000000000000   0.91275623718869070
    2.50000000000000000   0.85685696858437665
    2.56250000000000000   0.82612093576542467
    2.62500000000000000   0.83040268212058654
    2.68750000000000000   0.84507168637336227
    2.75000000000000000   0.86900534672485463
    2.81250000000000000   0.89575261784703952
    2.87500000000000000   0.93775749962857080
    2.93750000000000000   0.98532361307714322
    3.00000000000000000   1.03389558306228357
    3.06250000000000000   1.04685512016944737
    3.12500000000000000   1.01963385244055105
    3.18750000000000000   1.05128080884795105
    3.25000000000000000   1.19927808832982641
    3.31250000000000000   1.19156157746829283
    3.37500000000000000   1.13570750744760041
    3.43750000000000000   1.06161760307582198
    3.50000000000000000   0.99337596171473663
    3.56250000000000000   0.94352565749129358
    3.62500000000000000   0.91130798595051810
    3.68750000000000000   0.89670733034840511
    3.75000000000000000   0.89401582772866606
    3.81250000000000000   0.89945182070181517
    3.87500000000000000   0.91536032818410595
    3.93750000000000000   0.94402235417419544
    4.00000000000000000   0.97679321159385568
    4.06250000000000000   1.01465943921301816
    4.12500000000000000   1.04836682517722113
    4.18750000000000000   1.07863686316519769
    4.25000000000000000   1.09677989647875473
    4.31250000000000000   1.09412624498595190
    4.37500000000000000   1.07704550774929197
    4.43750000000000000   1.04098676850384142
    4.50000000000000000   0.99471069365925247
    4.56250000000000000   0.94562905297097777
    4.62500000000000000   0.92688827274477703
    4.68750000000000000   0.97400584913247679
    4.75000000000000000   0.97374738851675913
    4.81250000000000000   0.96978598729321619
    4.87500000000000000   0.97038437574915104
    4.93750000000000000   0.97910057768774528
  };
  \addplot[draw=none, domain=0.062500:4.937500, name path=B] {0};
  \addplot[teal!10] fill between[of=A and B];
  \addplot[help lines,dashed, domain=0:4.937500] {1};

\end{axis};
\end{tikzpicture}

%% file: res/spectrums/DeepPointCorrelation/radialSOT.tex
\begin{tikzpicture}
  \begin{axis}[
    width=10cm,
    height=5cm,
    ymin = 0.000000e+00,
    ymax = 4.200000e+00,
    xmin = 0.000000e+00,
    xmax=4.937500e+00,
    xtick align = outside,
  ]
  \addplot[teal, thick, name path=A] table {
    0.06250000000000000   0.00417175191061464
    0.12500000000000000   0.00384668063938082
    0.18750000000000000   0.00591417082370585
    0.25000000000000000   0.00742547305133273
    0.31250000000000000   0.00846665764172786
    0.37500000000000000   0.01116660968007977
    0.43750000000000000   0.02133886683245059
    0.50000000000000000   0.04370015539687030
    0.56250000000000000   0.08994444246281616
    0.62500000000000000   0.23731701656584026
    0.68750000000000000   0.48922148171897806
    0.75000000000000000   0.84908763838221524
    0.81250000000000000   1.11645730058149395
    0.87500000000000000   1.29839300862203766
    0.93750000000000000   1.30783638856205031
    1.00000000000000000   1.19701729793762346
    1.06250000000000000   1.14552318634004036
    1.12500000000000000   1.16913643764062014
    1.18750000000000000   1.19948383104719758
    1.25000000000000000   1.19003670896624536
    1.31250000000000000   1.17469553366390400
    1.37500000000000000   1.12235237374720676
    1.43750000000000000   1.06837727665247617
    1.50000000000000000   1.00264011727045088
    1.56250000000000000   0.96151671400233407
    1.62500000000000000   0.91325277350248701
    1.68750000000000000   0.90380307110242941
    1.75000000000000000   0.87482004352244469
    1.81250000000000000   0.90774025742205100
    1.87500000000000000   0.88822230847501460
    1.93750000000000000   0.89895569515708296
    2.00000000000000000   0.97278135565504908
    2.06250000000000000   1.00118095770821247
    2.12500000000000000   1.00757152323569299
    2.18750000000000000   1.02936575371904371
    2.25000000000000000   1.06344751142959071
    2.31250000000000000   1.06088549126270837
    2.37500000000000000   1.02402857433224970
    2.43750000000000000   0.99504396454351995
    2.50000000000000000   0.99563251142920361
    2.56250000000000000   0.96992570424302293
    2.62500000000000000   0.99202156135927566
    2.68750000000000000   1.02238987106070356
    2.75000000000000000   1.05024144779784923
    2.81250000000000000   1.04163685063551048
    2.87500000000000000   1.03005715495785921
    2.93750000000000000   1.02354305704211224
    3.00000000000000000   0.99552846613361301
    3.06250000000000000   0.98009282487694338
    3.12500000000000000   0.96128857124193767
    3.18750000000000000   0.96312968538112975
    3.25000000000000000   0.94700900017061207
    3.31250000000000000   0.97676657248254761
    3.37500000000000000   0.98239326726007292
    3.43750000000000000   1.00562548363641580
    3.50000000000000000   1.02577362005106565
    3.56250000000000000   1.03000383879215551
    3.62500000000000000   1.02283101964321710
    3.68750000000000000   1.01245843253480916
    3.75000000000000000   1.00056141274762678
    3.81250000000000000   0.99782576115680910
    3.87500000000000000   0.99067401009711997
    3.93750000000000000   0.99264107123211853
    4.00000000000000000   0.99899456553401733
    4.06250000000000000   1.00384216006123661
    4.12500000000000000   1.00954208450749849
    4.18750000000000000   1.00861809968802207
    4.25000000000000000   1.00494710243848040
    4.31250000000000000   0.99220098675270907
    4.37500000000000000   0.98820626323555572
    4.43750000000000000   0.98695234532596754
    4.50000000000000000   0.98567142685257303
    4.56250000000000000   0.99110131944597923
    4.62500000000000000   0.99933503224656006
    4.68750000000000000   1.00389245153007467
    4.75000000000000000   1.01375180028186063
    4.81250000000000000   1.01402905472502769
    4.87500000000000000   1.01647951910060264
    4.93750000000000000   1.00693683968050318
  };
  \addplot[draw=none, domain=0.062500:4.937500, name path=B] {0};
  \addplot[teal!10] fill between[of=A and B];
  \addplot[help lines,dashed, domain=0:4.937500] {1};

\end{axis};
\end{tikzpicture}

%% file: res/spectrums/DeepPointCorrelation/radialLDBN.tex
\begin{tikzpicture}
  \begin{axis}[
    width=10cm,
    height=5cm,
    ymin = 0.000000e+00,
    ymax = 4.200000e+00,
    xmin = 0.000000e+00,
    xmax=4.937500e+00,
    xtick align = outside,
  ]
  \addplot[teal, thick, name path=A] table {
    0.06250000000000000   0.00663235295522584
    0.12500000000000000   0.00880311286327612
    0.18750000000000000   0.01261274950790637
    0.25000000000000000   0.01589651872914696
    0.31250000000000000   0.02025209211297143
    0.37500000000000000   0.02441696626193747
    0.43750000000000000   0.02962065972940702
    0.50000000000000000   0.04571549710932868
    0.56250000000000000   0.06013155662005763
    0.62500000000000000   0.08645927997233299
    0.68750000000000000   0.12429416702660415
    0.75000000000000000   0.20816143897213571
    0.81250000000000000   0.44295956661435670
    0.87500000000000000   0.84352344469444107
    0.93750000000000000   1.82072422504323916
    1.00000000000000000   2.51713061477287825
    1.06250000000000000   2.64049444155452528
    1.12500000000000000   1.98816508226249722
    1.18750000000000000   1.28712408281969193
    1.25000000000000000   0.92044645711595163
    1.31250000000000000   0.65442556139413477
    1.37500000000000000   0.55825427017756801
    1.43750000000000000   0.51676027114225775
    1.50000000000000000   0.52013745771240971
    1.56250000000000000   0.55051373525599290
    1.62500000000000000   0.61958056598942313
    1.68750000000000000   0.76219636652580802
    1.75000000000000000   0.90019775222655329
    1.81250000000000000   1.01802122793410299
    1.87500000000000000   1.14806538692946436
    1.93750000000000000   1.30017642221459107
    2.00000000000000000   1.34225636436410678
    2.06250000000000000   1.36625679129189659
    2.12500000000000000   1.26003934529351813
    2.18750000000000000   1.15517369706603956
    2.25000000000000000   1.00850977870556635
    2.31250000000000000   0.91377849702058256
    2.37500000000000000   0.85362067308806377
    2.43750000000000000   0.82683796749537142
    2.50000000000000000   0.83144394207593986
    2.56250000000000000   0.86175535763314570
    2.62500000000000000   0.92272200542773342
    2.68750000000000000   0.95305334665361452
    2.75000000000000000   1.02326411907680193
    2.81250000000000000   0.99638050229938768
    2.87500000000000000   1.05491916959199172
    2.93750000000000000   1.03465209378674117
    3.00000000000000000   1.03616658645983950
    3.06250000000000000   1.05873278104609425
    3.12500000000000000   1.06195109416039024
    3.18750000000000000   1.05802191126227663
    3.25000000000000000   1.02876841293671917
    3.31250000000000000   1.00268260577103585
    3.37500000000000000   0.97957708295000745
    3.43750000000000000   0.97342020606086688
    3.50000000000000000   0.96977487854351041
    3.56250000000000000   0.97566365218252005
    3.62500000000000000   0.97900138180248508
    3.68750000000000000   0.98332976571237074
    3.75000000000000000   0.98829987127388241
    3.81250000000000000   0.98890847271236237
    3.87500000000000000   0.99386737555896121
    3.93750000000000000   0.99672396497648885
    4.00000000000000000   1.00517200481085234
    4.06250000000000000   1.01073418794616754
    4.12500000000000000   1.01603288598772656
    4.18750000000000000   1.01639685598344820
    4.25000000000000000   1.01844627319911818
    4.31250000000000000   1.01320113331249151
    4.37500000000000000   1.00527563820457533
    4.43750000000000000   1.00070170626988442
    4.50000000000000000   0.99344503113741855
    4.56250000000000000   0.99263509074490497
    4.62500000000000000   0.98425761291119973
    4.68750000000000000   0.98210974933297446
    4.75000000000000000   0.98978601578687775
    4.81250000000000000   0.99275111650215830
    4.87500000000000000   1.00496443471027175
    4.93750000000000000   1.00303392113035739
  };
  \addplot[draw=none, domain=0.062500:4.937500, name path=B] {0};
  \addplot[teal!10] fill between[of=A and B];
  \addplot[help lines,dashed, domain=0:4.937500] {1};

\end{axis};
\end{tikzpicture}

%% file: res/spectrums/DeepPointCorrelation/radialOwenSpectre2d.tex
\begin{tikzpicture}
  \begin{axis}[
    width=10cm,
    height=5cm,
    ymin = 0.000000e+00,
    ymax = 4.200000e+00,
    xmin = 0.000000e+00,
    xmax=4.937500e+00,
    xtick align = outside,
  ]
  \addplot[teal, thick, name path=A] table {
    0.06250000000000000   0.00440827366743536
    0.12500000000000000   0.01431699113294803
    0.18750000000000000   0.04318320014593180
    0.25000000000000000   0.09408647038710849
    0.31250000000000000   0.15405059220392850
    0.37500000000000000   0.23679844091123611
    0.43750000000000000   0.33384978199440579
    0.50000000000000000   0.43758465987845901
    0.56250000000000000   0.52072892294556561
    0.62500000000000000   0.59346462047949089
    0.68750000000000000   0.65128901786720272
    0.75000000000000000   0.69686694062463506
    0.81250000000000000   0.74308552790443627
    0.87500000000000000   0.75138808695491799
    0.93750000000000000   0.76197705750624956
    1.00000000000000000   0.84221918396686346
    1.06250000000000000   0.82368930609059443
    1.12500000000000000   0.81066454052367731
    1.18750000000000000   0.78428494843200058
    1.25000000000000000   0.80482581161272604
    1.31250000000000000   0.82725207900973308
    1.37500000000000000   1.43699573897449451
    1.43750000000000000   2.61421890124475453
    1.50000000000000000   1.86418861511037859
    1.56250000000000000   1.41331146973504662
    1.62500000000000000   1.12053243259200208
    1.68750000000000000   0.94125454799238095
    1.75000000000000000   0.88471829361620991
    1.81250000000000000   0.79134669616379250
    1.87500000000000000   0.73645396546274322
    1.93750000000000000   0.70744386044543273
    2.00000000000000000   0.71726852063925195
    2.06250000000000000   0.73458922737491827
    2.12500000000000000   0.76079908501054372
    2.18750000000000000   0.80758316036482691
    2.25000000000000000   0.88684471519076036
    2.31250000000000000   0.94440959749863740
    2.37500000000000000   0.96771639664919751
    2.43750000000000000   0.92962256560907242
    2.50000000000000000   0.81685161959712227
    2.56250000000000000   0.98796498126860732
    2.62500000000000000   1.09162107413605258
    2.68750000000000000   1.18044938593594795
    2.75000000000000000   1.25577834834045476
    2.81250000000000000   1.33931912297245392
    2.87500000000000000   1.36348376228809465
    2.93750000000000000   1.27202508814988713
    3.00000000000000000   1.16597660633365430
    3.06250000000000000   1.05362811512051935
    3.12500000000000000   0.96249427765216444
    3.18750000000000000   0.88852299164500814
    3.25000000000000000   0.85132132522733384
    3.31250000000000000   0.82591265164829153
    3.37500000000000000   0.81265676968351175
    3.43750000000000000   0.81587405462940243
    3.50000000000000000   0.82750034498246616
    3.56250000000000000   0.85485351565752732
    3.62500000000000000   0.89457530786987438
    3.68750000000000000   0.93970923697320230
    3.75000000000000000   0.97723095098718848
    3.81250000000000000   1.00441288485099633
    3.87500000000000000   1.01588141134201693
    3.93750000000000000   1.02960053417351216
    4.00000000000000000   1.07488531534235787
    4.06250000000000000   1.12146416789078529
    4.12500000000000000   1.15649870373738572
    4.18750000000000000   1.17299686633316691
    4.25000000000000000   1.16921888141781127
    4.31250000000000000   1.13717643023192472
    4.37500000000000000   1.08772725819431182
    4.43750000000000000   1.03407135318225074
    4.50000000000000000   0.98136894259401108
    4.56250000000000000   0.93461854206409023
    4.62500000000000000   0.90313717259289106
    4.68750000000000000   0.88942119382727081
    4.75000000000000000   0.87673196260912589
    4.81250000000000000   0.87971728350518952
    4.87500000000000000   0.89463965831061310
    4.93750000000000000   0.91791137517050037
  };
  \addplot[draw=none, domain=0.062500:4.937500, name path=B] {0};
  \addplot[teal!10] fill between[of=A and B];
  \addplot[help lines,dashed, domain=0:4.937500] {1};

\end{axis};
\end{tikzpicture}

%% file: res/spectrums/Final/radialPoisson.tex
\begin{tikzpicture}
  \begin{axis}[
    width=10cm,
    height=5cm,
    ymin = 0.000000e+00,
    ymax = 4.200000e+00,
    xmin = 0.000000e+00,
    xmax=4.937500e+00,
    xtick align = outside,
  ]
  \addplot[blue, thick, name path=A] table {
    0.06250000000000000   0.07186046083405796
    0.12500000000000000   0.06985682715935830
    0.18750000000000000   0.07359020964976193
    0.25000000000000000   0.08216673411545546
    0.31250000000000000   0.09228518117268267
    0.37500000000000000   0.11093572951757032
    0.43750000000000000   0.13816667206010833
    0.50000000000000000   0.18009901388007821
    0.56250000000000000   0.23629590869427985
    0.62500000000000000   0.31913987887828732
    0.68750000000000000   0.43364019759918676
    0.75000000000000000   0.60334448537231844
    0.81250000000000000   0.81686099902658849
    0.87500000000000000   1.07907710518509670
    0.93750000000000000   1.29196578870408518
    1.00000000000000000   1.43579810757282589
    1.06250000000000000   1.40481507682236262
    1.12500000000000000   1.31095679088527239
    1.18750000000000000   1.20114507016135352
    1.25000000000000000   1.09765821458174506
    1.31250000000000000   1.01891033262603692
    1.37500000000000000   0.96574101751548913
    1.43750000000000000   0.94107726657153157
    1.50000000000000000   0.92634857207240506
    1.56250000000000000   0.93182342622862924
    1.62500000000000000   0.92805042206721700
    1.68750000000000000   0.94658594469992818
    1.75000000000000000   0.95185946496685325
    1.81250000000000000   0.98545004170702721
    1.87500000000000000   0.98798714873273463
    1.93750000000000000   1.00143531940200248
    2.00000000000000000   1.01182699136322807
    2.06250000000000000   1.01912156880866700
    2.12500000000000000   1.01321766380760447
    2.18750000000000000   1.00521809325467149
    2.25000000000000000   1.00606808462286890
    2.31250000000000000   1.00736899310167427
    2.37500000000000000   1.01041129378105765
    2.43750000000000000   1.00393625350696158
    2.50000000000000000   0.99820182361486265
    2.56250000000000000   0.99847408777143365
    2.62500000000000000   0.99138108471752195
    2.68750000000000000   0.99366428698452325
    2.75000000000000000   1.00468348692709752
    2.81250000000000000   1.00009334921145654
    2.87500000000000000   0.99999799801891032
    2.93750000000000000   0.99921584526761509
    3.00000000000000000   1.00408632547975540
    3.06250000000000000   0.99942003565293402
    3.12500000000000000   1.00205343350528975
    3.18750000000000000   1.00019759491288296
    3.25000000000000000   1.00154320166645783
    3.31250000000000000   0.99802878076653190
    3.37500000000000000   0.99566040570517478
    3.43750000000000000   1.00598461564718122
    3.50000000000000000   1.00039239290108140
    3.56250000000000000   0.99682632175522645
    3.62500000000000000   0.99770797234542663
    3.68750000000000000   0.99964058712256176
    3.75000000000000000   0.99941222630735504
    3.81250000000000000   0.99868544039757523
    3.87500000000000000   0.99727003127231106
    3.93750000000000000   1.00138561810770343
    4.00000000000000000   0.99838665777927771
    4.06250000000000000   1.00227953672156689
    4.12500000000000000   1.00400724418755272
    4.18750000000000000   0.99787387938434724
    4.25000000000000000   1.00313118621215569
    4.31250000000000000   0.99744828039645206
    4.37500000000000000   1.00004663363393753
    4.43750000000000000   1.00083055734672244
    4.50000000000000000   0.99732995721122020
    4.56250000000000000   1.00031404699875281
    4.62500000000000000   1.00343730132053799
    4.68750000000000000   1.00015491613753449
    4.75000000000000000   1.00159750683897997
    4.81250000000000000   1.00142662941053096
    4.87500000000000000   1.00058992517465550
    4.93750000000000000   0.99811864053669153
  };
  \addplot[draw=none, domain=0.062500:4.937500, name path=B] {0};
  \addplot[blue!10] fill between[of=A and B];
  \addplot[help lines,dashed, domain=0:4.937500] {1};

\end{axis};
\end{tikzpicture}

%% file: res/spectrums/Final/radialGBN.tex
\begin{tikzpicture}
  \begin{axis}[
    width=10cm,
    height=5cm,
    ymin = 0.000000e+00,
    ymax = 4.200000e+00,
    xmin = 0.000000e+00,
    xmax=4.937500e+00,
    xtick align = outside,
  ]
  \addplot[blue, thick, name path=A] table {
    0.06250000000000000   0.00580700792812688
    0.12500000000000000   0.00742177596478517
    0.18750000000000000   0.00900610460639900
    0.25000000000000000   0.01133641487816517
    0.31250000000000000   0.01485462057995352
    0.37500000000000000   0.02212911240087953
    0.43750000000000000   0.03316005310815419
    0.50000000000000000   0.05332302457075218
    0.56250000000000000   0.09714300467243601
    0.62500000000000000   0.18303486271343775
    0.68750000000000000   0.34779546382919624
    0.75000000000000000   0.58908250156236408
    0.81250000000000000   0.90269480210431052
    0.87500000000000000   1.22425045396380838
    0.93750000000000000   1.49320787739407779
    1.00000000000000000   1.69304448976811361
    1.06250000000000000   1.57849535766408522
    1.12500000000000000   1.45057837300759407
    1.18750000000000000   1.26680390595405190
    1.25000000000000000   1.10389945857271687
    1.31250000000000000   0.97859187448252660
    1.37500000000000000   0.88799805117850950
    1.43750000000000000   0.83231533472343933
    1.50000000000000000   0.80955178940080830
    1.56250000000000000   0.81429459118915637
    1.62500000000000000   0.83573343647895748
    1.68750000000000000   0.87240285398535500
    1.75000000000000000   0.92223481677711550
    1.81250000000000000   0.96641529769622891
    1.87500000000000000   1.00007734491284261
    1.93750000000000000   1.03948382111606086
    2.00000000000000000   1.05440962187272569
    2.06250000000000000   1.06526059250725136
    2.12500000000000000   1.06966713559664139
    2.18750000000000000   1.05420808766017404
    2.25000000000000000   1.04092551593292693
    2.31250000000000000   1.03478384425828662
    2.37500000000000000   1.01616705445257827
    2.43750000000000000   0.99480137029672044
    2.50000000000000000   0.98906842479628732
    2.56250000000000000   0.97631702093988171
    2.62500000000000000   0.97376847565067637
    2.68750000000000000   0.97903297993030491
    2.75000000000000000   0.98870244952491237
    2.81250000000000000   0.98426503569323165
    2.87500000000000000   0.99466846138354215
    2.93750000000000000   0.99367925760083764
    3.00000000000000000   0.99323548693001973
    3.06250000000000000   1.01018448305704811
    3.12500000000000000   1.00766960438355913
    3.18750000000000000   0.99796217633667295
    3.25000000000000000   1.00874640077971289
    3.31250000000000000   1.00429927341292413
    3.37500000000000000   1.00806169662638934
    3.43750000000000000   1.00331849282609387
    3.50000000000000000   1.00299589734160954
    3.56250000000000000   0.99709783101382665
    3.62500000000000000   0.99550423090038975
    3.68750000000000000   0.99365335875802441
    3.75000000000000000   1.00271806394687757
    3.81250000000000000   1.00134395156924727
    3.87500000000000000   1.00174332351585416
    3.93750000000000000   1.00122473550680002
    4.00000000000000000   0.99576880441019477
    4.06250000000000000   1.00312146442922900
    4.12500000000000000   0.99823151120698062
    4.18750000000000000   0.99910384554203036
    4.25000000000000000   0.99746351175041459
    4.31250000000000000   1.00189120859197756
    4.37500000000000000   0.99763918643372773
    4.43750000000000000   1.00039916724742994
    4.50000000000000000   1.00210088366772543
    4.56250000000000000   1.00246573071781908
    4.62500000000000000   0.99646781870070456
    4.68750000000000000   1.00339221061192441
    4.75000000000000000   1.00195909503479630
    4.81250000000000000   0.99898572859929968
    4.87500000000000000   1.00030334122515985
    4.93750000000000000   0.99685956753175997
  };
  \addplot[draw=none, domain=0.062500:4.937500, name path=B] {0};
  \addplot[blue!10] fill between[of=A and B];
  \addplot[help lines,dashed, domain=0:4.937500] {1};

\end{axis};
\end{tikzpicture}

%% file: res/spectrums/Final/radialSOT.tex
\begin{tikzpicture}
  \begin{axis}[
    width=10cm,
    height=5cm,
    ymin = 0.000000e+00,
    ymax = 4.200000e+00,
    xmin = 0.000000e+00,
    xmax=4.937500e+00,
    xtick align = outside,
  ]
  \addplot[blue, thick, name path=A] table {
    0.06250000000000000   0.00055950324979640
    0.12500000000000000   0.00120488118845182
    0.18750000000000000   0.00225700766365275
    0.25000000000000000   0.00438562593874987
    0.31250000000000000   0.00822912771058475
    0.37500000000000000   0.01639762821587217
    0.43750000000000000   0.03527543065962418
    0.50000000000000000   0.06058221883082292
    0.56250000000000000   0.11119864010565811
    0.62500000000000000   0.23307652775852644
    0.68750000000000000   0.50538327816653572
    0.75000000000000000   0.85124553914297141
    0.81250000000000000   1.11521505388844400
    0.87500000000000000   1.27526974627762146
    0.93750000000000000   1.32121280719934986
    1.00000000000000000   1.24472607429660864
    1.06250000000000000   1.22347274962905939
    1.12500000000000000   1.17076385755757295
    1.18750000000000000   1.17950362222907734
    1.25000000000000000   1.17591251464976065
    1.31250000000000000   1.12630605602088729
    1.37500000000000000   1.08165903133883234
    1.43750000000000000   0.99993646381203161
    1.50000000000000000   0.96348609031854826
    1.56250000000000000   0.92841392586663474
    1.62500000000000000   0.90277348134728796
    1.68750000000000000   0.90553358706176534
    1.75000000000000000   0.91819081520854406
    1.81250000000000000   0.92471514261896426
    1.87500000000000000   0.93482831259356758
    1.93750000000000000   0.95012778197666936
    2.00000000000000000   0.97664140192423865
    2.06250000000000000   1.00059656998500102
    2.12500000000000000   0.99913930839776355
    2.18750000000000000   1.02925431339332052
    2.25000000000000000   1.03605758257093772
    2.31250000000000000   1.03458600205816986
    2.37500000000000000   1.03239023715914802
    2.43750000000000000   1.02682467451964787
    2.50000000000000000   1.02542958213172319
    2.56250000000000000   1.01607681335935296
    2.62500000000000000   1.01859978381767569
    2.68750000000000000   1.00825698057506385
    2.75000000000000000   1.00012280641561002
    2.81250000000000000   0.98773854735644273
    2.87500000000000000   0.98677573641435368
    2.93750000000000000   0.98969193548477963
    3.00000000000000000   0.98899972545143944
    3.06250000000000000   0.98515491365022068
    3.12500000000000000   0.98706504677431683
    3.18750000000000000   0.99330787967433032
    3.25000000000000000   1.00718822187624446
    3.31250000000000000   0.99492835688899850
    3.37500000000000000   1.00068009500379662
    3.43750000000000000   1.00004036255955975
    3.50000000000000000   1.00630828575404574
    3.56250000000000000   1.00153713930502963
    3.62500000000000000   1.00215196830492914
    3.68750000000000000   0.99761536675669060
    3.75000000000000000   1.00298461906122305
    3.81250000000000000   1.00538381335255589
    3.87500000000000000   1.00280306121610052
    3.93750000000000000   0.99376640054485055
    4.00000000000000000   1.00357923942363714
    4.06250000000000000   0.99615091263840572
    4.12500000000000000   0.99936561624345521
    4.18750000000000000   0.99956868045627367
    4.25000000000000000   1.00311150295846030
    4.31250000000000000   1.00025456566744064
    4.37500000000000000   0.99982049365904269
    4.43750000000000000   1.00267311389270986
    4.50000000000000000   0.99643327004969684
    4.56250000000000000   1.00131779832089141
    4.62500000000000000   1.00257325871029490
    4.68750000000000000   0.99800125497881487
    4.75000000000000000   0.99977177408470685
    4.81250000000000000   0.99946488412044920
    4.87500000000000000   1.00199407673470398
    4.93750000000000000   0.99628480637621875
  };
  \addplot[draw=none, domain=0.062500:4.937500, name path=B] {0};
  \addplot[blue!10] fill between[of=A and B];
  \addplot[help lines,dashed, domain=0:4.937500] {1};

\end{axis};
\end{tikzpicture}

%% file: res/spectrums/Final/radialLDBN.tex
\begin{tikzpicture}
  \begin{axis}[
    width=10cm,
    height=5cm,
    ymin = 0.000000e+00,
    ymax = 4.200000e+00,
    xmin = 0.000000e+00,
    xmax=4.937500e+00,
    xtick align = outside,
  ]
  \addplot[blue, thick, name path=A] table {
    0.06250000000000000   0.00379592156476718
    0.12500000000000000   0.00748555610200188
    0.18750000000000000   0.01077456257431368
    0.25000000000000000   0.01423871949239433
    0.31250000000000000   0.02011603267902613
    0.37500000000000000   0.02601155437716699
    0.43750000000000000   0.03293883450016970
    0.50000000000000000   0.04672229420643552
    0.56250000000000000   0.06540907802213618
    0.62500000000000000   0.09339757688775026
    0.68750000000000000   0.13970251370083980
    0.75000000000000000   0.23316718933172673
    0.81250000000000000   0.46409419208888425
    0.87500000000000000   0.95751887614061004
    0.93750000000000000   1.77918752863994367
    1.00000000000000000   2.61690331842634416
    1.06250000000000000   2.54416899974614319
    1.12500000000000000   1.87193776674929602
    1.18750000000000000   1.25480454066649849
    1.25000000000000000   0.86070141133896061
    1.31250000000000000   0.67169944559484951
    1.37500000000000000   0.57646988282110923
    1.43750000000000000   0.52203332442936434
    1.50000000000000000   0.53167098780564415
    1.56250000000000000   0.57987961775379959
    1.62500000000000000   0.65571791644281474
    1.68750000000000000   0.78611085691630445
    1.75000000000000000   0.90468369574506202
    1.81250000000000000   1.04312952147531601
    1.87500000000000000   1.16487245046970633
    1.93750000000000000   1.24830260499386458
    2.00000000000000000   1.33317057501988123
    2.06250000000000000   1.30381754496734970
    2.12500000000000000   1.24464884948492305
    2.18750000000000000   1.14629693235104924
    2.25000000000000000   1.04129514631674658
    2.31250000000000000   0.93714494335337450
    2.37500000000000000   0.88637204614933895
    2.43750000000000000   0.83280325632207097
    2.50000000000000000   0.84399096916651273
    2.56250000000000000   0.85160872195141901
    2.62500000000000000   0.88286172918953032
    2.68750000000000000   0.92479938630152092
    2.75000000000000000   0.97794068404651247
    2.81250000000000000   1.01037521595321733
    2.87500000000000000   1.05017702224459253
    2.93750000000000000   1.06392599597765725
    3.00000000000000000   1.08430731872823327
    3.06250000000000000   1.09097606443275885
    3.12500000000000000   1.07532466472056520
    3.18750000000000000   1.05129784091727174
    3.25000000000000000   1.01017465381711724
    3.31250000000000000   0.98835436361159534
    3.37500000000000000   0.97341925912704474
    3.43750000000000000   0.96355382154005287
    3.50000000000000000   0.95916643500367538
    3.56250000000000000   0.96325147420039969
    3.62500000000000000   0.96179016937514072
    3.68750000000000000   0.97736806006693422
    3.75000000000000000   0.98813474879982865
    3.81250000000000000   0.99928420301850918
    3.87500000000000000   1.01085522592655130
    3.93750000000000000   1.00785278361863684
    4.00000000000000000   1.02518710298080928
    4.06250000000000000   1.02300823069946345
    4.12500000000000000   1.01578686397811846
    4.18750000000000000   1.00665835160489991
    4.25000000000000000   1.00665255488118310
    4.31250000000000000   1.00218016998922366
    4.37500000000000000   0.99469603179328259
    4.43750000000000000   0.99972400115132243
    4.50000000000000000   0.98507032296728869
    4.56250000000000000   0.99033935740724255
    4.62500000000000000   1.00172353696853000
    4.68750000000000000   0.99010435508808270
    4.75000000000000000   0.99447859648565040
    4.81250000000000000   1.00636189972473900
    4.87500000000000000   0.99899480211347136
    4.93750000000000000   0.99879825644185449
  };
  \addplot[draw=none, domain=0.062500:4.937500, name path=B] {0};
  \addplot[blue!10] fill between[of=A and B];
  \addplot[help lines,dashed, domain=0:4.937500] {1};

\end{axis};
\end{tikzpicture}

%% file: res/spectrums/Final/radialOwen.tex
\begin{tikzpicture}
  \begin{axis}[
    width=10cm,
    height=5cm,
    ymin = 0.000000e+00,
    ymax = 4.200000e+00,
    xmin = 0.000000e+00,
    xmax=4.937500e+00,
    xtick align = outside,
  ]
  \addplot[blue, thick, name path=A] table {
    0.06250000000000000   0.00337148861143083
    0.12500000000000000   0.01858217097428586
    0.18750000000000000   0.05174712114039245
    0.25000000000000000   0.11286541314757630
    0.31250000000000000   0.19614504050725667
    0.37500000000000000   0.30057554922600543
    0.43750000000000000   0.42016086203908387
    0.50000000000000000   0.54154927830436062
    0.56250000000000000   0.65302753311713302
    0.62500000000000000   0.75725237027901460
    0.68750000000000000   0.84084408680366529
    0.75000000000000000   0.92015214616055707
    0.81250000000000000   0.93682685756190853
    0.87500000000000000   0.95501928580106310
    0.93750000000000000   0.97467183234080579
    1.00000000000000000   1.19356693115213486
    1.06250000000000000   0.98982063940640908
    1.12500000000000000   1.00759409804464761
    1.18750000000000000   1.00237373279367370
    1.25000000000000000   1.00829922844364273
    1.31250000000000000   0.99944766421643516
    1.37500000000000000   0.99676922071594964
    1.43750000000000000   0.99176686223260679
    1.50000000000000000   0.99705007862524375
    1.56250000000000000   0.99230351191693300
    1.62500000000000000   1.00032489054327756
    1.68750000000000000   1.00341694155926731
    1.75000000000000000   0.99951550007992862
    1.81250000000000000   0.99990942537591188
    1.87500000000000000   0.99726951951248610
    1.93750000000000000   0.99516688717159196
    2.00000000000000000   1.00769788420333106
    2.06250000000000000   0.99994757002775725
    2.12500000000000000   1.00028093234195614
    2.18750000000000000   1.00302327161977045
    2.25000000000000000   1.00719655374898998
    2.31250000000000000   0.99295162493870903
    2.37500000000000000   1.00272277284279165
    2.43750000000000000   1.00890330137647699
    2.50000000000000000   1.00045871527612018
    2.56250000000000000   1.00307699874102374
    2.62500000000000000   1.00461019527995332
    2.68750000000000000   0.99541049976817764
    2.75000000000000000   0.99325312101755736
    2.81250000000000000   1.00407510783144516
    2.87500000000000000   1.00307348510079586
    2.93750000000000000   0.99392852192742276
    3.00000000000000000   0.99235781559587377
    3.06250000000000000   1.00494492645096067
    3.12500000000000000   1.00026248243820515
    3.18750000000000000   0.99671561891247296
    3.25000000000000000   1.00100898723034937
    3.31250000000000000   0.99999197865999156
    3.37500000000000000   1.00477860890282811
    3.43750000000000000   0.99300001931138371
    3.50000000000000000   1.00143982924506969
    3.56250000000000000   1.00181842007074695
    3.62500000000000000   0.99872357329162254
    3.68750000000000000   1.00362481455914820
    3.75000000000000000   1.00295180340432233
    3.81250000000000000   0.99401929496482877
    3.87500000000000000   1.00242103296733420
    3.93750000000000000   1.00400127707467690
    4.00000000000000000   0.99770852649997743
    4.06250000000000000   0.99805611474464184
    4.12500000000000000   1.00102299971726727
    4.18750000000000000   1.00231789968004148
    4.25000000000000000   0.99558434594372125
    4.31250000000000000   0.99536573461276712
    4.37500000000000000   0.99801124869534630
    4.43750000000000000   1.00114534282546241
    4.50000000000000000   1.00007367187719720
    4.56250000000000000   1.00250592813935757
    4.62500000000000000   1.00223300581690222
    4.68750000000000000   1.00168595906868929
    4.75000000000000000   1.00305467777918245
    4.81250000000000000   0.99787494775481389
    4.87500000000000000   1.00008156171573526
    4.93750000000000000   0.99911383347645510
  };
  \addplot[draw=none, domain=0.062500:4.937500, name path=B] {0};
  \addplot[blue!10] fill between[of=A and B];
  \addplot[help lines,dashed, domain=0:4.937500] {1};

\end{axis};
\end{tikzpicture}

%% file: res/spectrums/Final/radialR1.tex
\begin{tikzpicture}
  \begin{axis}[
    width=10cm,
    height=5cm,
    ymin = 0.000000e+00,
    ymax = 4.200000e+00,
    xmin = 0.000000e+00,
    xmax=4.937500e+00,
    xtick align = outside,
  ]
  \addplot[blue, thick, name path=A] table {
    0.06250000000000000   0.00172616930271942
    0.12500000000000000   0.00201878333412632
    0.18750000000000000   0.00231435440968324
    0.25000000000000000   0.00284720303383969
    0.31250000000000000   0.00355109427031496
    0.37500000000000000   0.00473374868671113
    0.43750000000000000   0.00637784518453479
    0.50000000000000000   0.00877905163309644
    0.56250000000000000   0.01209059225447641
    0.62500000000000000   0.01714386505354283
    0.68750000000000000   0.02399821330195838
    0.75000000000000000   0.03760843870798115
    0.81250000000000000   0.05735805743792433
    0.87500000000000000   0.09115499397567096
    0.93750000000000000   0.15460994748005191
    1.00000000000000000   0.27372514380067120
    1.06250000000000000  12.81695453508812399
    1.12500000000000000   0.20451657960832920
    1.18750000000000000   0.13479684395822084
    1.25000000000000000   0.09660939498704632
    1.31250000000000000   0.08077195201254996
    1.37500000000000000   0.07470261157433966
    1.43750000000000000   0.07480659886949417
    1.50000000000000000   0.08166297346904455
    1.56250000000000000   0.09212305411767939
    1.62500000000000000   0.10639958616530988
    1.68750000000000000   0.13365300624547682
    1.75000000000000000   0.22318572269842821
    1.81250000000000000   4.83149959084364866
    1.87500000000000000   2.32938771004899703
    1.93750000000000000   0.26028103622845827
    2.00000000000000000   0.23230092605065622
    2.06250000000000000   0.37129420243726069
    2.12500000000000000   5.75693147844748498
    2.18750000000000000   0.33360335246950090
    2.25000000000000000   0.19823481056602829
    2.31250000000000000   0.15191704441904402
    2.37500000000000000   0.14115106495102767
    2.43750000000000000   0.14093068322687155
    2.50000000000000000   0.14840712456182259
    2.56250000000000000   0.16630273808318113
    2.62500000000000000   0.19983328281138005
    2.68750000000000000   0.29946334912063127
    2.75000000000000000   0.55171026410965052
    2.81250000000000000   7.83655023374490600
    2.87500000000000000   0.69738802161938551
    2.93750000000000000   0.33957717170945728
    3.00000000000000000   0.26177699686897221
    3.06250000000000000   0.26211513132271697
    3.12500000000000000   0.37085935674332715
    3.18750000000000000   2.60565422477895092
    3.25000000000000000   1.28853731072332023
    3.31250000000000000   0.27857196662446732
    3.37500000000000000   0.22426594020653171
    3.43750000000000000   0.21944779546304688
    3.50000000000000000   0.23555863567386087
    3.56250000000000000   0.29380201537933459
    3.62500000000000000   0.39524235295870114
    3.68750000000000000   2.18567807426884286
    3.75000000000000000   1.34648965059588499
    3.81250000000000000   3.11120029334400572
    3.87500000000000000   2.67869524543072179
    3.93750000000000000   0.56694056200307741
    4.00000000000000000   0.34415652719842593
    4.06250000000000000   0.29808891772149482
    4.12500000000000000   0.29908367241604450
    4.18750000000000000   0.40432005790384068
    4.25000000000000000   1.74252880390104314
    4.31250000000000000   1.05040971817701112
    4.37500000000000000   0.39245972063520129
    4.43750000000000000   0.32927610790209833
    4.50000000000000000   0.37329618496466616
    4.56250000000000000   0.49249285125289211
    4.62500000000000000   2.38385182359814651
    4.68750000000000000   2.05543900990154427
    4.75000000000000000   0.75591554227612345
    4.81250000000000000   1.19546072408612813
    4.87500000000000000   1.77243174994741337
    4.93750000000000000   1.90101281769096220
  };
  \addplot[draw=none, domain=0.062500:4.937500, name path=B] {0};
  \addplot[blue!10] fill between[of=A and B];
  \addplot[help lines,dashed, domain=0:4.937500] {1};

\end{axis};
\end{tikzpicture}

%% file: res/spectrums/DeepPointCorrelation/radialr1-2d.tex
\begin{tikzpicture}
  \begin{axis}[
    width=10cm,
    height=5cm,
    ymin = 0.000000e+00,
    ymax = 4.200000e+00,
    xmin = 0.000000e+00,
    xmax=4.937500e+00,
    xtick align = outside,
  ]
  \addplot[teal, thick, name path=A] table {
    0.06250000000000000   1.71390302169094300
    0.12500000000000000   0.77346104906250668
    0.18750000000000000   2.51855003425564750
    0.25000000000000000   1.21221158766404113
    0.31250000000000000   0.45476413996192150
    0.37500000000000000   0.56379797332422210
    0.43750000000000000   0.39003777892884572
    0.50000000000000000   0.30632989098722657
    0.56250000000000000   0.28420633341270113
    0.62500000000000000   0.25119290627434798
    0.68750000000000000   0.22590186301603801
    0.75000000000000000   0.20003960376273344
    0.81250000000000000   0.17485085143024831
    0.87500000000000000   0.15053531187365793
    0.93750000000000000   0.12586580973186029
    1.00000000000000000   0.15996319238272064
    1.06250000000000000   0.24863280949250402
    1.12500000000000000   0.34547980631577940
    1.18750000000000000   0.57164071627638913
    1.25000000000000000   0.98637922686525037
    1.31250000000000000   1.94713495960598371
    1.37500000000000000   4.21396497411176085
    1.43750000000000000   2.96317182501014864
    1.50000000000000000   1.80574229031153721
    1.56250000000000000   1.36265363622512625
    1.62500000000000000   1.09195110951197005
    1.68750000000000000   0.85002934544460973
    1.75000000000000000   0.71532598696928240
    1.81250000000000000   0.62146689683290568
    1.87500000000000000   0.53961066491219345
    1.93750000000000000   0.48277271957767237
    2.00000000000000000   0.44128605296268975
    2.06250000000000000   0.41556177721159676
    2.12500000000000000   0.41554040793262903
    2.18750000000000000   0.45584078584574023
    2.25000000000000000   0.56536627261442707
    2.31250000000000000   0.75782904982782195
    2.37500000000000000   1.00466809367095711
    2.43750000000000000   1.21565155493033683
    2.50000000000000000   1.31705223990884157
    2.56250000000000000   1.32842983367199707
    2.62500000000000000   1.32731927092728674
    2.68750000000000000   1.39161821300338562
    2.75000000000000000   1.54653839403058924
    2.81250000000000000   1.55728864132160094
    2.87500000000000000   1.36527409390875887
    2.93750000000000000   1.19024124162052813
    3.00000000000000000   1.04890459125404734
    3.06250000000000000   0.92870875734244707
    3.12500000000000000   0.82749354306065559
    3.18750000000000000   0.75170989883714145
    3.25000000000000000   0.68951212818179575
    3.31250000000000000   0.64238680192275444
    3.37500000000000000   0.61970269615249851
    3.43750000000000000   0.62251273753351655
    3.50000000000000000   0.66619277680609201
    3.56250000000000000   0.76043070437660676
    3.62500000000000000   0.90714850579526574
    3.68750000000000000   1.08076103000851220
    3.75000000000000000   1.23164846975393338
    3.81250000000000000   1.30662463118364780
    3.87500000000000000   1.30531424037315813
    3.93750000000000000   1.24740018866170632
    4.00000000000000000   1.18019640585536489
    4.06250000000000000   1.12028623354689105
    4.12500000000000000   1.09997292297802352
    4.18750000000000000   1.09896827324209578
    4.25000000000000000   1.07628242705757393
    4.31250000000000000   1.03346468396762980
    4.37500000000000000   0.98501667136179205
    4.43750000000000000   0.93597784158227315
    4.50000000000000000   0.88697952924381973
    4.56250000000000000   0.84760392291796016
    4.62500000000000000   0.82113905447387492
    4.68750000000000000   0.80969615468438694
    4.75000000000000000   0.82164516161300105
    4.81250000000000000   0.85204698066132745
    4.87500000000000000   0.90354741904759628
    4.93750000000000000   0.97102736752374963
  };
  \addplot[draw=none, domain=0.062500:4.937500, name path=B] {0};
  \addplot[teal!10] fill between[of=A and B];
  \addplot[help lines,dashed, domain=0:4.937500] {1};

\end{axis};
\end{tikzpicture}

%% file: res/spectrums/DeepPointCorrelation/radialr1-1d.tex
\begin{tikzpicture}
  \begin{axis}[
    width=10cm,
    height=5cm,
    ymin = 0.000000e+00,
    ymax = 4.200000e+00,
    xmin = 0.000000e+00,
    xmax=4.937500e+00,
    xtick align = outside,
  ]
  \addplot[blue, thick, name path=A] table {
    0.06250000000000000   0.19917158479240221
    0.12500000000000000   0.12817539033459660
    0.18750000000000000   0.22220359461189568
    0.25000000000000000   0.20614181219246938
    0.31250000000000000   0.19787969785003129
    0.37500000000000000   0.28575504779779559
    0.43750000000000000   2.86024900509719116
    0.50000000000000000   2.00951341028309205
    0.56250000000000000   0.53395831205115452
    0.62500000000000000   2.09103786299198635
    0.68750000000000000   0.39123671852159458
    0.75000000000000000   1.52885741472507108
    0.81250000000000000   0.56526033769360495
    0.87500000000000000   1.45529441578423513
    0.93750000000000000   1.49191269225969969
    1.00000000000000000   0.61609486305796979
    1.06250000000000000   2.33908503712888294
    1.12500000000000000   1.42561865912129115
    1.18750000000000000   0.59079835695241123
    1.25000000000000000   1.58610840372045558
    1.31250000000000000   0.76704511054282631
    1.37500000000000000   1.98167805352188031
    1.43750000000000000   0.51276909133898407
    1.50000000000000000   0.67649820154705043
    1.56250000000000000   1.58932397446069040
    1.62500000000000000   1.38206849004251486
    1.68750000000000000   0.99295015841731360
    1.75000000000000000   0.52720330449023978
    1.81250000000000000   1.59180912980360878
    1.87500000000000000   0.86989537555795515
    1.93750000000000000   1.57022153739559389
    2.00000000000000000   0.76861031076373654
    2.06250000000000000   0.83294021486352454
    2.12500000000000000   0.79252563020379085
    2.18750000000000000   1.00254429182246296
    2.25000000000000000   0.92194123922076587
    2.31250000000000000   0.86181863377287327
    2.37500000000000000   0.73157059255743839
    2.43750000000000000   0.91319794213309036
    2.50000000000000000   0.90174637056444773
    2.56250000000000000   0.86366017107324289
    2.62500000000000000   0.78809842395295582
    2.68750000000000000   0.98716827530125062
    2.75000000000000000   0.89295730040886812
    2.81250000000000000   0.82083885109291121
    2.87500000000000000   0.88448485185242753
    2.93750000000000000   0.89286003351882370
    3.00000000000000000   0.83852767281256113
    3.06250000000000000   0.84539701781569709
    3.12500000000000000   0.85306464440174423
    3.18750000000000000   0.87441102220333722
    3.25000000000000000   0.85682915541209470
    3.31250000000000000   0.85935289413434457
    3.37500000000000000   0.84014818154248272
    3.43750000000000000   0.86175829660031555
    3.50000000000000000   0.88101554033061447
    3.56250000000000000   0.87509598402909128
    3.62500000000000000   0.86929298430025692
    3.68750000000000000   0.87822698440132896
    3.75000000000000000   0.90814736520824235
    3.81250000000000000   0.90034366798755794
    3.87500000000000000   0.90648341889293449
    3.93750000000000000   0.88204548319725173
    4.00000000000000000   0.91789482531946232
    4.06250000000000000   0.92164445194840316
    4.12500000000000000   0.93256694331387791
    4.18750000000000000   0.92989833532768973
    4.25000000000000000   0.94276493027755648
    4.31250000000000000   0.95183048363356293
    4.37500000000000000   0.96336722448742085
    4.43750000000000000   0.97202676899068863
    4.50000000000000000   0.98021815021382541
    4.56250000000000000   0.98590754299626415
    4.62500000000000000   0.99343656276596570
    4.68750000000000000   0.99743600981083647
    4.75000000000000000   1.00293312354585784
    4.81250000000000000   1.01172610711310385
    4.87500000000000000   1.02329957566684682
    4.93750000000000000   1.01130934964957220
  };
  \addplot[draw=none, domain=0.062500:4.937500, name path=B] {0};
  \addplot[blue!10] fill between[of=A and B];
  \addplot[help lines,dashed, domain=0:4.937500] {1};

\end{axis};
\end{tikzpicture}